\documentclass[bibyear]{aa}
\usepackage[utf8]{inputenc}
\usepackage{booktabs}
\usepackage{rotating}
\usepackage{float}
\usepackage[caption = false]{subfig}
\usepackage{graphicx}
\usepackage{natbib}
\usepackage{txfonts}
\usepackage{amsmath}
\usepackage{amsfonts}
\usepackage{amssymb}
\bibpunct{(}{)}{;}{a}{}{,}
\usepackage[english]{babel}
\usepackage{longtable}
\usepackage[dvipsnames]{xcolor}
\usepackage{hyperref}
\usepackage{booktabs}
\colorlet{myurlcolor}{violet}
\colorlet{myallcolor}{MidnightBlue}
\hypersetup{
  linkcolor  = myallcolor,
  citecolor  = myallcolor,
  urlcolor   = myurlcolor,
  colorlinks = true,
}

\begin{document}
\authorrunning{Karamehmetoglu et al.}
\titlerunning{A population of Type Ibc supernovae with massive progenitors}
\title{A population of Type Ibc supernovae with massive progenitors}
\subtitle{broad lightcurves not uncommon in (i)PTF}

\author{
E. Karamehmetoglu\inst{1,2}\orcid{0000-0001-6209-838X}
\and J.~Sollerman\inst{2}\orcid{0000-0003-1546-6615}
\and F.~Taddia \inst{1,2}\orcid{0000-0002-2387-6801}
\and C.~Barbarino \inst{2}
\and U.~Feindt \inst{3}
\and C.~Fremling \inst{4} \orcid{0000-0002-4223-103X}
\and A.~Gal-Yam \inst{5}\orcid{0000-0002-3653-5598}
\and M.~M. Kasliwal \inst{4}\orcid{0000-0002-5619-4938}
\and T.~Petrushevska \inst{6}
\and S.~Schulze \inst{3}\orcid{0000-0001-6797-1889}
\and M.~D. Stritzinger\inst{1}\orcid{0000-0002-5571-1833}
\and E.~Zapartas \inst{7}\orcid{0000-0002-7464-498X}
}

\institute{
  Department of Physics and Astronomy, Aarhus University, Ny Munkegade 120, DK-8000 Aarhus C, Denmark\\ (\email{emir.k@phys.au.dk}) 
\and
Department of Astronomy, The Oskar Klein Centre, Stockholm University, AlbaNova, 106 91 Stockholm, Sweden.
\and Department of Physics, The Oskar Klein Centre, Stockholm University, AlbaNova, 106 91 Stockholm, Sweden. 
\and Division of Physics, Mathematics, and Astronomy, California Institute of Technology, Pasadena, CA 91125, USA 
\and Department of Particle Physics and Astrophysics, Weizmann Institute of Science, 76100 Rehovot, Israel 
\and Centre for Astrophysics and Cosmology, University of Nova Gorica, Vipavska 11c, 5270 Ajdov\v{s}\u{c}ina, Slovenia 
\and IAASARS, National Observatory of Athens, Vas. Pavlou and I. Metaxa, Penteli, 15236, Greece. 
}
 \date{Received date; Accepted date } 

\abstract{If high-mass stars ($\gtrsim \text{20--25}~M_\sun$) are the progenitors of stripped-envelope (SE) supernovae (SNe), their massive ejecta should lead to broad, long-duration lightcurves. Instead, literature samples of SE~SNe have reported relatively narrow lightcurves corresponding to ejecta masses between $\text{1--4}~M_\sun$ that favor intermediate-mass progenitors ($\lesssim \text{20--25}~M_\sun$). Working with an untargeted sample from a single telescope to better constrain their rates, we searched the Palomar Transient Factory (PTF) and intermediate-PTF (iPTF) sample of SNe for SE~SNe with broad lightcurves. Using a simple observational marker of $g$- or $r$-band lightcurve stretch compared to a template to measure broadness, we identified eight significantly broader Type~Ibc SNe after applying quantitative sample selection criteria. The lightcurves, broad-band colors, and spectra of these SNe are found to evolve more slowly relative to typical Type~Ibc SNe, proportional with the stretch parameter. Bolometric lightcurve modeling and their nebular spectra indicate high ejecta masses and nickel masses, assuming radioactive decay powering. Additionally, these objects are preferentially located in low-metallicity host galaxies with high star-formation rates, which may account for their massive progenitors, as well as their relative absence from the literature. Our study thus supports the link between broad lightcurves (as measured by stretch) and high-mass progenitor stars in SE~SNe with independent evidence from bolometric lightcurve modeling, nebular spectra, host environment properties, and photometric evolution. 

In the first systematic search of its kind using an untargeted sample, we use the stretch distribution to identify a higher than previously appreciated 
fraction of SE~SNe with broad lightcurves (${\sim}13\%$). Correcting for Malmquist and lightcurve duration observational biases, we conservatively estimate that a minimum of ${\sim} 6\%$ of SE~SNe are consistent with high-mass progenitors. This result has implications for the progenitor channels of SE~SNe including late stages of massive stellar evolution, the oxygen fraction in the universe, and the formation channels for stellar-mass black holes.}


\maketitle

\section{Introduction}

Core-collapse (CC) of massive stars, partially or fully stripped of their hydrogen and/or helium envelopes, are thought to give rise to Type~Ibc, and Type~IIb supernova (SN) explosions. Type~Ib SNe lack hydrogen, Type Ic SNe lack hydrogen and helium, while Type~IIb SNe first show hydrogen, then later transition to show mostly helium in their spectra. These observational classes, based on the optical spectra (see \citealp{Filippenko1997} and \citealp{GalYam2017}), along with the rarer broad-lined Type~Ic (Ic-BL) SNe, often related to gamma-ray bursts, are jointly referred to as stripped-envelope (SE) SNe\footnote{This definition excludes the exotic hydrogen-poor superluminous SNe (SLSNe-I), which can be observationally distinguished from SE~SNe \citep{GalYam2019}.}.

The nature of the progenitors of SE~SNe has long been debated. The favored progenitor scenario of Type~Ibc SNe has traditionally been solitary massive Wolf-Rayet (WR) stars stripped of their hydrogen and/or helium envelopes by strong line-driven winds \citep[e.g.,][]{Conti1975}. Massive WR stars need to have $M_{ZAMS} \gtrsim \text{20--25}~M_\sun$ to form a SE~SN \citep{Smith2011,Smith2014,Smartt2015}, where $M_{ZAMS}$ is the zero-age main-sequence mass of the progenitor star. Such massive stripped stars, if they explode as SE~SNe, should be recognized by their broad lightcurves \citep[e.g.,][]{Yoon2015}. They should stay luminous over a longer time period as the energy slowly diffuses out of their more massive ejecta, as compared to a less massive progenitor\footnote{Assuming successful CC SN remnant masses are not too dissimilar.}. In a simple radioactive decay powered diffusion lightcurve \citep{Arnett1982}, both higher mass or lower velocity ejecta lead to broader bolometric lightcurves. For example, Type Ic-BL SNe are associated with higher ejecta masses (and more massive stars) than normal Type Ibc SNe \citep{Taddia2019b}, but their bolometric lightcurves are only slightly broader, if at all, as a consequence of their high ejecta velocities. At relatively similar ejecta velocities, however, broader lightcurves are expected for more massive progenitors\footnote{For stripped stars, progenitor sizes and density profiles are not expected to be hugely different even between increasingly more massive progenitors, hence the expectation of a straightforward relationship between mass and lightcurve broadness.}. 

Observations of the lightcurves of SE~SNe reveal mainly narrower peaked lightcurves, compatible with lower ejecta masses \citep[e.g.,][]{Drout2011,Cano2013,Taddia2015,Lyman2016,Prentice2016a,Prentice2018,Taddia2018d,Barbarino2021}. This suggests that the progenitors of SE~SNe are not as massive as initially thought, and that they could rather come from relatively lower-mass stars ($M_{ZAMS}$  $\lesssim \text{20--25}$~$M_\sun$) stripped of their envelope by a binary companion. This result from lightcurve studies is also consistent with direct detection searches of SE~SN progenitors, which have failed to find any evidence for high-mass stars (\citealp{Eldridge2013,Smartt2015}, but see \citealp{VanDyk2018}), as well as with the relatively high rates of SE~SNe (${\sim}30\%$ of CC SNe; see e.g., \citealp{Smith2011,Smith2014,Shivvers2017,Graur2017}). Further evidence for lower-mass progenitors has been found in studies of nebular spectra, which can probe the pre-explosion mass when compared to model spectra \citep[e.g.,][]{Fransson1987,Jerkstrand2013,Jerkstrand2015}. The combined results of the above mentioned studies have been interpreted to favor the binary progenitor scenario for most SE~SNe, since binary Roche-Lobe overflow can strip the envelope of stars with 
$M_{ZAMS}$ $\lesssim \text{20--25}$~$M_\sun$ before core collapse, while steady-state stellar winds cannot \citep{Smith2014,Yoon2017a,Beasor2022}.

Meanwhile, studies of hydrogen-rich Type II SNe, which make up ${\sim} \text{60--70}\%$ of CC SNe \citep{Shivvers2017}, have found them to be associated with red super-giant progenitors that have $M_{ZAMS} \lesssim \text{20--25}$~$M_\sun$ \citep{Smartt2009}. This combined observational evidence from Type II and SE~SNe was noticed by \citet{Smartt2015}, who made a case that there is a surprising lack of high-mass stars in the SN record. Based just on progenitor direct detection searches, \citet{Smartt2015} estimated the probability for this being a chance occurrence to be small, and argued that observational biases did not seem likely to account for the missing stars.

While a majority of SE~SNe seem to have relatively lower mass progenitors (located in interacting binaries), it has been difficult to establish just what fraction of SE~SNe could have more massive progenitors. At the time \citet{Smartt2015} made his case for the missing high-mass stars, only a few Type Ibc SE~SNe with broad lightcurves, and thus potentially arising from massive stars, had been studied in detail in the literature: e.g., Type Ib SN~2005bf \citep{Folatelli2006} and Type Ic SN~2011bm \citep{Valenti2012}. Since then, several other objects have been added to this list\footnote{While this is not an exhaustive list, these SNe were studied with a particular attention paid to their broad lightcurves and potentially massive star origins.}, including Type Ic SN iPTF15dtg \citep{Taddia2016,Taddia2019a}, Type Ic SN iPTF11mnb \citep{Taddia2018e}, the first broad Type IIb SN~2013bb \citep{Prentice2018}, Type Ib SN~2016coi \citep[][although it may be related to relativistic Type Ic-BL SNe]{Terreran2019}, and the Type Ib SN LSQ13abf \citep{Stritzinger2020}. In the sample papers of \citet{Lyman2016} and \citet{Prentice2018}, these above-mentioned broad lightcurve SE SNe make up only one and two events out of samples sizes of ${\sim}30$ and ${\sim}50$, respectively. Thus, the fraction of SE~SNe with a high-mass origin has not been meaningfully constrained so far. Moreover, literature samples have often been drawn from targeted surveys that can be biased, and until recently the field suffered from a lack of high quality lightcurves and spectra for a large number of SNe from an untargeted sample. 

The Palomar Transient Factory \citep[PTF;][]{Rau2009,Law2009} and its successor the intermediate-PTF (iPTF) were untargeted surveys conducted on largely the same instrument, which combined have spectroscopically classified over 200 SE~SNe. This large and untargeted sample offers an opportunity to investigate the fraction of SE~SNe with high-mass progenitors. In this paper, we perform template fits to the lightcurves of this large sample and identify several broad candidates (six Type Ic, two Type Ib, and six Type IIb SNe). 

The goals of this paper are to present all of the SE~SNe with broad lightcurves that have been found in the combined PTF and iPTF databases, study the properties of this sample, investigate the link between lightcurve broadness and massive progenitors, and to ultimately estimate the fraction of SE~SNe with broad lightcurves that are potentially coming from massive stars. The spectra of the full sample of (i)PTF SE SNe were presented in \citet{Fremling2018}, while the lightcurves of the Type Ic sample were presented in \citet{Barbarino2021}. We present new photometry and spectroscopy of the eight broad lightcurve Type Ibc SNe PTF09dfk, PTF10inj, PTF11bov (SN 2011bm), PTF11mnb, iPTF11rka, iPTF15dtg, iPTF16flq, and iPTF16hgp. Note that iPTF11bov (SN 2011bm) was studied by \citet{Valenti2012}, but we have new iPTF data for the same object, in addition to what was already published by \citet{Taddia2016}. iPTF15dtg was previously studied by \citet{Taddia2016,Taddia2019a}, PTF11mnb by \citet{Taddia2018e}, and 
PTF11rka by \citet{Pian2020}. 

The paper is organized as follows: in Sect.~\ref{sec:thesample} we outline the sample selection steps used to find the SNe with broad lightcurves starting from the initial 220 SE~SNe. Detailed description of each step of the procedure can be found in Appendix~\ref{sec:sampleselection}. We also describe the selection of the final uniform sample of Type Ibc SNe with broad lightcurves, while highlighting the possibility of a similar investigation into Type IIb SNe. In the following sections, this sample of 8 broad Type Ibc SNe is studied in detail. In Sect.~\ref{sec:data}, photometry and lightcurve fits are presented and colors are compared to similar data of SE SNe in the literature, as well as used to estimate host extinction. The spectra are presented in Sect.~\ref{sec:spec}. We measure line velocities from photospheric phase spectra and calculate line fluxes in nebular phase spectra. In Sect. \ref{sec:model} we construct pseudo-bolometric lightcurves of these transients and obtain explosion parameters using lightcurve models. In Sect. \ref{sec:hosts} we study the environments of these particular SNe and estimate their metallicities. We investigate the fraction of SE~SNe in (i)PTF with broad lightcurves in Sect. \ref{sec:biases} and attempt to correct for observational biases. We also demonstrate the promise of our method by applying it to Type IIb SNe (Appendix~\ref{sec:IIb}) in order to broaden the discussion from Type Ibc to all SE~SNe, but a detailed study of the Type IIb SN sample is beyond the scope of this work. Finally, our results are discussed in Sect.~\ref{sec:discussion}. As we have applied some methodology that has not been widely utilized in the field, this paper makes extensive use of Appendices to explain details of said methods. 

\section{The sample \label{sec:thesample}}
The (i)PTF (combined PTF and iPTF) SN sample, observed from the start of PTF in 2009 to the end of iPTF in 2017, numbers 897 CC SNe among which we identified 220 spectroscopically classified SE~SNe\footnote{Classification differences with the comprehensive (i)PTF sample of \citet{Schulze2021} are enumerated in Appendix~\ref{sec:SNCLASSIFY}.} (SNe of either Type Ib, Ic, Ib/c, or Ic-BL). Using this large database of SNe, we attempt to find SE~SNe with broad lightcurves as measured by fitting a stretched template to their photometry. By using a reproducible statistical approach, we proceed to place a meaningful quantitative constraint on the fraction of broad lightcurve SE~SNe. 

In order to draw from a uniform sample we first focus on the Type Ibc\footnote{Type~Ibc refers to Type Ib and Ic together, while Type Ib/c refers to those SNe where a secure spectral identification as either type was not possible.} SNe, whose lightcurves and spectra behave very similarly, ignoring the Type IIb SNe for now. Since the spectra of Type Ib, Ic, and Ib/c SNe are very similar, a stretched lightcurve template in a single band is likely to capture a real difference in lightcurve energetics independent of differences due to spectral lines. Based on this logic, the final selection of broad Type Ibc SNe are studied in detail to investigate their nature and confirm our \textit{ansatz} that SE~SNe coming from high-mass stars should show broader than average lightcurves.

After demonstrating our method on Type Ibc SNe, we return to Type IIb SNe in Appendix~\ref{sec:IIb}, since they also represent a large minority of SE~SNe. For Type IIb SNe, we are only interested in the statistical properties of the sample combined with Type Ibc SNe. The last remaining major sub-type, Type Ic-BL SNe, only make up ${\sim}3\%$ of CC SNe, and the (i)PTF sample of them was studied in detail by \citet{Taddia2019b}. We exclude them due to their rarity, relativistic ejecta, and association of some with highly energetic gamma-ray bursts. We thus limited our sample to SE~SNe classified as typical SNe Ibc using their spectra. 

\subsection{Data collection}
The (i)PTF SE~SN sample is comprised of 58 Type IIb, 45 Type Ib, 62 Type Ic, 18 Ib/c, and 37 Type Ic-BL SNe (see Appendix~\ref{sec:SNCLASSIFY}). Their full lightcurves were constructed in $g$ and $r$ band using the PTF survey data from the Palomar 48-inch (P48) survey telescope as well as data from other telescopes such as the Palomar 60-inch (P60), the Nordic Optical Telescope (NOT), and the Las Cumbres Observatory (LCO) global telescope network \citep{Brown2013}. The main survey telescope for (i)PTF, P48, primarily observed in the Mould $R$ band with a typical average cadence of 2 observations every 3 nights. P48 data were calibrated to SDSS \citep[Sloan Digital Sky Survey;][]{York2000} stars and filters (specifically $g$ and $r$), and the Mould $R$-band filter of (i)PTF is similar but somewhat redder than the SDSS $r$ band according to \citet{Ofek2012}. 

The spectral classification and monitoring was done by the (i)PTF survey collaboration using telescopes such as the P60 and Palomar 200-inch on Palomar Mountain, the Keck and Gemini on Hawaii, as well as the NOT and Telescopio Nazionale Galileo (TNG) on Canary Islands, among others \citep[see][for a full list]{Fremling2018}. The spectra were all reduced using standard procedures with pipelines developed for their respective instruments and uploaded to a central marshal by (i)PTF team members.

\subsection{Selecting the Broad Sample}

\begin{figure}[!ht]
    \centering
    \includegraphics[width=\linewidth]{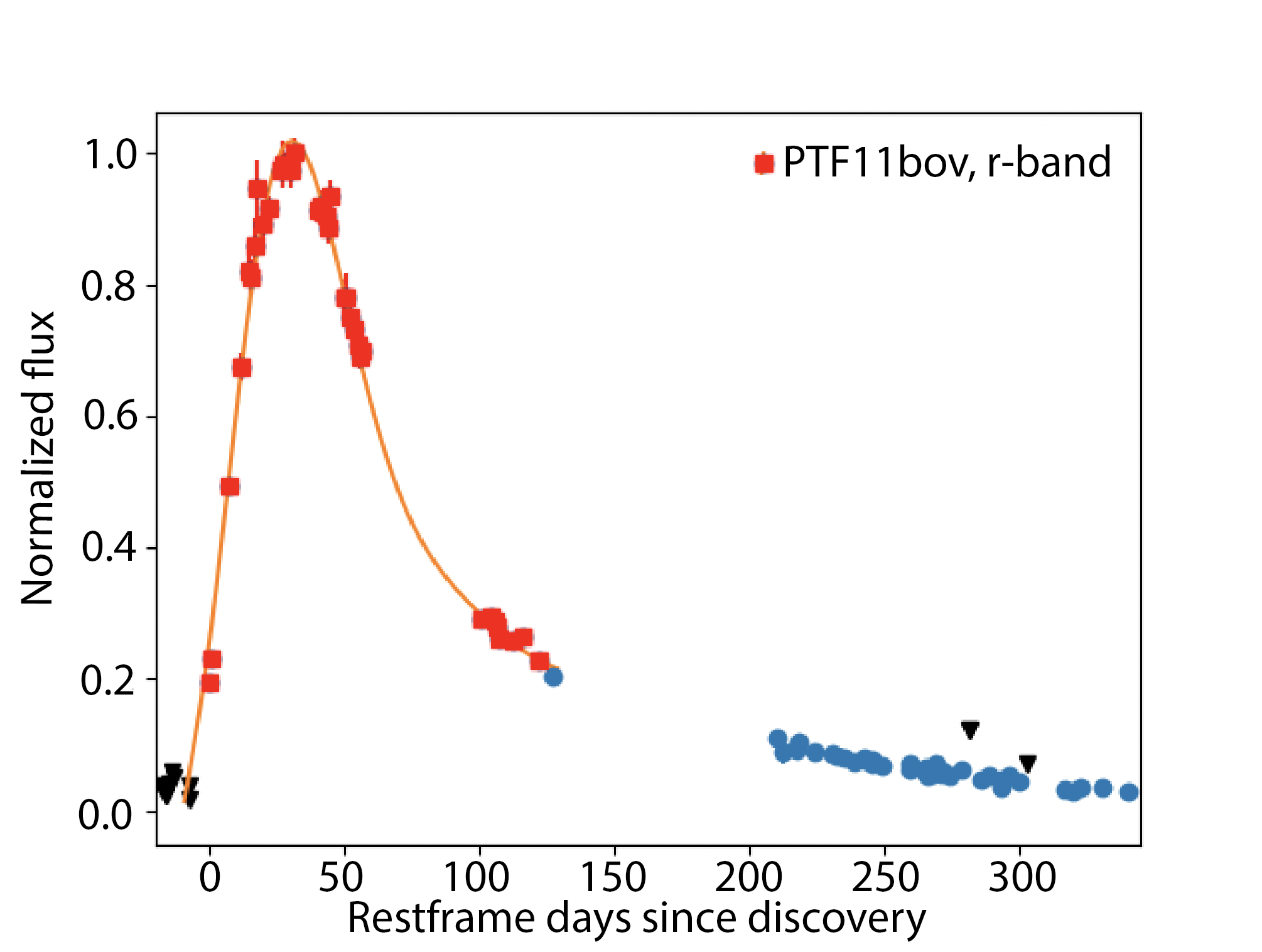}\\
    \includegraphics[width=\linewidth]{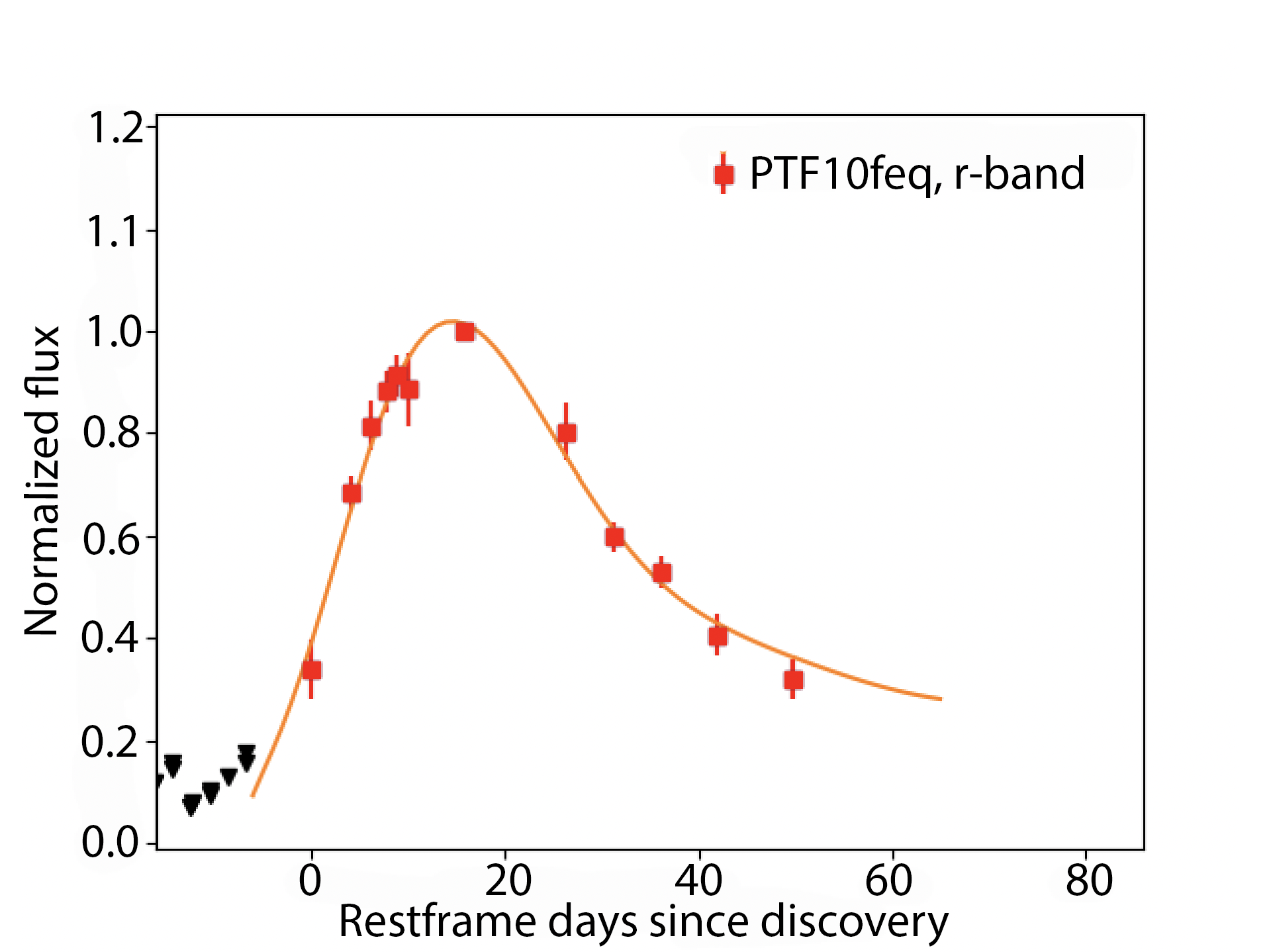}
    \caption{Example stretched-template fits to lightcurves of PTF11bov: a broad SE~SN, and PTF10feq: an ordinary SE~SN. Red squares are included in the fitting, black triangles indicate upper limits, and the orange line shows the fitted stretched-template lightcurve, with stretch of 2.04 and 1.05, respectively. The lightcurves are plotted normalized to their brightest epoch.} 
    \label{fig:ex_temp_fits}
\end{figure}
We draw our samples of Type Ibc and Type IIb SNe from the (i)PTF surveys following the methodology described in Appendix~\ref{sec:sampleselection}, but which we also briefly outline below. To find our sample of SE~SNe with broad lightcurves, we first constructed a well observed sub-sample of SE~SNe. The various cuts used to select the final broad SE~SN sample are listed in Table~\ref{tab:cuts}. Eight SNe were removed due to poor photometry. 

\begin{figure*}
    \centering
    \includegraphics[width=\linewidth]{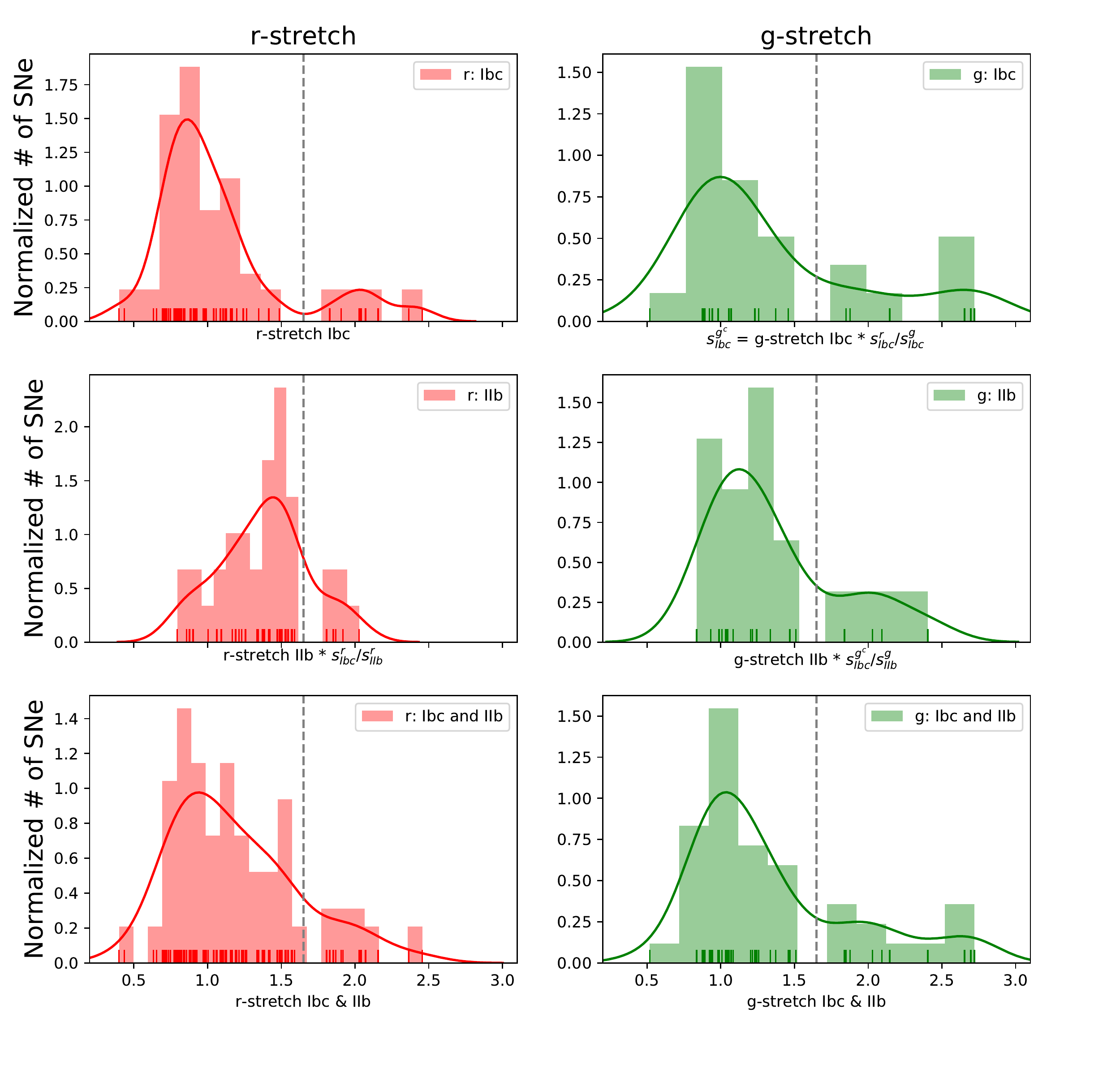}
    \caption{\textit{Top:} Type Ibc SNe: normalized histogram of stretch parameter measured in $r$ (left) and $g$ (right panels) bands. A kernel density estimate using a Gaussian kernel, which shows an estimate of the continuous density, has been over-plotted as a line of matching color. Each measurement is also indicated with a tick mark. All values have been corrected to the space of Type Ibc $r$-band template stretch values by deriving a stretch correction factor to the Type Ibc $r$-band template by fitting the other templates (see text). Stretch = 1.65 is marked with a grey dashed-line. Plotted normalized with respect to the total counts and bin width. Results from Appendix~\ref{sec:IIb} have been added to construct the same for \textit{Middle:} Type IIb SNe, \textit{Bottom:} combined Type Ibc and IIb SNe.} 
    \label{fig:hist_norm_stretch}
\end{figure*}

We constructed our sample using reproducible and numerical criteria, from absolute magnitude lightcurves, using a template fitting approach. Two example template fits are shown in Fig.~\ref{fig:ex_temp_fits}, for a broader and a typical width lightcurve, respectively. We performed the template fitting process for the (i)PTF sample with visual verification of the fits to obtain the distribution of lightcurve broadness in $g$ and $r$ bands, measured as the stretch parameter. The resulting distributions of the stretch parameter are plotted in Fig.~\ref{fig:hist_norm_stretch}.

The distribution of stretch values for Type Ibc SNe seem to indicate the presence of two populations. The majority of SE~SNe cluster around a stretch value of 1.0, with a secondary distribution located at high broadness values ($\gtrsim1.5$ in stretch value). We tested this apparent bimodality using two approaches, Gaussian Mixture Models (GMM) and K-Means clustering. Using the Bayesian information criteria (BIC) to evaluate our GMM fits, exactly two clusters are most preferred, and K-Means algorithm exactly picks out the same two clusters we identify by eye as the most tightly bound clustering possible. After statistically verifying its existence, we used the results of these tests to derive our final broad sample. Using a hard stretch cut-off of 1.65 to bisect the two populations, all SNe with stretch $\geq$ 1.65 are labeled as broad, while the remaining are labeled as ordinary. Within one sigma standard error in stretch, all SNe belong to one of the two groups (none cross the boundary). The details of each step and further statistical tests can be found in Appendix~\ref{sec:sampleselection}.

\begin{table}
    \centering
    \begin{tabular}{c|ccc}
        Selection               &   \# of SNe remaining & &  \\
        \hline
        (i)PTF CC-SN sample     &   897 & &\\
        SE~SNe                  &   220 & Ibc & IIb\\
        \hline
        Only Type Ibc/IIb       &   183 & 125 & 58\\ 
        Cut bad photometry   &   174 & 118 & 56 \\ 
        Cut no peak             &   114 & 73 & 41\\
        Cut on templatability   &   107 & 68 & 39 \\
        \hline 
        $r$ band (total)          &   98 & 62 & 36 \\
        $g$ band (total)          &   42 & 24 & 18 \\
        $g$ band (only)           &   9  & 6  & 3  \\
        \hline
        $S_r >1.65$ & 13 & 8 & 5 \\
        $S_g >1.65$ & 10 & 6 & 4 \\ 
        $S_{g,r} >1.65$ & 14 & 8 & 6 \\
        \hline
    \end{tabular}
    \caption{The number of SNe left after each step of sample selection. A stretch cut of $1.65$ normalized to the $r$-band Type Ibc template (Sect.~\ref{sec:sampleselect}) is used. $S_{g,r}$ meaning that we have a stretch value in $g$ or $r$. \label{tab:cuts}. Here, SE~SNe means only Type Ib, Type Ic, Type Ib/c, Type IIb, or Type Ic-BL.}
\end{table}

\subsubsection{The broad sample}
Based on the above cut-off and classification, there are 8 Type Ibc SNe in the broad sample out of a total of 68 Type Ibc SNe. The fraction of broad Type Ibc SNe in the (i)PTF is thus $8/68\approx12^{+4}_{-4}\%$ when considering both bands. Individually, the $r$ band shows $8/62\approx13^{+5}_{-4}\%$ and the $g$ band shows $6/24\approx25^{+6}_{-7}\%$. The uncertainties are $90\%$ Poissonian confidence intervals. These fractions are subject to caveats and observational biases, which we tackle in Sect.~\ref{sec:biases}.

\begin{table*}
    \centering
    \small
    \setlength\tabcolsep{3.0pt}
\begin{tabular}{llllllllllll}
\toprule
SN & Type &         RA &       Dec &      z &     DM & E($B-V$)$_\text{MW}$ & E($B-V$)$_\text{host}$ &   Max Epoch &  Exp. Epoch &  $s_r$ & $s_g$  \\
	&	  &	(deg)		   &	(deg)	   &		& (mag)	 & (mag)			  & (mag)		  		 &		(JD)   &  (JD) & & \\
\midrule
 PTF09dfk &   Ib &  347.30593 &   7.80429 &  0.016 &  34.20 &      0.0458 &  $0.28 \pm 0.07$ &  2455096.39 &  $2455053.8 \pm 1.3$ &  $2.36 \pm 0.08$ &               -- \\
 PTF10inj &   Ib &  238.73781 &  53.77195 &  0.064 &  37.34 &      0.0100 &  $0.20 \pm 0.12$ &  2455378.22 &  $2455339.9 \pm 0.5$ &  $1.91 \pm 0.06$ &  $2.17 \pm 0.04$ \\
 PTF11bov &   Ic &  194.22475 &  22.37448 &  0.022 &  34.90 &      0.0289 &  $0.06 \pm 0.02$ &  2455682.97 &  $2455645.0 \pm 0.5$ &  $2.04 \pm 0.05$ &  $2.23 \pm 0.08$ \\
 PTF11mnb &   Ic &    8.55522 &   2.80873 &  0.060 &  37.15 &      0.0157 &  $0.06 \pm 0.01$ &  2455860.18 &  $2455802.2 \pm 1.2$ &  $2.46 \pm 0.16$ &  $1.54 \pm 0.12$ \\
 PTF11rka &   Ic &  190.18696 &  12.88927 &  0.074 &  37.63 &      0.0295 &  $0.21 \pm 0.05$ &  2455932.87 &  $2455896.0 \pm 8.8$ &  $2.16 \pm 0.09$ &               -- \\
iPTF15dtg &   Ic &   37.58355 &  37.23519 &  0.052 &  36.84 &      0.0549 &        ${\sim0}$ &  2457368.64 &  $2457330.2 \pm 0.4$ &  $2.07 \pm 0.01$ &  $1.76 \pm 0.16$ \\
iPTF16flq &   Ic &    7.15223 &  -1.55092 &  0.060 &  37.14 &      0.0190 &        ${\sim0}$ &  2457647.67 &  $2457616.1 \pm 2.2$ &  $2.03 \pm 0.10$ &  $2.21 \pm 0.17$ \\
iPTF16hgp &   Ic &    3.02672 &  32.19747 &  0.081 &  37.77 &      0.0382 &  $0.06 \pm 0.01$ &  2457710.99 &  $2457679.7 \pm 0.5$ &  $1.83 \pm 0.12$ &  $1.51 \pm 0.05$ \\
\bottomrule
\end{tabular}
    \caption{The Type Ibc broad sample. Epoch of maximum light is estimated from $r$-band template fits. Explosion epoch is estimated in Sect.~\ref{sec:bololc} via Arnett model fits. The stretch parameter in each band, i.e. the broadness of the lightcurve compared to a template, is also reported as measured in Sect. \protect\ref{sec:LCfit_init}. The host extinction estimate is derived in Sect.~\ref{sec:colors}, and is not given when consistent with zero.}
    \label{tab:SNINFO}
\end{table*}

\section{Analysis of 8 broad-lightcurve SE~SNe \label{sec:data}}
In the rest of the analysis sections, we focus on the sample of 8 broad SE~SNe selected using single band-stretch (Sect. \ref{sec:thesample}), which are tabulated in Table~\ref{tab:SNINFO}. We return to the question of the fraction of broad SE~SNe in Sect. \ref{sec:biases}. Individual figures displaying the observed photometry, spectroscopy, as well as the lightcurve interpolation of each SN can be found in Appendix~\ref{sec:supportingfigures}. Unless noted otherwise, phases in the rest of the paper are given with respect to epoch of $r$-band maximum. 

\subsection{Photometry of the broad sample}
We present photometry in $u,B,g,r,i,z$ bands taken with the 48-inch Samuel Oschin telescope \citep[P48;][]{Rahmer2008}, the Palomar 60-inch telescope \citep[P60;][]{Cenko2006}, both located at Palomar Observatory in California, and with the NOT \citep{Djupvik2009}. 

The P48 data were reduced using PTFIDE \citep{masci2017} with template subtraction. The P60 and NOT data were reduced with FPipe \citep{Fremling2016} also using template subtraction. The observed lightcurves are plotted since discovery in Fig.~\ref{fig:apparentmags}. The photometry from the P48 telescope was sometimes shifted by a small constant in order to match the lightcurve from the P60\footnote{This was also done for the ordinary (i)PTF SE~SNe when there was an obvious shift. However, it was only necessary for a few SNe.}, since the telescopes do not use exactly the same filters. Since the color information comes from the multiband photometry of the P60 telescope, this does not significantly impact colors. 

\begin{figure*}[h]
 \includegraphics[width=\linewidth]{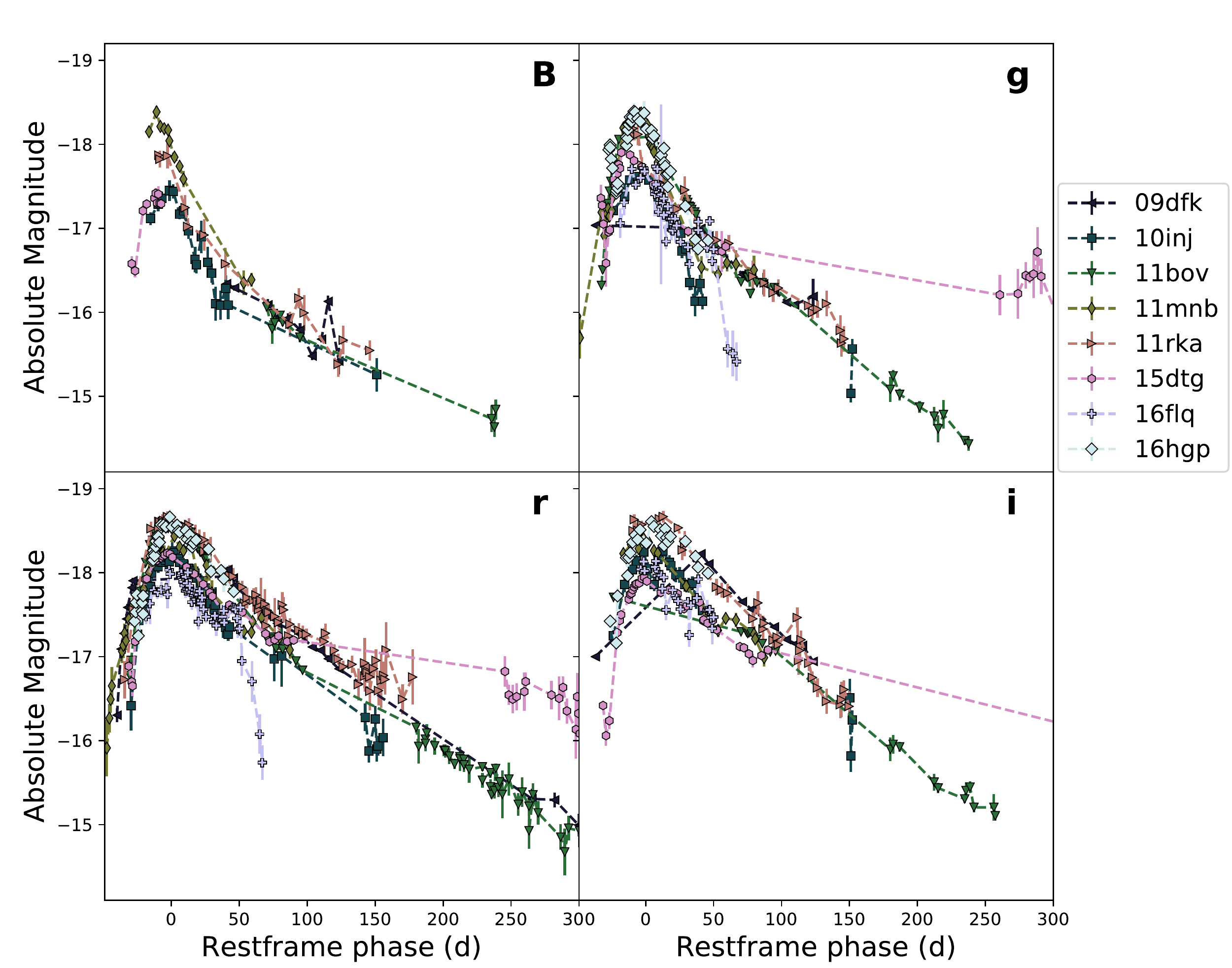}
    \caption{Absolute magnitude  $Bgri$-band lightcurves for our sample. Note that where relevant, the plotted color for each SN is preserved in subsequent figures.}
    \label{fig:absmag}
\end{figure*}

Absolute magnitude photometry for the broad SNe is calculated in the same way as for the full (i)PTF SE~SN sample (Appendix~\ref{sec:absmag0}). In addition, K-corrections calculated from the spectral sequence were applied as detailed in Sect.~\ref{sec:Kcorr}. The absolute magnitudes in $Bgri$ for the broad SN sample are plotted in Fig.~\ref{fig:absmag}.

Extinction due to host is derived in Sect.~\ref{sec:colors} by comparison to intrinsic color templates presented in \citet{Stritzinger2018b}. The E(B$-$V)$_\text{host}$ for the SNe where we detect a color-excess are listed in Table \ref{tab:SNINFO}. Since the estimated extinction is not high, and since the uncertainties are large (Sect.~\ref{sec:colors}), we do not correct for this host extinction in our analysis. Our approach is supported by the finding in \citet{Taddia2016,Taddia2018e} that PTF11mnb and iPTF15dtg do not show significant reddening as indicated by a lack of narrow Na~ID lines at the redshift of the host galaxy, which is typically used as an indicator of significant host extinction.

\begin{figure}
\includegraphics[width=\linewidth]{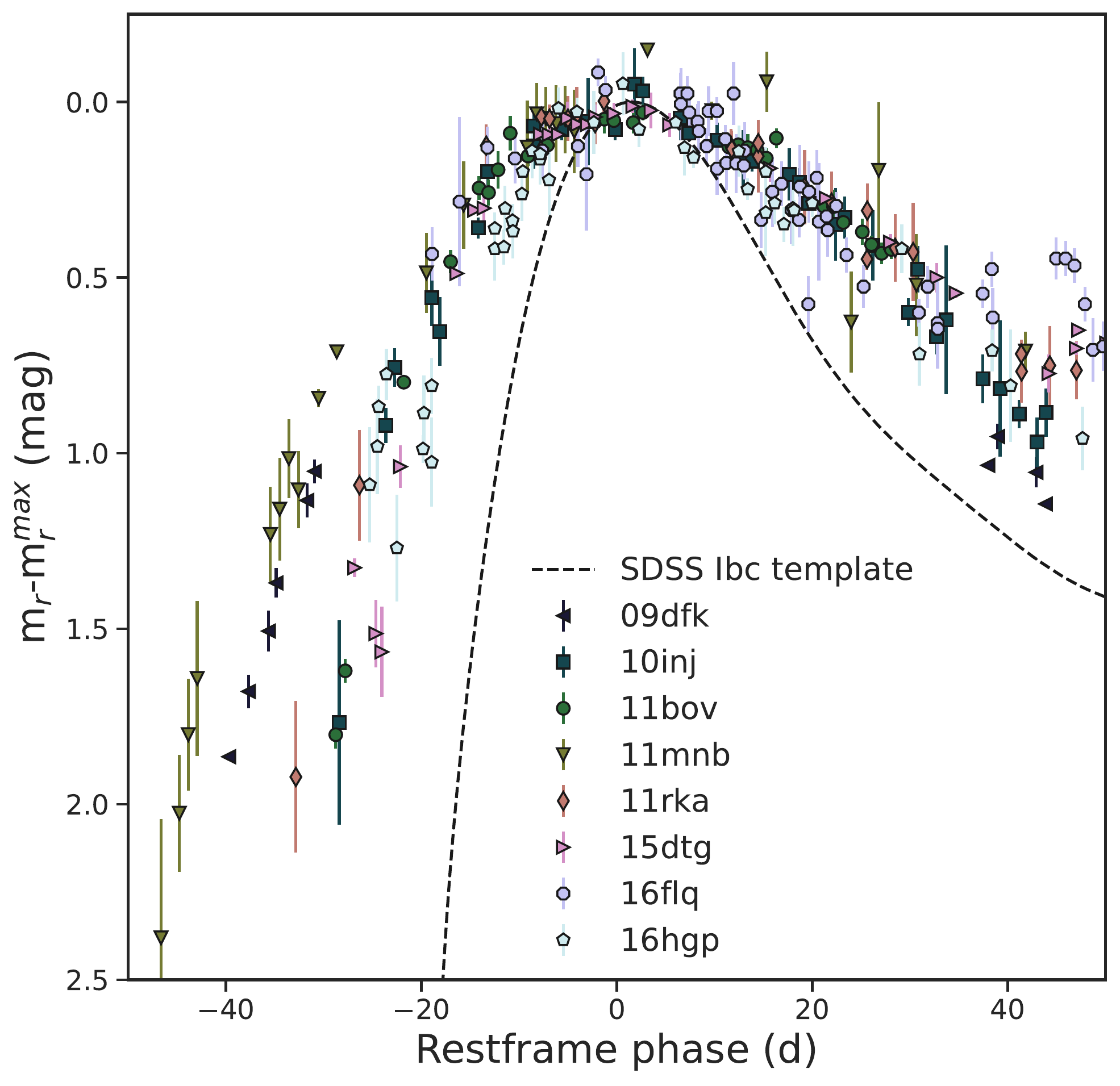}
\caption{Restframe apparent magnitude in $r$ band, normalized to the peak of the Type Ibc template lightcurve \citep{Taddia2015} in the same band. 
All eight lightcurves are broader than the template.}
\label{fig:comparebroadness}
\end{figure}

Unsurprisingly, the photometric evolution of our sample is unique compared to ordinary SE~SNe. As illustrated by Fig.~\ref{fig:comparebroadness}, not only do the lightcurves rise and decline over a longer time-scale but there also seems to be undulations or evidence of multi-peakness in some of the broad SNe. For the case of PTF11mnb this was studied in detail by \citet{Taddia2018e}. We also see similar behavior in iPTF16flq and iPTF16hgp.

\subsubsection{K-corrections for the broad sample \label{sec:Kcorr}}

K-corrections were calculated by comparing to synthetic photometry\footnote{To compute synthetic photometry we made use of \texttt{synphot}; \url{http://ascl.net/1811.001}.} obtained from Milky Way (MW) de-reddened optical spectra, and interpolated using a low-order polynomial assumed to be constant outside the interpolation region. Since the majority of our SNe were located in low-luminosity galaxies (Sect. \ref{sec:hosts}) we could do this without significant host contribution affecting the results. The spectra were absolute calibrated with the interpolated $r$-band LCs but not warped.

The SNe in our sample all have redshifts below $z=0.08$. Thus, generally, K-corrections were on the order of the photometric uncertainties and $<0.2$~mag in absolute value. We found a relatively good agreement between our K-corrections and those calculated from template Type Ibc spectra. We show this for the K-corrections for the $r$ band in Fig.~\ref{fig:kcorr}.  

\subsection{Lightcurve interpolation \label{sec:LCfits}}

Due to the uneven sampling of the lightcurves, interpolation was required. The traditional SE~SN lightcurve shapes have analytic approximations which can be used for this purpose \citep[e.g.,][]{Contardo2000,Bazin2011}. However, we chose the SNe in our sample on the basis of the uniqueness of their lightcurves.
Several of our objects also have non-standard lightcurves that potentially have multiple peaks \citep{Taddia2018e}. Therefore, we do not assume that our lightcurves can be fit with analytical approximations that were derived from ordinary SE~SN lightcurves. We instead use the more robust Gaussian Process (GP) regression to perform our lightcurve fits. By using GP regression and learning the shape and noise from the data, our fits are agnostic to assumptions of lightcurve shape and regularity.

To fit the early peak(s), a radial basis kernel was used, while the late-time linear decline was modeled with a linear combination of a linear and bias kernel (see e.g., \citealp{Papadogiannakis2019,Vincenzi2019,Karamehmetoglu2021}). We truly avoid any assumptions of a functional form by using a non-informative zero mean. 

To fully utilize our multiband lightcurves, we implemented a machine learning approach called multi-task learning applied to GP regression \citep{Bonilla2008} to derive our interpolation models. The multi-task method \citep{Caruana1997} allows us to constrain the fits using knowledge of the lightcurve shape from different bands simultaneously, which is useful when data are lacking in some bands, while avoiding overfitting. These fits (interpolation models) are used in the rest of the paper for the lightcurve analysis.

\subsection{Colors \label{sec:colors}}

\begin{figure*}
    \centering
    \includegraphics[width=\linewidth]{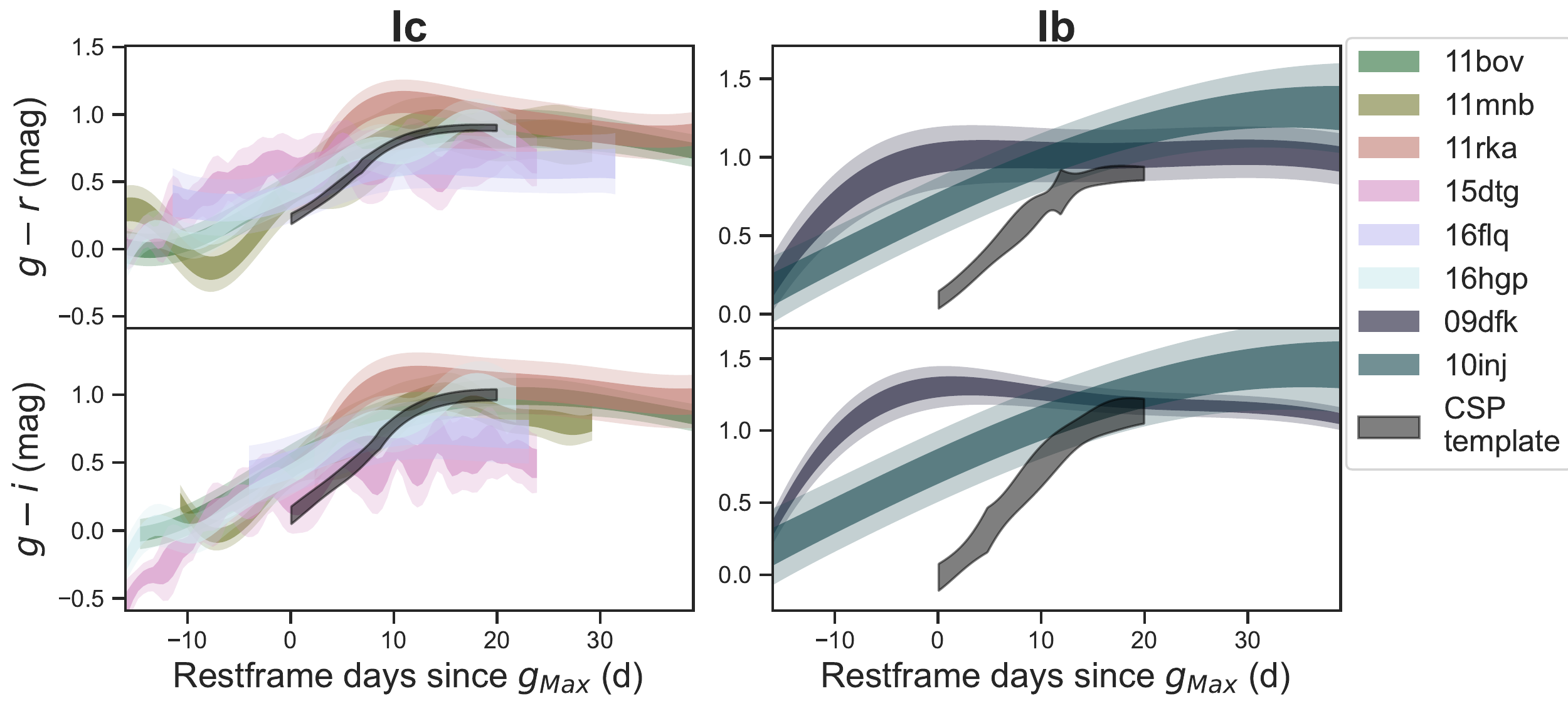}
    \caption{$g-r$ and $g-i$ restframe color curves, divided by inverse of $r$-band stretch parameter for comparison to the CSP Type Ic (left) and Type Ib (right) color templates (in grey) from \citet{Stritzinger2018b}.}
    \label{fig:colors_all}
\end{figure*}

Using the lightcurve fits from Sect.~\ref{sec:LCfits}, we calculate colors for our sample. Broadly speaking, the color evolution of objects in our sample is similar to that of other SE~SNe, except for the fact that it is more gradual. For example, the $g-i$ color peaks around $40$–$60$~d for our SNe, while $g-i$ is seen to peak around day 20 in the sample of ordinary SE~SNe from \citet{Stritzinger2018a}. To illustrate this, for each SN we compress the lightcurve by the inverse of its stretch value in the $r$ band. Then we plot SE~SN color templates from the Carnegie Supernova Project \citep[CSP;][]{Stritzinger2018b} alongside the color curves of our SNe in Fig.~\ref{fig:colors_all}. The SN color curves become much more similar to the template color curves in these figures after the stretch correction, indicating that the color evolution of the broad sample really is slower compared to that of ordinary SE~SNe, but otherwise seem to be similar.

In the literature sample studies \citep[e.g.,][]{Drout2011,Stritzinger2018b,Prentice2018}, the $g-r$ or $V-R$ color evolution of SE~SNe starts red at very early times, evolves to its bluest slightly before peak, and then evolves back to the red before flattening out. The same evolution is seen in our SNe, although with quite a bit of scatter. This scatter is seen to be minimal shortly after peak in low-extinction SE~SNe \citep{Drout2011,Stritzinger2018b}. By minimizing the scatter at 10 days past peak \citep{Drout2011} or in the region of 0--20 days past peak \citep{Stritzinger2018b}, it is possible to estimate the likely host extinction suffered by our SNe. However, these relationships were derived for ordinary SE~SNe with a typical color evolution. 

For our broad SNe, we used the fitting color template method of \citet{Stritzinger2018b}, but with the compressed SN colors and compare them to the template color as illustrated in Fig.~\ref{fig:colors_all}. Most of our SNe match the $g-r$ and $g-i$ color of the template relatively well, likely indicating that they are not significantly extincted. We calculated the weighted average magnitude difference between the template and the compressed SN color between 0--20 days past peak, following \citet{Stritzinger2018b}. We converted this color excess to E($B-V$)$_\text{host}$ using a \citet[][F99]{Fitzpatrick1999} extinction law with R$_\text{V}$=3.1. 

By varying our assumptions, we find that uncertainties with this method can be as large as $0.1-0.2$~mag due to the assumed R$_\text{V}$, filter effective-wavelength differences, and the magnitude of compressing of the color curves. In Table~\ref{tab:SNINFO}, we provide the weighted average E($B-V$)$_\text{host}$ derived via $g-r$, $g-i$, and $r-i$, and the error as the unbiased estimate of the weighted standard deviation. Correcting for host extinction would increase the bolometric flux at peak by an average of $4\%$ (excluding those with negligible E($B-V$)$_\text{host}$) or a maximum of $9\%$ for PTF09dfk. Due to the uncertainties involved in using this method on the broad sample, and since extinction seems to be low for all but PTF09dfk, we in the end do not correct for host extinction. 

It is perhaps not surprising that the overall color evolution of the broad SNe are similar to that of other SE~SNe, since we chose our sample from among spectroscopically normal Type Ibc SNe, but with broad lightcurves. The similarity between ordinary and broad SE~SNe we see in color reflects the similarity in spectral energy distributions (SEDs) between the groups. However, the fact that a simple stretch of the color templates seem to fit the broad SNe suggests that not only are the spectra normal and similar, but they also evolve at the approximate pace set by the lightcurve stretch. 

\section{Spectroscopy \label{sec:spec}}
As part of the follow-up campaign during (i)PTF, a total of 56 spectra of these eight objects were obtained as listed in Table~\ref{tab:speclog}. A spectral sequence for each SN is shown in Figs.~\ref{fig:spec1}, \ref{fig:spec2}, \ref{fig:spec3}, and \ref{fig:spec4}. Spectra not already released in previous papers are made available via WISeREP\footnote{https://wiserep.weizmann.ac.il \citep{yaron2012}.}.
All spectra were  reduced using
pipelines specific to each telescope and instrument and using standard data reduction methods for optical spectroscopy. The list of telescopes and instruments was collated by \citet{Fremling2018}.

The collection includes photospheric spectra for all SNe and nebular phase spectra for four objects\footnote{Some nebular features are also visible in the other four objects.}, as well as three spectra of host galaxies. We identify lines commonly seen in SE~SNe such as \ion{Fe}{ii} around ${\sim}5000$~\AA{}, \ion{O}{i} $\lambda\lambda 7772, 7774, 7775$, and the \ion{Ca}{ii} near-Infrared (NIR) triplet. We also see \ion{He}{i} in spectra of the Type Ib SNe 
as well as \ion{Na}{iD} and possibly \ion{Si}{ii} in Type Ic SNe. In the nebular spectra, we particularly see strong [\ion{O}{i}] $\lambda\lambda 6300, 6364$ as well as [\ion{Ca}{ii}] $\lambda\lambda 7319, 7324$. We measure line velocities from photospheric phase spectra in Sect.~\ref{sec:linevels}
and nebular line fluxes in Sect.~\ref{sec:nebular}.

\subsection{Line velocities \label{sec:linevels}}
To break the degeneracy in our modeling in Sect. \ref{sec:model}, we require an estimate of the characteristic ejecta velocity. Typically, the velocity of the absorption component of \ion{Fe}{II} $\lambda 5169$ around peak is used as an estimate \citep[see e.g.,][]{Taddia2018d}. In addition, we also measure the velocity of the absorption minima of \ion{Fe}{ii} $\lambda\lambda 4924, 5018$, \ion{He}{i} $\lambda 5876$, \ion{Na}{i} $\lambda\lambda 5890, 5896$, \ion{O}{i} $\lambda\lambda 7772, 7774, 7775$, and \ion{Si}{ii} $\lambda 6355$ where appropriate. For iPTF16hgp, the \ion{Si}{ii} feature has significantly higher velocities than other lines in the SN and could be incorrectly identified \citep[see][for alternatives]{Parrent2016}.

\begin{figure*}
    \centering
    \includegraphics[width=\linewidth]{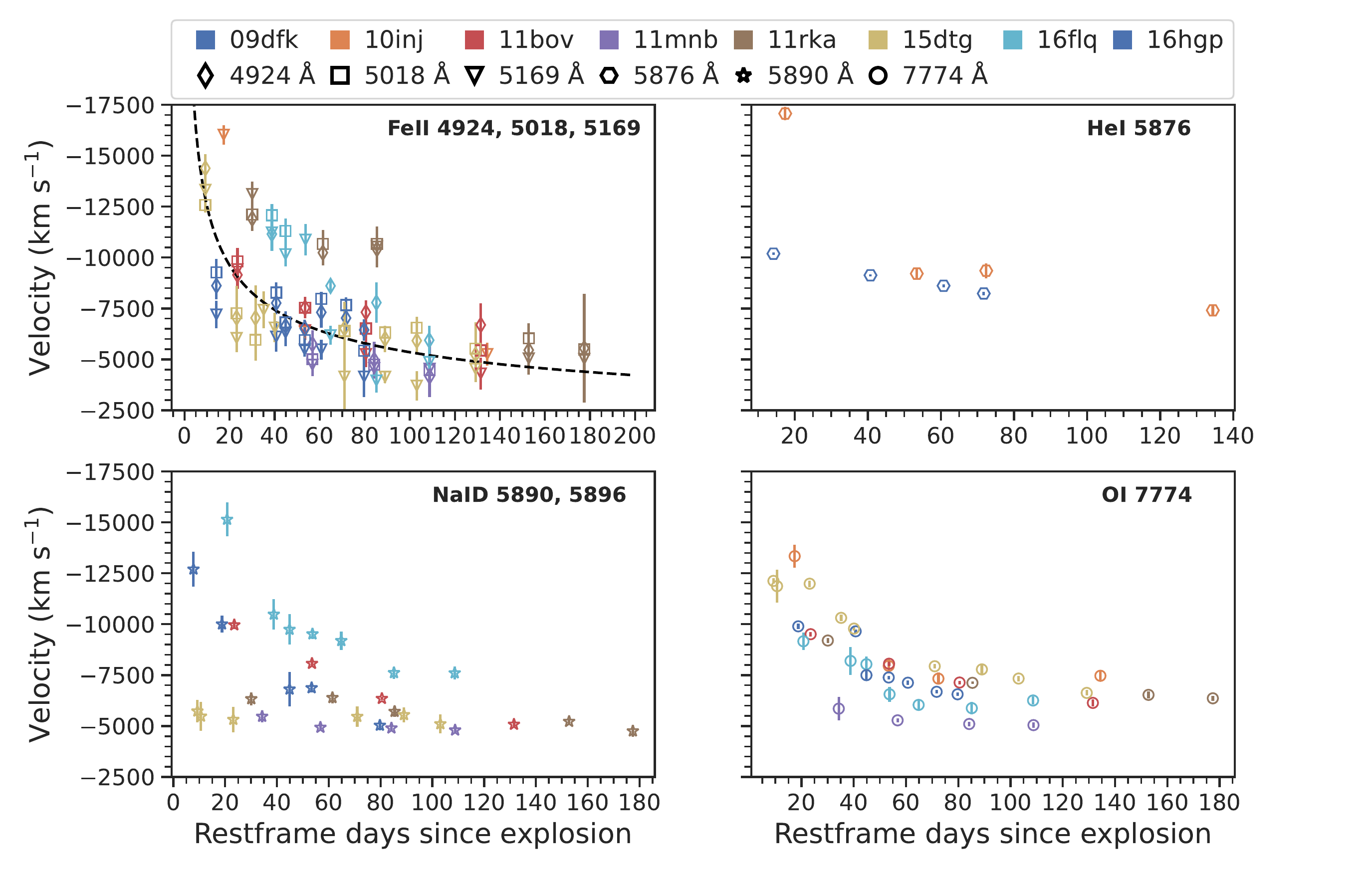}
    \caption{Line velocity as measured via the absorption minimum of the P-Cygni profile for \ion{Fe}{ii} $\lambda\lambda 4924, 5018, 5169$ (upper left), \ion{He}{i} $\lambda 5876$ (upper right), \ion{Na}{i} $\lambda\lambda 5890, 5896$ (lower left), and \ion{O}{i} $\lambda\lambda 7772, 7774, 7775$ (lower right).
    A power-law with $\alpha=-0.41$, used to find peak \ion{Fe}{ii} velocities, is over-plotted with a black dashed line.}
    \label{fig:linevels}
\end{figure*}

The restframe spectra which have been corrected for MW extinction as well as cleaned of narrow emission lines from the host galaxy are converted into velocity scale at rest wavelength of the line. Then, a continuum subtracted cut-out of the region around the line is fit with a Gaussian absorption profile. The minimum of this profile (mean of the Gaussian) is taken as the velocity. 500 realizations of this Gaussian fit is performed by varying end-points of the line profile randomly within $\pm50$~\AA{} to obtain the best fit and associated uncertainty. 
In addition, the spectrum is smoothed via boxcar smoothing of several increasingly larger windows and the whole process above is repeated for each smoothed spectrum in order to estimate the error that can be introduced by smoothing. These uncertainties are only statistical from the fitting procedure and do not account for blending or other systematics\footnote{We tried to avoid making a measurement if a feature was difficult to identify from a blend.}. 
Our results are shown in Fig.~\ref{fig:linevels}, tabulated in Table~\ref{tab:linevels}, and each ion is separately discussed below.

\subsubsection{\ion{Fe}{ii} \label{sec:FeII}}

Since the \ion{Fe}{ii} $\lambda 5169$ feature can be weak, blended, or hard to identify due to noise, we use the triple absorption feature formed by \ion{Fe}{ii} $\lambda\lambda 4924, 5018, 5169$ to increase our measurement reliability. We attempt to measure each of these ions separately although comparison to synthetic spectra show that at velocities near $10^4$~km~s$^{-1}$ this feature can be a blend \citep{Branch2002}. The absorption trough of \ion{Fe}{ii} $\lambda 5018$ is especially impacted by this. \ion{Fe}{ii} $\lambda 4924$ can be significantly affected by H$\beta$ \citep[which might be present even in Type Ibc SNe, see e.g.,][]{Parrent2016,Fremling2018}, possibly increasing the measured velocity with our method, as well as by emission from \ion{Mg}{ii} and \ion{He}{i}. Meanwhile, \ion{Fe}{ii} $\lambda 5169$ is blended with other Fe lines to the red side (often forming a characteristic ``W'' feature). Finally, strong host galaxy emission lines, especially forbidden transitions of [\ion{O}{iii}] around ${\sim}5000$~\AA{}, also make this measurement more difficult in some spectra.

We identify \ion{Fe}{ii} absorption features in many of our photospheric spectra and plot the velocities obtained in the upper left panel of Fig.~\ref{fig:linevels}. \ion{Fe}{ii} should be a good tracer of the photospheric velocity \citep{Dessart2005}, which in turn provides an estimate of the characteristic ejecta velocity when measured at lightcurve maximum. In order to obtain the velocity at peak for each SN, the \ion{Fe}{ii} velocities are fit using a power-law $v \propto t^\alpha$, with $\alpha=-0.41$, which was found to be a good fit for (i)PTF SE~SNe \citep{Barbarino2021}, in agreement with previous literature result of \citet{Taddia2018d}. Figure~\ref{fig:linevels} shows that the general shape of this power-law is indeed a good match to the observed \ion{Fe}{ii} velocities. The velocity at phase zero (lightcurve peak) is used in Sect.~\ref{sec:model} as an estimate for the characteristic ejecta velocity and is tabulated in Table~\ref{tab:fitting}.   

\subsubsection{\ion{He}{i} and \ion{Na}{iD}}

All SE~SNe show an absorption blueward of $~5900$~\AA{}, which is interpreted to either be \ion{He}{i} $\lambda 5876$ (possibly blended with \ion{Na}{iD} $\lambda\lambda 5890, 5986$) in Type Ib and IIb SNe \citep[e.g.,][]{Ergon2015a}, or \ion{Na}{iD} $\lambda\lambda 5890, 5986$ in Type Ic SNe (since they lack other \ion{He}{i} features). The velocities of \ion{He}{i} $\lambda 5876$ are shown in the upper right panel of Fig.~\ref{fig:linevels}. When measuring \ion{He}{i}, we also checked for the presence of other \ion{He}{i} lines, such as $\lambda\lambda 4471, 6678, 7065$, to increase our confidence in the line identification. For the \ion{Na}{iD} line velocities plotted in the lower left panel of Fig.~\ref{fig:linevels}, we assume the rest wavelength to be $5890$~\AA.

\subsubsection{\ion{O}{i}}

The velocities of \ion{O}{i} $\lambda\lambda 7772, 7774, 7775$ are plotted in the lower right panel of Fig.~\ref{fig:linevels}. For this ion, there could be contamination from \ion{Ca}{ii} as well as possible telluric absorption at low redshift. As seen in Fig.~\ref{fig:velsbysn}, \ion{O}{i} absorption is consistently located at lower velocity compared to \ion{He}{i} absorption in our sample.

\subsubsection{Velocity evolution}
The absorption minima of \ion{Fe}{ii} are located at higher velocity than that of \ion{O}{i} at early times in several objects: PTF10inj and iPTF15dtg (only at very early times), PTF11rka and iPTF16flq (out to many weeks past peak). In PTF10inj the \ion{O}{i} lines evolve to have higher velocities by lightcurve peak, while this happends already ${\sim}\text{two}$~weeks before peak in iPTF15dtg. This crossover seems to occur three months after peak in PTF11rka, and one to two months after peak in iPTF16flq.

Since \ion{Fe}{ii} is easier to excite, the velocity crossover in \ion{O}{i} and \ion{Fe}{ii} can be understood either by primordial iron tightly tracing the outer layers of the receding photosphere in a spherically symmetric explosion, or as a consequence of ejecta asymmetry, as was seen in SN 1987A \citep[e.g.,][]{Larsson2013}, or jets \citep{Piran2017}. 

\subsection{Line flux of \ion{O}{i} and \ion{Ca}{ii} in nebular spectra \protect\protect\label{sec:nebular}}

We measure the late-time luminosity of [\ion{O}{i}] $\lambda\lambda 6300, 6364$ and [\ion{Ca}{ii}] $\lambda\lambda 7291, 7324$ in our nebular-phase spectra where these lines are clearly visible, and use the flux ratio $L_{[\ion{Ca}{ii}]}/L_{[\ion{O}{i}]}$ as an indicator of the progenitor mass. Lower values of the ratio are associated with higher core mass \citep[see][]{Fransson1987,Fransson1989,Terreran2019,Fang2019,Fang2022}, although also see \citet{Dessart2021,Prentice2022,Ergon2022}. For ordinary SE~SNe, just the [\ion{O}{i}] line strength in nebular spectra normalized by energy deposition can be used. Typically energy deposition from standard nickel powering is assumed at late epochs, but this assumption is more strained for our SNe given their non-standard broad lightcurves.

Following the procedure outlined in \citet{Fang2019}, we establish a linear continuum from the left edge of the [\ion{O}{i}] feature to the right edge of the [\ion{Ca}{ii}] in the continuum-subtracted spectrum. We fit the [\ion{O}{i}] $\lambda\lambda 6300, 6364$ feature using a double Gaussian profile centered on the rest wavelength of the lines, having the same standard deviation, and an amplitude (flux) ratio of 3:1, which is expected on theoretical grounds. Any excess flux in the red part of this complex is thought to be coming from [\ion{N}{ii}]. [\ion{Ca}{ii}] is measured by integrating the entire complex first, then subtracting the contribution from \ion{Fe}{ii}, as was done by \citet{Fang2019}. We follow their method exactly and integrate from the blue edge of the complex to 7155~\AA{}, then double this amount and assume it to be the flux contribution from \ion{Fe}{ii}, which we subtract from the total flux of the complex to obtain the [\ion{Ca}{ii}] flux. For iPTF15dtg, a more detailed study of the nebular [\ion{O}{i}] $\lambda\lambda 6300, 6364$ lines was performed by \citet{Taddia2019a}.

The literature values from \citet{Fang2019} were obtained from a single spectrum per SN, taken between 200 and 300 days past peak (mean phase and standard deviation of $216 \pm 37$~d). In order to make a comparison, we also produce a ``Gold sample'' consisting of a clearly nebular spectrum (visible nebular lines and $>100$~d after peak) that was taken closest in time to 200--300 days since peak. Our Gold sample consists of a spectrum chosen with these criteria from each of PTF09dfk, PTF11bov, PTF11rka, and iPTF15dtg, with a mean phase and standard deviation of $269 \pm 101$~d. We also import a comparison sample from \citet{Terreran2019}, using the measured value closest in epoch to the mean of our Gold sample (with a resulting mean phase and standard deviation of $270 \pm 59$~d). We only take the Type Ibc SNe into this sample, and note that \citet{Terreran2019} use a slightly different method (without explicitly accounting for the contribution from [\ion{N}{ii}] and \ion{Fe}{ii}). Additionally, they do not have a representative sample, but purposefully pick several objects with low values of this ratio to highlight the unique nature of SN~2016coi.

\begin{table}
    \centering
    \begin{tabular}{c|c}
        Source & Mean $L_{[\ion{Ca}{ii}]}/L_{[\ion{O}{i}]}$ \\
        \toprule
        \citet{Terreran2019} & $0.66 \pm 0.29$ \\
        \citet{Fang2019} & $0.80 \pm 0.51$ \\
        Gold sample & $0.53\pm0.21$ \\
        Broad SE~SNe & $0.64 \pm 0.19$
    \end{tabular}
    \hrule
    \caption{Mean and standard deviation of $L_{[\ion{Ca}{ii}]}/L_{[\ion{O}{i}]}$ for our broad SE~SNe compared to literature values.}
    \label{tab:nebular}
\end{table}

We tabulate the mean and standard deviation of $L_{[\ion{Ca}{ii}]}/L_{[\ion{O}{i}]}$ measured for our broad SE~SNe with values from \citet{Fang2019} and 
\citet{Terreran2019} 
in Table~\ref{tab:nebular}. For our Gold sample we obtain a mean $L_{[\ion{Ca}{ii}]}/L_{[\ion{O}{i}]} = 0.53 \pm 0.21$. If we also include the values measured from the latest spectrum for the remaining 4 SNe, (PTF10inj, PTF11mnb, iPTF16flq, and iPTF16hgp), this value becomes $0.64 \pm 0.19$ for the entire broad sample. However, those 4 spectra are not fully nebular and thus suffer from contamination. 
\begin{figure}
    \centering
    \includegraphics[width=\linewidth]{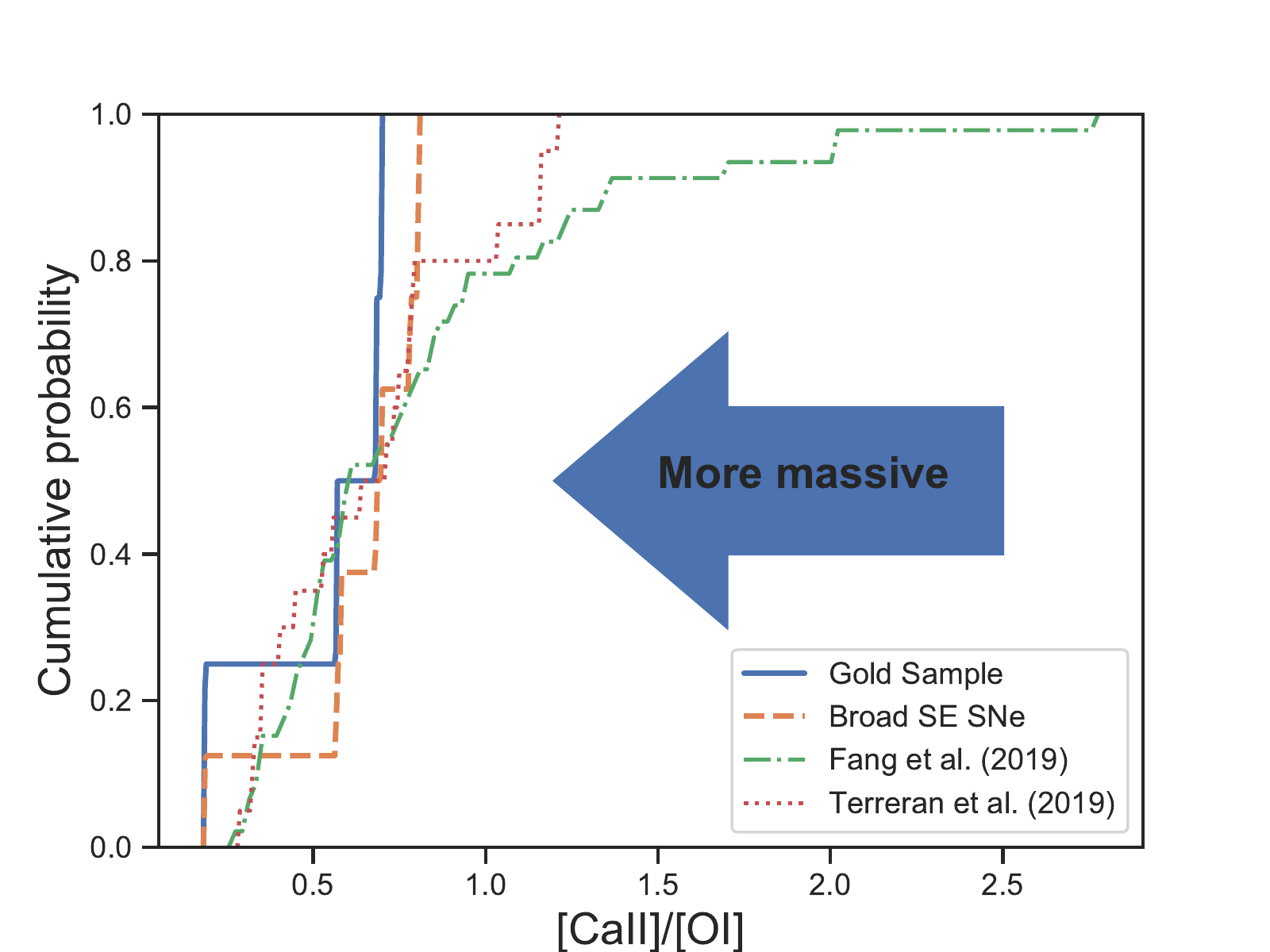}
    \caption{Cumulative distribution of the ratio of [\ion{Ca}{ii}]/[\ion{O}{i}] flux in nebular spectra from our Gold sample and total sample compared to literature samples from \citet{Fang2019,Terreran2019}. Lower flux ratio is associated with more massive progenitors.}
    \label{fig:CaIIOI}
\end{figure}

We show a comparison of the measurements for our broad SE~SNe with the same literature samples in Fig.~\ref{fig:CaIIOI}. The cumulative distribution of our SNe is sharply peaked at lower values of $L_{[\ion{Ca}{ii}]}/L_{[\ion{O}{i}]}$, which is associated with more massive progenitors in models and via comparison to other independent methods of estimating the progenitor mass. We note that all of our SNe have $L_{[\ion{Ca}{ii}]}/L_{[\ion{O}{i}]} < 0.81$.

From these comparisons we conclude that the broad SE~SNe seem to prefer generally low values of $L_{[\ion{Ca}{ii}]}/L_{[\ion{O}{i}]}$, similar or lower than the comparison sample of \citet{Terreran2019}. However, the differences of the means are not statistically significant for any sample, probably indicating that these samples also have broad or massive SE~SNe. \citet{Fang2019} have found a trend between lightcurve broadness and lower values of $L_{[\ion{Ca}{ii}]}/L_{[\ion{O}{i}]}$, which our results also seem to support. As noted by \citet{Terreran2019}, models also show a lower $L_{[\ion{Ca}{ii}]}/L_{[\ion{O}{i}]}$ with larger nickel mixing. Additionally, $L_{[\ion{Ca}{ii}]}/L_{[\ion{O}{i}]}$ ratio shows that the Broad SE~SNe are not Ca-rich events. 
\section{Bolometric modeling \label{sec:model}}

The bolometric properties of our broad sample can be used to derive progenitor and SN explosion parameters, such as ejecta and nickel mass. In this section, we construct pseudo-bolometric lightcurves for our SNe and investigate these parameters via semi-analytic lightcurve modeling using the \citet{Arnett1982} method. Then, we compare our ejecta and nickel mass distributions with those from the literature. 

\subsection{Bolometric lightcurves \label{sec:bololc}}
Bolometric lightcurves of PTF11mnb and iPTF15dtg calculated from empirical bolometric correction estimates of \citet{Lyman2014} were found to be in agreement with those from other methods, such as SED fitting, by  \citet{Taddia2016,Taddia2018e,Taddia2019a}. The \citet{Lyman2014} method estimates the bolometric correction as a function of color, which was shown to be a good surrogate for the full SED in their reference SNe. In fact, SN~2011bm, i.e. PTF11bov, was in the reference sample used to derive this correction. The benefit of this method for our SNe is that a bolometric correction based on the color is independent of the phase. 

As shown in Sect.~\ref{sec:colors}, the color evolution of our SNe is similar to ordinary SE~SNe and seems to account for the high stretch naturally (since the color evolution is equally stretched out). Since we lack sufficient multiband coverage on many epochs for our SNe, and since our colors seem to be well behaved compared to ordinary SE~SNe, we use the empirical relation for $g-r$ from \citet{Lyman2014} to calculate the bolometric correction for our SN lightcurves.

The bolometric lightcurves plotted in Fig.~\ref{fig:bolomags} used the $g-r$ color to calculate a bolometric correction and applied it to the $g$-band lightcurve fits from 
Sect.~\ref{sec:LCfits}. \citet{Lyman2014} found that this empirical correction based on the $g-r$ color had a systematic uncertainty of 0.076 mag, which we propagate together with the error in the lightcurve fits and color.

Where we have good data, the error is ${\sim}50\%$ larger than the systematic uncertainty while it becomes dominated by the uncertainty in the color via the lightcurve fits for epochs further from good data. \citet{Lyman2014} found that early-cooling lightcurves had a different empirical correction. We do not fit the earliest phases of the bolometric lightcurves that show a possible early cooling behavior during the fitting.

Within uncertainties, we find the shape and brightness of our bolometric lightcurves to be consistent with a blackbody fit and extrapolation to the photometry, especially around the peak where the lightcurve fitting is performed. In addition to the works mentioned above, our lightcurves are consistent with pseudo-bolometric lightcurves of the same SNe in previous studies \citep{Valenti2012, Prentice2018, Pian2020}.

\subsection{Lightcurve properties \label{sec:LCprop}}

\begin{figure}[t]
\begin{center}
    \includegraphics[width=0.95\linewidth]{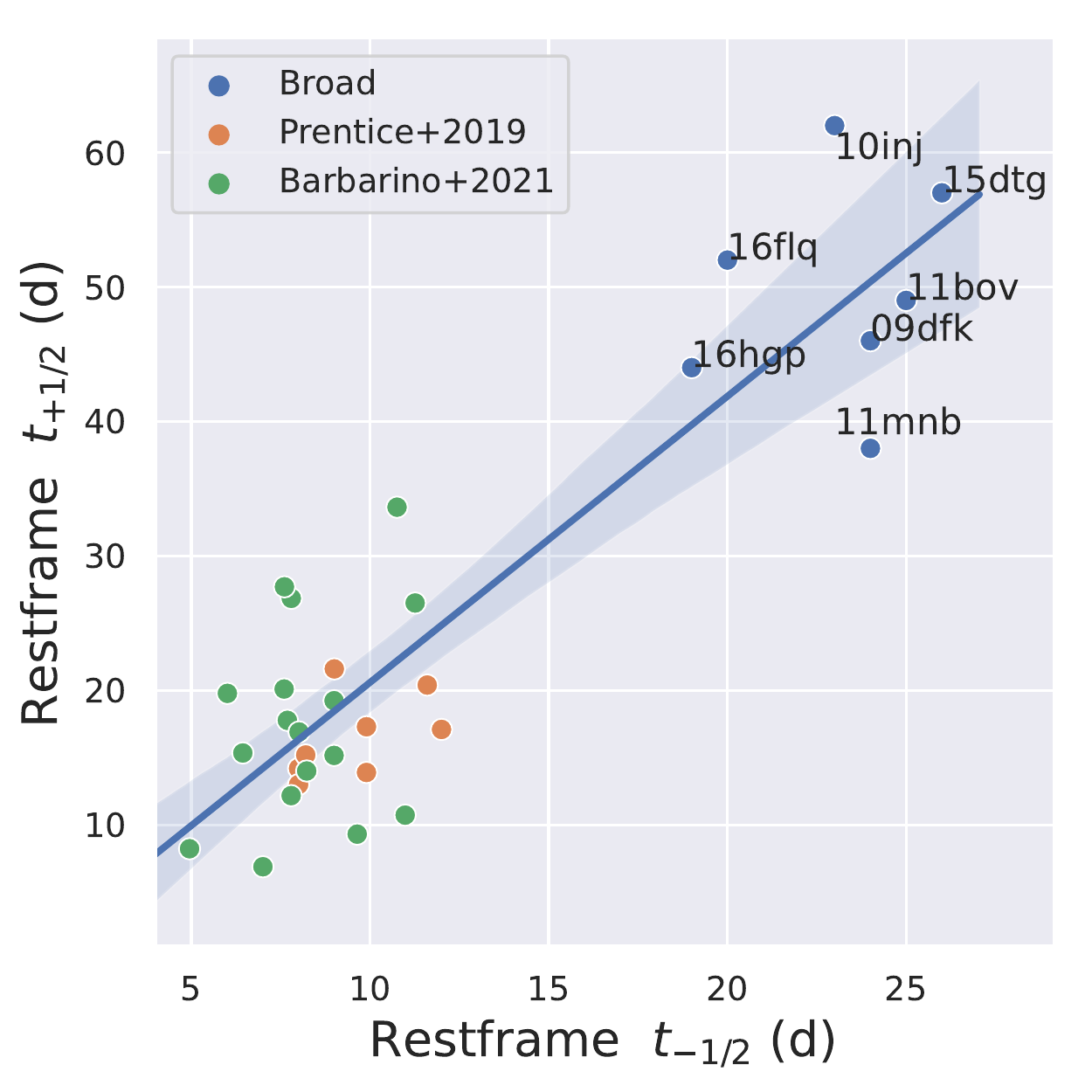}\\
    \includegraphics[width=0.99\linewidth]{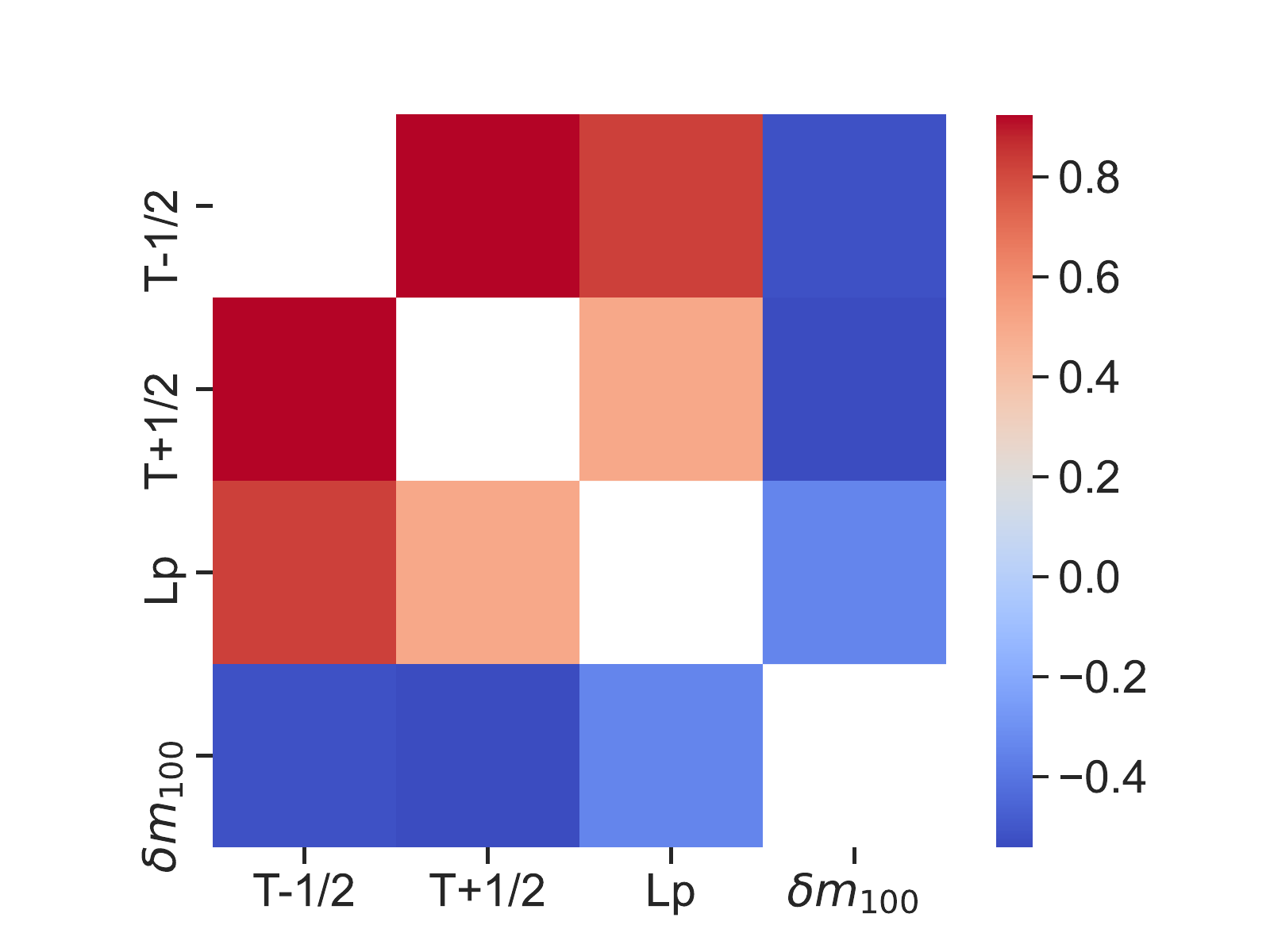}
    \caption{\textit{Top:} Bolometric lightcurve properties of the broad sample compared to SE~SNe from \citet{Prentice2018} and \citet{Barbarino2021}. \textit{Bottom:} A correlation heatmap between the broad sample and the literature SE~SNe from \citet{Prentice2018} with darker shades representing a stronger correlation with red for a positive and blue for a negative correlation.}
    \label{fig:LCprop}
    \end{center}
\end{figure}

We compare the bolometric lightcurve properties of our sample to those of \citet{Prentice2018}. We measure $t_{\pm1/2}$, $L_p$, and $\delta m_{100}$, where $t_{\pm1/2}$ is the restframe time it takes for the lightcurve to rise or decline to half of the peak luminosity, $L_p$ is the peak luminosity, and $\delta m_{100}$ is the linear decline around 100 days past peak in magnitudes per day (mag d$^{-1}$). In the same way, we calculate $t_{\pm1/2}$, $L_p$ for the Type Ic sample of \citet{Barbarino2021} using their published bolometric lightcurves, where the lightcurves allow us to do so. However, it was not possible to reliably calculate $\delta m_{100}$ for this sample. 

The result of our comparison is shown in Fig.~\ref{fig:LCprop}. In the upper panel, we plot the comparison of rise and decline times to both literature samples. As expected, we find that our objects are broader as parametrized by $t_{\pm1/2}$. In the lower panel, we provide a correlation heatmap between all four lightcurve properties, using combined sample of our broad SNe and the SE~SNe of \citet{Prentice2018}, but excluding the purely Type Ic Sample of \citet{Barbarino2021}. We find a strong positive correlation between $t_{-1/2}$ and $t_{+1/2}$ when we add our objects to the sample of \citet{Prentice2018}, which seems to support our approach of using a single lightcurve stretch as a method of finding broad SE~SNe. There is also a relatively weaker correlation between these parameters and $L_p$, in congruence with our finding in 
Sect.~\ref{sec:biases} that the broad SE~SNe have slightly brighter peaks on average. Finally, we notice a strong (negative) correlation between $t_{+1/2}$ and $\delta m_{100}$, indicating that the broader SE~SNe also decline more slowly on longer timescales. In the above comparisons, $t_{-1/2}$ or $\delta m_{100}$ were excluded from the combined sample if the lightcurve did not respectively extend early or late enough.

\subsection{Arnett model fits}

For SE~SNe powered by the decay of radioactive $^{56}$Ni, the semi-analytical model of \citet{Arnett1982} can be used to estimate ejecta and nickel masses. In implementing the equations, we follow the working example presented by \citet{Cano2013} and numerically fit our bolometric lightcurves around peak. This simple Arnett model has often been used in the SE~SN literature and allows us to make meaningful comparisons to other studies. We discuss its limitations for our use case in Sect.~\ref{sec:multipm}. 

In our modeling we assume a constant effective opacity of $\kappa = 0.07~\text{g}~\text{cm}^{-2}$. The ejecta mass ($M_{ej}$) is estimated assuming a constant density ejecta with $M_{ej}=\frac{10E_k}{3V^2}$ where $E_k$ is the kinetic energy of the explosion and $V$ is the characteristic ejecta velocity, for which we use the values derived in Sect.~\ref{sec:FeII} from \ion{Fe}{ii} velocities. We fit this model to our SN lightcurves to obtain the nickel mass and ejecta mass. For the explosion epoch, a first guess is derived from the template fitting in Sect.~\ref{sec:LCfit_init}, but is then allowed to vary slightly during the fitting procedure, while respecting the discovery epochs.

\begin{figure}
    \centering
    \includegraphics[width=\linewidth]{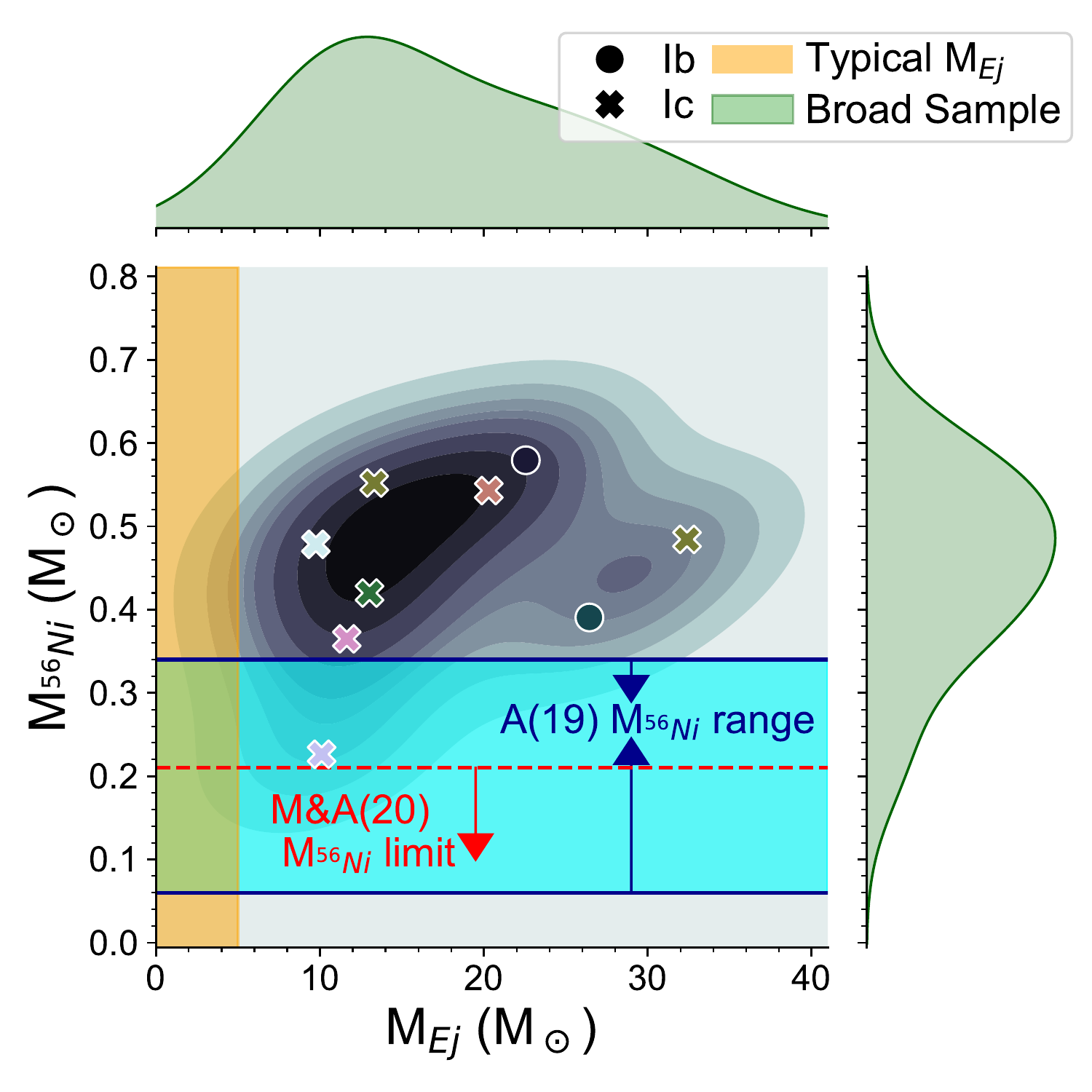}
    \caption{Ejecta and nickel masses obtained from the simple Arnett fitting. The colors of the SNe are the same as in previous figures. The margins show the distribution of the individual parameter as a KDE (ejecta mass on top, nickel mass on the right). The lower ejecta mass fit of PTF11mnb is from a fit to the peak only (blue line in Fig.~\ref{fig:bolomags}). The orange region shows the typical range of ejecta masses for ordinary literature SE~SNe (Sect.~\ref{sec:ejectamass}). The red dashed line labeled M\&A(20) is the upper limit of the nickel mass distribution of SE~SNe in \citet{Meza2020}, discussed in Sect.~\ref{sec:nickel}. The cyan region with blue arrows labeled A(19) shows the mean and standard deviation of literature Type Ibc nickel masses in \citet{Anderson2019}, see Sect.~\ref{sec:nickelmass}.}
    \label{fig:EjNi}
\end{figure}

The resulting fits were visually verified and are plotted in Fig.~\ref{fig:bolomags}, where we show both weighted and unweighted fits. We did not fit the early cooling part in iPTF15dtg and iPTF16hgp. The ejecta and nickel masses from the weighted fits are plotted in Fig.~\ref{fig:EjNi} and all parameters are tabulated in Table~\ref{tab:fitting}.

For all but PTF11mnb, lightcurves near peak are well fit by the Arnett model. For PTF11mnb, we also performed a separate fit to the main peak alone (blue line in Fig.~\ref{fig:bolomags}), which yielded a slightly higher nickel mass but a significantly lower ejecta mass. This second fit is also included in Fig.~\ref{fig:EjNi}. \citet{Taddia2018e} proposed that either a double distribution of nickel or impact from an additional powering mechanism, such as the spin-down of a rapidly rotating magnetar, can explain the lightcurve of PTF11mnb. While the ejecta and nickel masses in the latter case are dependent on the assumed and uncertain additional powering mechanism, in the former case \citet{Taddia2018e} found a good model fit with $7.8~M_\sun$ of ejecta and $0.59~M_\sun$ of $^{56}$Ni. These values are somewhat similar to our estimate in Table~\ref{tab:fitting} from fitting only the peak.

Similar to PTF11mnb, iPTF16flq also seems to show evidence of multiple peaks in its bolometric lightcurve. iPTF16flq possibly also has a more rapid late-time decline than the other SNe. Such lightcurves cannot be adequately described by a simple Arnett model, but we use it here to provide a rough estimate of the nickel and ejecta mass required to power the main peak.

\subsection{Ejecta Mass \label{sec:ejectamass}}

In Fig.~\ref{fig:EjNi}, the distribution of ejecta and nickel masses for our broad sample of SE~SNe are plotted. Our SNe all have ejecta masses ${\gg}5~M_\sun$, whereas virtually all previous sample studies of SE~SNe have found ejecta masses in the range of 1--5~$M_\sun$ (orange region in Fig.~\ref{fig:EjNi}). It thus seems that broad lightcurves are in fact a good way to find SE~SNe with large ejecta masses.

Within the constraints of the model, the errors on the ejecta masses are dominated by the uncertainty in the explosion epoch and the uncertainty associated with defining a characteristic ejecta velocity from the observations. In addition, the uncertainties involved with constructing the bolometric lightcurve play a role. However, an even bigger unknown is whether the simple Arnett model is the adequate way to model SE~SNe (see Sect.~\ref{sec:multipm}). 
We limit our conclusions to stating that within the assumptions of the Arnett model, the broad SE~SNe have much larger ejecta masses than the average SE~SN. However, PTF11mnb, iPTF15dtg, and possibly iPTF16flq show us that this might not be the full story. In Sect.~\ref{sec:multipm}, we discuss how additional powering mechanisms may play a confounding role.

\subsection{Nickel mass \label{sec:nickelmass}}

The weighted mean and standard deviation of the nickel mass for our sample is $0.42 \pm 0.08$ $M_{\sun}$. Similar to the ejecta masses, the nickel masses of our sample are also significantly larger than the average for SE SNe in the literature. In a meta analysis of nickel masses, \citet{Anderson2019} found that the mean for Type Ibc SNe was ${0.20} \pm {0.14}$ $M_{\sun}$ with a median of ${\sim}0.16$ $M_{\sun}$, so the nickel masses of our sample are more than one sigma above that mean. However, the ratio of nickel mass to ejecta mass in our sample is similar to what has been reported previously \citep{Lyman2016}, since the ejecta masses of our sample are also correspondingly larger. We note that the addition of $<10\%$ bolometric flux from host extinction would only marginally increase our nickel mass estimates. However, uncertainty from distance and host extinction would be larger than the purely statistical uncertainty given in Table~\ref{tab:fitting}.

\section{Host environment properties \label{sec:hosts}}

The host environment of a SE~SN can provide independent evidence of progenitor properties, including mass. Therefore, we compared the hosts galaxy properties of our broad SN sample with the host galaxy properties of the (i)PTF CC SN sample. 

\subsection{The unique environments of iPTF SE~SNe with broad lightcurves} 

 After identifying the hosts, galaxy properties were obtained via SED fitting to data ranging from far Ultra-Violet to IR, which was obtained from various public surveys. Details of the SED fitting and data collection, as well as analysis of the host galaxies of the entire (i)PTF CC-SN sample can be found in \citet{Schulze2021}. Here, we specifically selected the hosts of the 8 broad SE~SNe and compared their brightness, mass, and specific star formation rate (sSFR) to the typical SE~SN host.

\begin{figure}
    \centering
    \includegraphics[width=\linewidth]{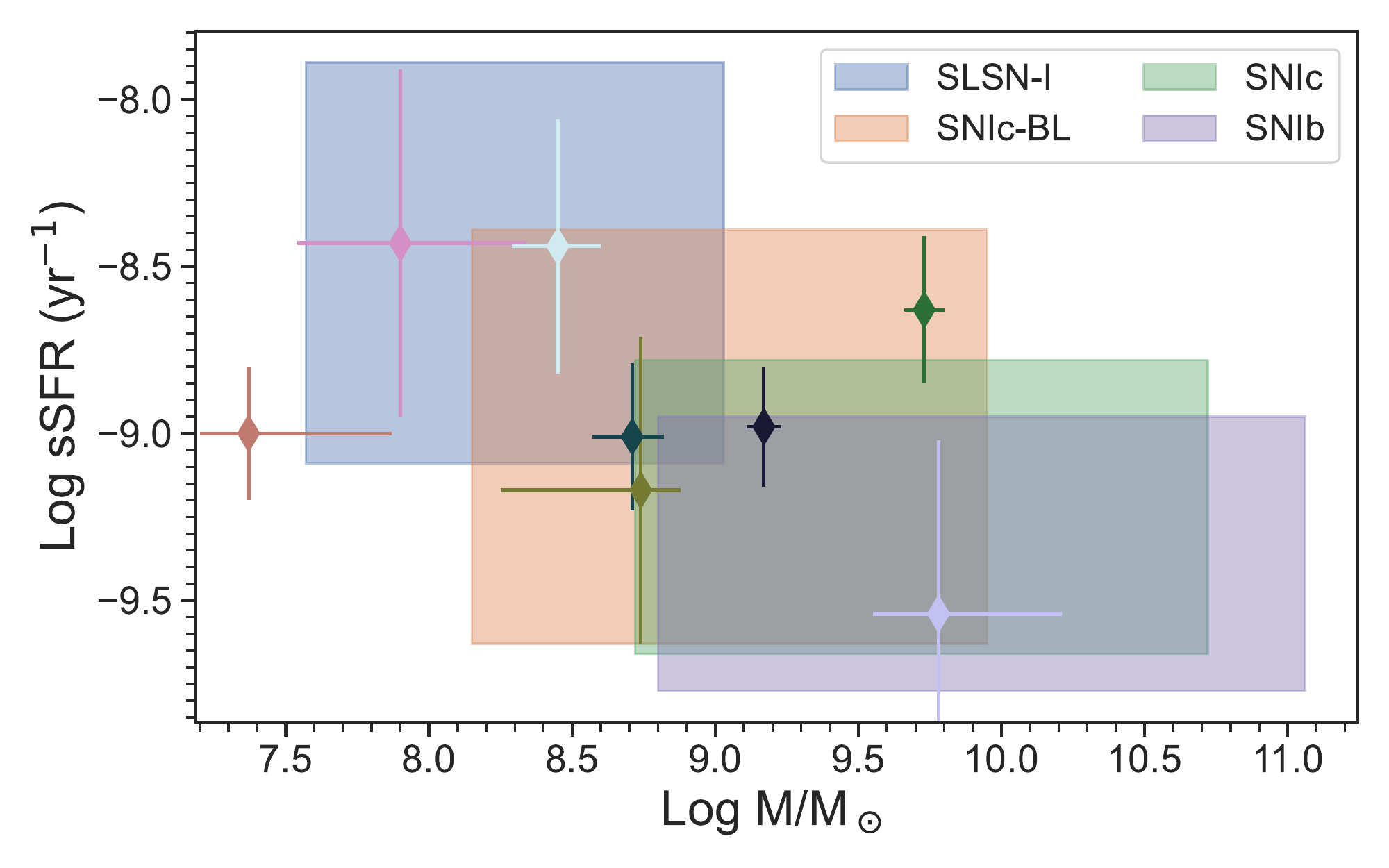}
    \caption{Log M/$M_\sun$ vs. Log sSFR for the broad sample compared to the typical range for the host galaxies of Type Ib, Ic, Ic-BL SNe, and SLSNe-I, obtained from the combined (i)PTF sample of CC-SN host galaxies \citep{Schulze2021}. Several members of the broad sample seems to prefer lower mass and higher sSFR galaxies, more similar to SLSN-I hosts than Type Ibc SN hosts.}
    \label{fig:host_MvsSSFR}
\end{figure}

Our results are plotted in Fig.~\ref{fig:host_MvsSSFR}. The 8 broad iPTF SE~SNe seem to favor lower-mass galaxies with higher sSFR, as compared to the average SE~SN. Many lie in the range of values that are typical of the host galaxies of SLSNe-I and Type Ic-BL SNe, and none are in very massive galaxies with $\gtrsim 10^{10}~M_\sun$. Both SLSNe and SNe Ic-BL are known to be located in lower metallicity environments compared to ordinary SE~SNe \citep[e.g.,][]{Galbany2016,Schulze2018,Taddia2019b}. 

In agreement with the above, several of the 8 iPTF SE~SNe with broad lightcurves are located in low-luminosity host galaxies. Low-luminosity galaxies are likely characterized by low metallicity, while ordinary SE~SNe favor luminous hosts, characterized by solar or super-solar metallicities \citep{Anderson2010,Arcavi2010,Leloudas2011,Modjaz2011a,Sanders2012a}. 

\subsection{Metallicity}

\begin{figure}
    \centering
    \includegraphics[width=\linewidth]{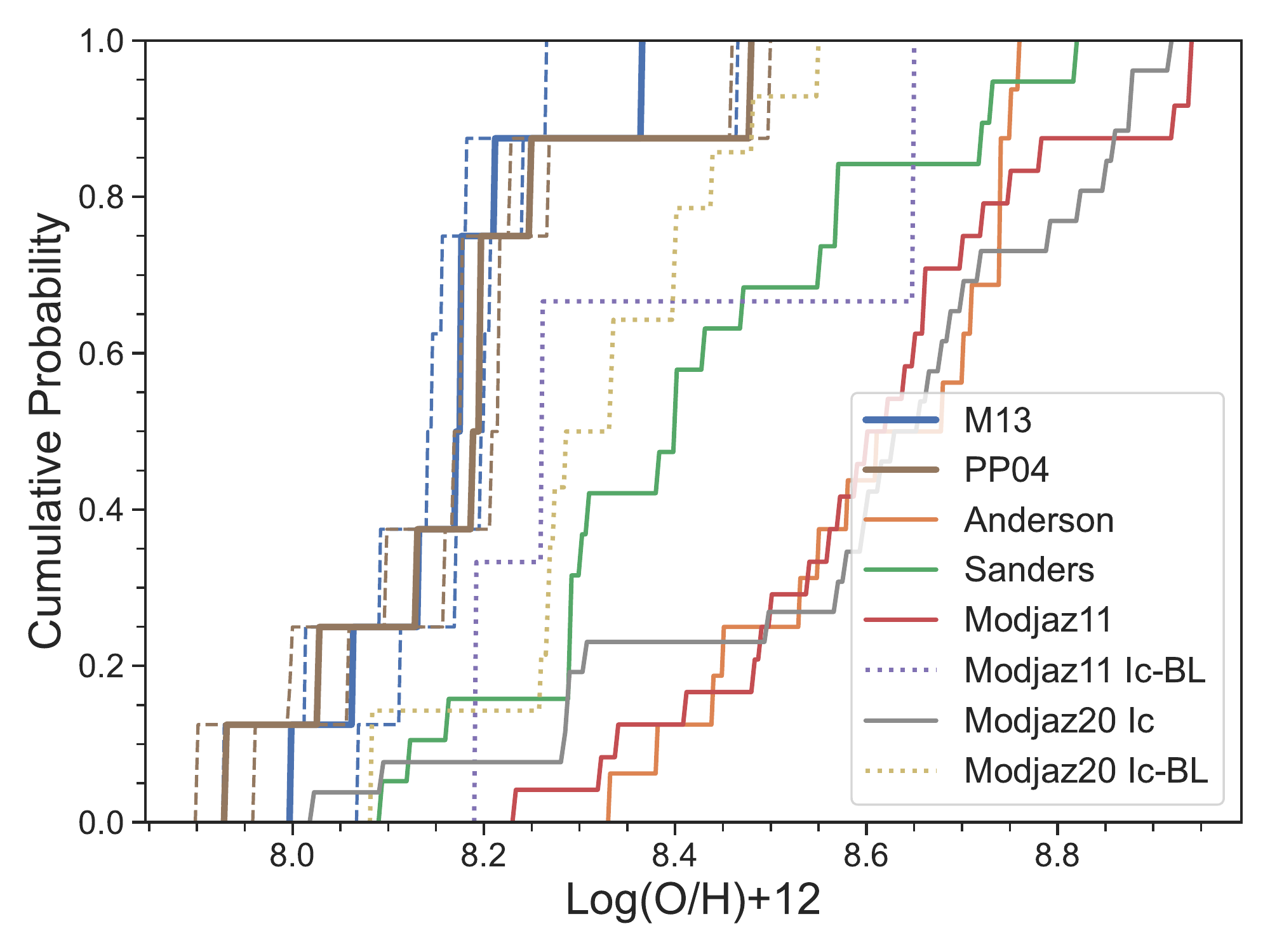}
    \caption{Cumulative distribution of the metallicity of our broad SE~SNe, measured using the PP04 (included for comparison to the literature) and more up to date M13 method. Literature samples of Type Ibc metallicities from \citet{Anderson2010,Modjaz2011a,Sanders2012a,Modjaz2020} are also shown for comparison, which were calculated using the PP04 method. We see a clear statistically-significant difference between the metallicities of our sample compared to the average SE~SN host metallicity.}
    \label{fig:hostmetal}
\end{figure}

Late-time spectra of our broad lightcurve SE~SNe, either SN spectra affected by host contamination or  host galaxy spectra, allow estimating the metallicities at the SN location using the O3N2 method of \citet[][PP04]{Pettini2004} as updated by \citet[][M13]{Marino2013}. The results are presented in Fig.~\ref{fig:hostmetal} and in Table~\ref{tab:metal}. The average metallicity is log(O/H)+12 $\sim8.2$~dex with none above ${\sim}8.4$~dex, which makes all of the host galaxies subsolar. This average metallicity is lower than for ninety percent of the SE~SN hosts measured in an untargeted sample of normal SE~SNe \citep{Sanders2012a}. For comparison, in Fig.~\ref{fig:hostmetal} we also plot the metallicities measured by \citet{Sanders2012a}, \citet{Anderson2010}, \citet{Modjaz2011a}, and \citet{Modjaz2020}. The sample for \citet{Modjaz2020} was Type Ic and Ic-BL SNe from PTF. As it turns out, they also measured host metallicity of PTF11bov and PTF11rka using PP04 and found $8.29 \pm 0.01$~dex and $8.02 \pm 0.03$~dex, respectively, (with purely statistical error bars). For PTF11bov our measurements are consistent within uncertainties. For PTF11rka, they are nearly consistent, separated by only $0.02$~dex.

Clearly, the broad SE~SNe are located at much lower metallicity, and the difference is statistically significant. The metallicity of our objects is similar to, or even lower than, the metallicity of the literature Type Ic-BL SN sample from \citet{Modjaz2011a} and the PTF Type Ic-BL SN sample from \citet{Modjaz2020}. Type Ic-BL SNe have been shown to favor very low metallicity environments ($8.3-8.4$~dex, \citealp{Galbany2016}) with high SFR, which matches the characteristics of the host galaxies of our sample as well.

\citet{Galbany2016} studied the literature metallicities from targeted and untargeted searches. They found that untargeted searches on average yielded somewhat lower metallicities. For Type Ibc SNe specifically, they found the mean metallicity to be $8.4$~dex for untargeted searches and $8.5$~dex for targeted searches. Our broad SE~SNe are located in very low metallicity galaxies that are all below the mean metallicities mentioned above. This might help explain why such broad lightcurve SE~SNe have been relatively absent from the literature, which is largely made up of SNe from targeted searches.  

\section{The fraction of broad Type Ibc SNe from the (i)PTF \label{sec:biases}}
Using single band stretch as a simple observable we calculated that ${\sim12}^{+4}_{-4}\%$ of (i)PTF Type Ibc were broad. In comparison, previous literature samples found a smaller fraction of 3--4\% \citep{Lyman2016,Prentice2018}\footnote{Since these number include Type IIb SNe, we make an apples to apples comparison in Appendix~\ref{sec:IIb}, after discussing the fraction of broad Type IIb SNe using our stretch-based identification method.}. However, these are biased observed fractions. The real unbiased fraction is of interest since our preceding analysis indicates that broadness is a good proxy for ejecta mass via multiple independent links between broad SE~SNe and high mass progenitors. However, even if they were all peculiar due to other reasons, their relative rate compared to ordinary SE~SNe is still valuable for constraining their origins. 

The two main sources of observational bias which significantly bias the results are the lightcurve duration bias, which affects sample selection step, and the Malmquist bias, which is a well known survey bias of magnitude limited surveys. Ideally, we are interested in quantifying the impact of all biases as a function of stretch value. We use novel methodology in each case to correct for the bias and present our results in Table~\ref{tab:IbcFrac} as the broad fraction after each bias correction step. Finally, other sources of bias, which are largely found to be insignificant for measuring the stretch distribution of our sample, are briefly discussed in Sect.~\ref{sec:otherbias}.

\begin{table}[]
    \centering
    \begin{tabular}{c|ccc}
        \toprule
        band     & observed & duration  & Malmquist  \\
        \midrule
        $r$ & $0.129^{+0.05}_{-0.04}$ & $0.112^{+0.05}_{-0.04}$ & $0.064^{+0.03}_{-0.03}$ \\ [8pt]
        $g$ & $0.250^{+0.06}_{-0.07}$ & $0.226^{+0.06}_{-0.07}$ & $0.187^{+0.06}_{-0.07}$ \\ [8pt]
        combined & $0.118^{+0.04}_{-0.04}$ & $0.102^{+0.04}_{-0.04}$ & $0.059^{+0.03}_{-0.03}$\\
        \bottomrule
    \end{tabular}
    \caption{The fraction of broad Type Ibc SNe after correcting for observational biases, reported with 90\% Poissonian uncertainties.}
    \label{tab:IbcFrac}
\end{table}

\subsection{Lightcurve duration bias \label{sec:lcd}}

Calculating the relative fraction of broad versus ordinary SE~SNe requires correcting for the observational bias that slowly evolving lightcurves can be more readily observed \citep[e.g.,][]{Kasliwal2012,Karamehmetoglu2017}, which we will call the lightcurve duration bias. Since we have relatively strict sample selection criteria based on lightcurve quality, this also means that broader objects have a higher chance of passing our criteria. 

We use the survey simulation tool \emph{simsurvey} \citep{Feindt2019} to simulate ${\sim}10^5$ model Type Ibc SN lightcurves in the redshift range of $z = 0.01-0.2$ and with a range of $0.5-3.0$ in lightcurve stretch. The simulated lightcurves were generated with a typical average iPTF cadence of two observations every three nights, minus the nights lost with realistic weather data measured for (i)PTF between 2009 to 2016. All epochs in the simulated lightcurve brighter than the typical iPTF magnitude limit (${\sim}21$ mag on dark nights, ${\sim}20.5$ mag on gray nights) were tested with the sample selection criteria to derive the final fraction of SNe passing the ``templatability`` criteria in a given stretch bin. We used the templatability criteria in Sect.~\ref{sec:thesample}, except that we here know the exact peak of the simulated lightcurve and do not have to derive this via fitting.

\begin{figure}
    \centering
    \includegraphics[width=0.95\linewidth]{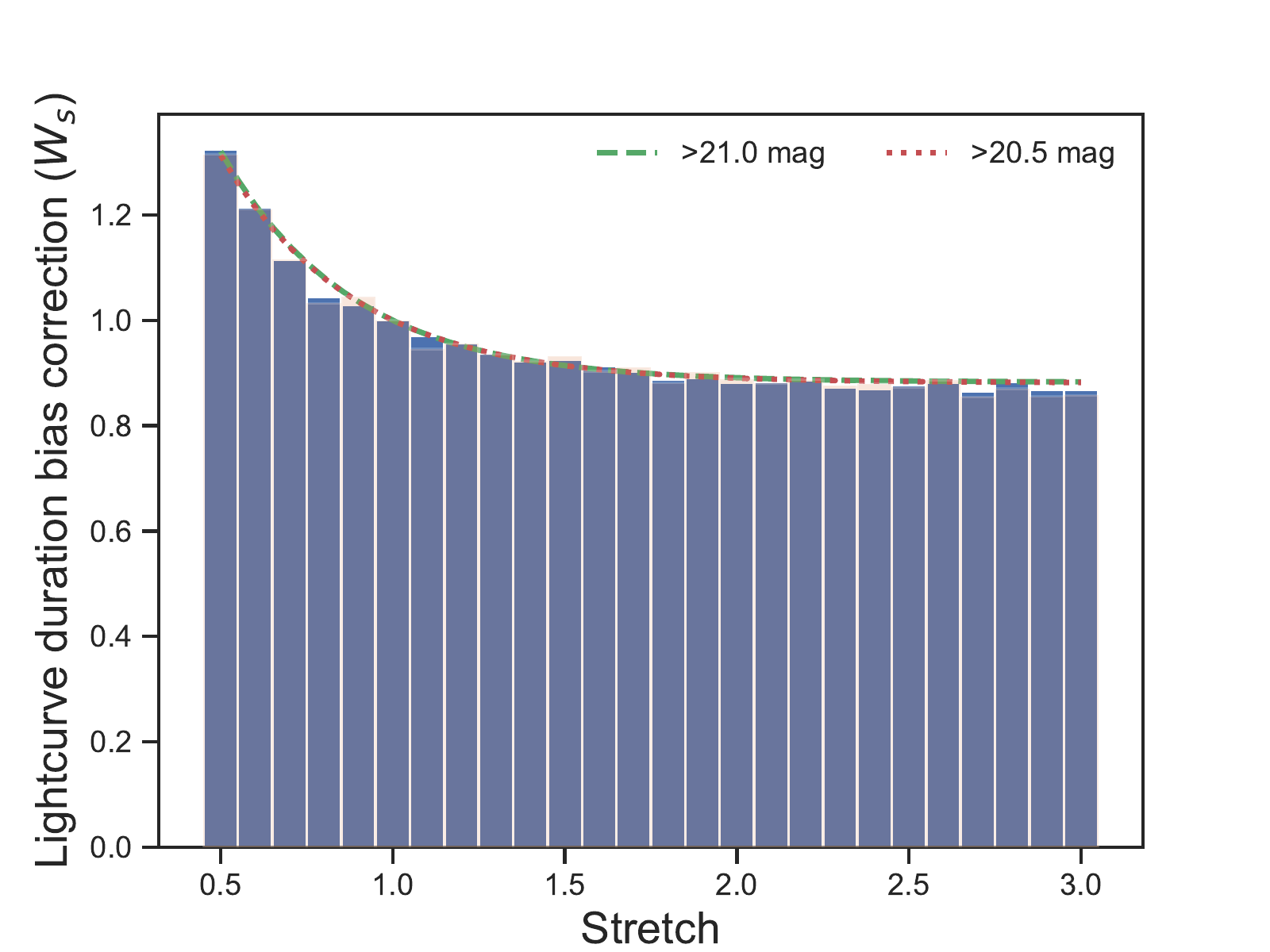} \\
    \includegraphics[width=\linewidth]{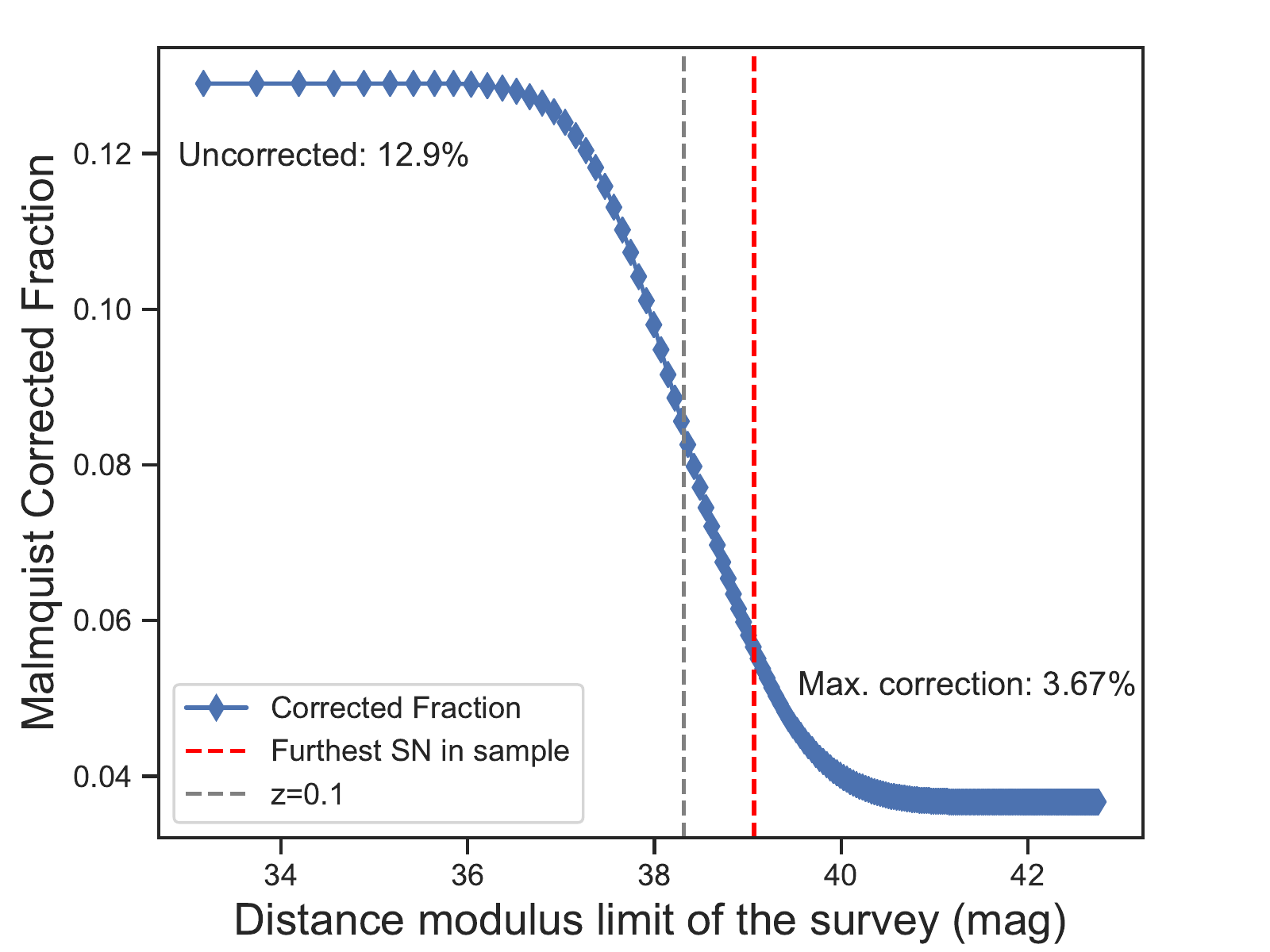}
    \caption{\emph{Top}: Weights $W_s=1/P_s$ that correct for the bias in detection and survey selection criteria as a function of the stretch parameter from ${\sim}10^5$ simulations of model Type Ibc lightcurves, while including realistic iPTF cadence and downtime (see Sect.~\ref{sec:lcd}). \emph{Bottom}: The impact of the Malmquist bias on the Type Ibc $r$-band broad fraction estimated from truncated regression fits to the broad and ordinary samples. Changing the redshift limit has a significant impact on the Malmquist bias (see Sect.~\ref{sec:textmalm}).}
    \label{fig:biases}
\end{figure}

To correct for this bias, we calculate a correction weight $W_s$ as a function of stretch ($s$) and plot the results in the top panel of Fig.~\ref{fig:biases}. If we represent the fraction of SNe that were detected and passed the survey selection criteria at a given $s$ as $P_s$, then $W_s=(1/P_s)$. $W_s$ can be used as a weight representing the proportion of SNe that should have been in our sample at that stretch value, but was missed due to the lightcurve duration bias. To obtain $W_s$ as a function of stretch, we fit our results using an exponential function of the form:

\begin{equation}
\label{eq:ws}
\centering
    W_s=e^{-\frac{s-s_0}{\tau}}+c.
\end{equation}

\noindent Where $s_0, \tau,$ and $c$ are fitted constants producing the green and red lines in the top panel of Fig.~\ref{fig:biases}, plotted after normalizing $W_s(1) = 1$ to yield the bias corrected relative occurrence in the sample. The values of the constants for different nominal (i)PTF cutoff values and their normalizations are reported in Table~\ref{tab:lcd}.

We find that a SN that is twice as broad as the template ($s\approx2$) is ${\sim}12\%$ more likely to be detected and pass our sample criteria as the template ($s\equiv1$). Since none of the lightcurves lacking a peak can pass our sample selection criteria, this procedure should also account for those SNe removed in step-zero of sample selection. Finally, in order to correct for this bias, we weight each SN in the broad and ordinary categories by $W_s$, based on its stretch value. When combining $r$ and $g$ bands, the $r$-band stretch value is preferred when available. We found that varying the magnitude cut-off limit of the survey by $\pm0.5$ mag did not have an appreciable effect on the relative weight attached to each stretch value within purely statistical uncertainties. 

Overall, the effect is small and only changes the rate from 12\% to 10\%, meaning that the typical cadence of (i)PTF was appropriate for not missing too many of the more ordinary or rapidly evolving SNe. We note that our lightcurve duration bias was calculated for the typical $r$-band cadence which was similar year to year for (i)PTF. Thus, it could be underestimated for the more inconsistently observed $g$ band.

\subsection{Malmquist bias \label{sec:textmalm}}

Any magnitude limited survey will suffer from an important observational bias referred to as the Malmquist bias \citep{1920MeLuF..96....1M}. Faint and more distant objects will be excluded from the sample, thereby biasing the distribution to be richer in more luminous events, which can be observed out to a further distance. As pointed out by \citet{Meza2020} and \citet{Ouchi2021}, broad, high-mass, large-nickel SE~SNe will be over-represented due to the Malmquist bias. The presence of this bias in the $r$-band lightcurves of Type Ibc (i)PTF SE~SNe is shown in Appendix~\ref{sec:malmappendix}. The bias comes from the fact that there is a weak but significant correlation between peak brightness and lightcurve broadness (see Sect.~\ref{sec:LCprop}). 

Since we do not know the luminosity function of SE~SNe as a function of stretch, we created a method\footnote{We note that our method can be used to correct for the effects of the Malmquist bias on an unrelated third variable measured in any magnitude limited survey in a very simple way.} to correct for the Malmquist bias, details of which can be found in Appendix \ref{sec:malmappendix}. The basic outline is as follows. We estimate the luminosity functions of the broad and ordinary samples using truncated regression models. Using these and a nominal survey detection cut-off ($21.0$ mag for iPTF), we calculate the detection probability as a function of survey distance (redshift). Finally, we integrate the probability of detection in each of the ordinary and broad groups out to the distance limit of the SN sample. On its own, without the lightcurve duration bias, this correction leads to a bias-free fraction of $0.074^{+0.04}_{-0.03}$ for the $r$ band and $0.21^{+0.06}_{-0.06}$ for the $g$ band, reported with $90\%$ confidence intervals.

As shown in the bottom panel of Fig.~\ref{fig:biases}, the effect of the Malmquist bias correction is sensitive to the maximum redshift allowed into the sample. At low $z_{max}$, almost all SNe in both the ordinary and broad groups would be bright enough to be detected and included in the sample. At high $z_{max}$, the opposite would be true and none would be included, making the Malmquist correction reach a maximum. Lowering this maximum to $z=0.1$ by excluding the 5 SNe above this value, the Malmquist bias corrected fraction instead becomes $10^{+4}_{-4}\%$ (after removing these SNe from all previous steps). While we can estimate the Malmquist bias for any given redshift limited sub-sample, it is not possible for us to pick this $z_{max}$ \emph{a posteriori} without biasing the outcome. Therefore, we use the redshift of the furthest SN in our sample at $z_{max}=0.138$ to find a conservatively large Malmquist bias correction.

Correcting also for the Malmquist bias brings the broad fraction from ${\sim}12\%$ to ${\sim}6\%$. Although we calculated the Malmquist correction based on the $r$ band data, the relative similarity of the luminosity functions in either band and the well known similarity in $g-r$ color displayed by SE~SNe around peak (Sect. \ref{sec:colors}) means that the estimate should be valid for both bands.

\subsection{Human (follow-up) bias}

Ultimately, the human element to the (i)PTF scanning process remains the main problem, since there could be complex human biases based on the particular scanner working on a day. In fact, other than the duration bias, the $g$ band broad fraction being much larger than the $r$ band one is likely due to the human follow-up bias. This is due to the fact that triggering $g$-band observations was significantly influenced by human decision making, since data acquisition was more automatic in that band. 

\citet{2017ApJS..230....4F} have calculated the transient source detection efficiency of the PTF, which should account for some of these human biases. However, their method only considers single detections of a source, and does not consider the biases from follow-up, multiple detections in a night, lightcurve shape, and non-detection history. We will not attempt to correct for these hard to quantify biases with the (i)PTF. More promising are approaches which remove the human bias, such as the strategy followed by ZTF Bright Transient Survey \citep{Perley2020,Fremling2020}, which can be used in the future to constrain the impact of this bias and improve our results. In a similar vein, the magnitude of the human follow-up bias should be more limited in the primary survey band ($r$). Combined with the larger sample size and faster cadence, we prefer the broad fraction in the $r$ band as a less biased estimate. 

\subsection{Corrected fraction of broad Type Ibc SNe \label{sec:otherbias}}

After conservative bias corrections, most significantly from the Malmquist bias, the fraction of broad Type Ibc SNe in (i)PTF is $\approx6^{+3}_{-3}\%$ in the $r$ band and $\approx19^{+6}_{-7}\%$ in the $g$ band. However, since the $g$-band fraction suffers from a follow-up bias, and since our bias corrections were calculated for the more numerous $r$ band, the $r$-band and combined broad fractions are less biased, and we focus on them in the rest of the discussion. It is worth reiterating that we could be underestimating the broad fraction by over-correcting for the Malmquist bias, since the corrected fraction could be as high as ${\sim}10\%$ when using a very reasonable but slightly lower $z_{max}=0.1$ in the Malmquist bias correction (Sect.~\ref{sec:textmalm}).

In reaching our result, we also considered several other biases but concluded that their impacts were minimal. These include: 1) template fitting errors, 2) feature selection (appropriateness of using single-band stretch), 3) cosmic evolution, 4) K-corrections, 5) extinction, and 6) correlated biases. We ruled out 1) by the fact that, within one sigma, standard errors of the stretch yield the same label, and by the fact that all SNe with multiband data have the same label in both bands. For 2), the fact that single-band template stretch provides good visual fits, independently recovers numerous broad and rapidly-evolving literature SNe, and is found to be just as good as using rise or decline times (Sect.~\ref{sec:LCprop}), argues for our feature selection method. For 3), cosmic evolution as a possible bias is not supported due to the redshift distribution of our sample having a relatively low limit and being mixed between the broad and ordinary SE~SNe. For 4), since only a rapid evolution of K-corrections around peak can influence stretch, we verified that K-corrections calculated from template Type Ibc spectra\footnote{Obtained from Nugent, P. (\url{https://c3.lbl.gov/nugent/nugent_templates.html}) based on the work in \citet{NugentTemplate}.} were not rapidly varying around peak. In the redshift range of our sample, the calculated K-corrections vary by maximum ${\sim}0.01$~mag for 10 days around peak, below even measurement uncertainties. Furthermore, correlations between K-corrections and stretch are found to be below the standard error of fitted stretch values. 5) Similarly to K-corrections, host or MW extinction do not bias the measured stretch due to being constants, nor do the broad SNe show evidence of peculiar or strong extinction, as evidenced by their colors (Sect. \ref{sec:colors}) and spectra (Sect. \ref{sec:spec}). Finally for 6), we tested correlations using the shared parameter of brightness, which is used in both bias correction methods and is actually the primary \emph{cause} of the Malmquist bias. We found that varying the brightness when calculating the lightcurve duration bias had a minimal effect, and thus believe independence to be a reasonable assumption due to minimal correlations.

\section{Discussion \label{sec:discussion}}
Motivated by the success of our method in discovering a sample of broad Type Ibc SNe, we also tested applying it to Type IIb SNe (see Appendix~\ref{sec:IIb}). Moving forward, we include Type IIb SNe when we refer to the combined broad SE~SNe sample to aid literature comparisons. However, since adding Type IIb SNe did not significantly change our results (see Table \ref{tab:broadFrac}), the discussion would still be valid for just Type Ibc SNe. 
\subsection{Corrected fraction of broad SE~SNe \label{sec:corrfrac_disc}}
\begin{table}[]
    \centering
    \begin{tabular}{c|ccc}
        \toprule
        band     & observed & duration  & Malmquist  \\
        \midrule
        $r$ & $0.133^{+0.04}_{-0.03}$ & $0.119^{+0.04}_{-0.03}$ & $0.064^{+0.03}_{-0.02}$\\ [8pt]
        $g$ & $0.238^{+0.04}_{-0.05}$ & $0.218^{+0.05}_{-0.05}$ & $0.129^{+0.04}_{-0.04}$\\ [8pt]
        combined & $0.131^{+0.03}_{-0.03}$ & $0.117^{+0.03}_{-0.03}$ & $0.063^{+0.02}_{-0.02}$ \\ 
        \bottomrule
    \end{tabular}
    \caption{The fraction of broad SE~SNe (Type Ibc and Type IIb combined) after correcting for observational biases, reported with 90\% Poissonian uncertainties.}
    \label{tab:broadFrac}
\end{table}

We estimate the fraction of broad SE~SNe in (i)PTF as $13^{+4}_{-3}\%$ in the $r$ band and $24^{+4}_{-5}\%$ in the $g$ band. After all bias corrections, these become $6^{+3}_{-2}\%$ and $13^{+4}_{-4}\%$, respectively (Table~\ref{tab:broadFrac}). Similar to SNe Ibc, we consider the $r$-band estimate to be more reliable. As discussed in Sect.~\ref{sec:textmalm}, if the survey is instead limited to $z=0.1$, this estimate goes up to ${\sim}10^{+3}_{-3}\%$ in the $r$ band due to the significantly decreased role of the Malmquist bias, so the real fraction of broad SE~SNe could be as large as ${\sim}10\%$.

In comparison, the uncorrected fraction of broad Type Ibc and IIb SNe in the samples of \citet{Lyman2016} and \citet{Prentice2018} were $3^{+7}_{-3}\%$ and $4^{+5}_{-3}\%$, respectively (with 90\% Poissonian confidence intervals). The respective authors drew their samples from the literature and did not attempt to calculate any corrections for observational biases. Since correcting for observational biases from Sect.~\ref{sec:biases} would lower the literature broadness fraction, these estimates can be treated as upper limits. 

While our observed and corrected fractions are consistent with the literature upper limits at the 90\% confidence level, this work establishes a conservative minimum that, at the $90\%$ confidence limit, $>4\%$ of SE~SNe are broad, unlike the $>0\%$ and $>1\%$ as found in the biased values from the literature. We conclude that we have found the fraction of broad SE~SNe in (i)PTF to be larger than previously appreciated in the literature. 
The discrepancy suggests that the (i)PTF and literature studies sample different SN populations, and the inconspicuous 
low-metallicity environments of the broad sample might play a role. 

\subsection{Lightcurve broadness as a proxy for ejecta mass \label{sec:broadproxy}}

Following the Arnett model with a diffusion approximation, the association of lightcurve broadness with ejecta mass is expected on theoretical grounds, as long as the ejecta velocities between SE~SNe are not too different. The diffusion timescale ($\tau_m$) effectively sets the lightcurve width and $\tau_m \propto (M_{ej}/V_{ph})^{1/2}$ \citep{Arnett1982,Cano2013}. Therefore, an increase in the diffusion time and hence lightcurve broadness requires an increase in the ratio of ejecta mass to the characteristic ejecta velocity. 

Our bolometric lightcurves are broader and slightly brighter than the average for a SE~SN (Fig.~\ref{fig:LCprop}). Meanwhile, the velocities we measure are similar to what has been found for more ordinary SE~SNe. For example, the \ion{Fe}{ii} velocity in the nearby Type Ib SN iPTF13bvn evolves from $10^4$~km~s$^{-1}$ a week after peak to about half of that two months after. This mirrors the \ion{Fe}{ii} velocity evolution of our broad sample although they peak later and the evolution occurs more gradually (Fig.~\ref{fig:linevels}). In our bolometric Arnett modeling, we find an expected straightforward relationship between high-stretch Type Ibc SNe with larger than 10~$M_\sun$ of ejecta (Fig.~\ref{fig:EjNi}). The distribution of ejecta masses of our broad sample is significantly higher than what has been deduced for most SE~SNe in the literature (they are ${\sim}6$--$10~M_\sun$ more massive).

The fact that a simple stretch of the CSP SE~SN color templates seems to fit the broad SE~SNe (Fig.~\ref{fig:colors_all}) suggests that not only are the spectra of our sample generally similar to those of an ordinary SE~SN, but they also evolve at the approximate pace set by the lightcurve stretch. Our spectral analysis in Sect.~\ref{sec:spec}, also revealed major similarities between the spectra of broad and ordinary SE~SNe, modulated by stretch.
This behavior is most naturally explained if the broad SE~SNe have higher ejecta masses but are otherwise similar to ordinary SE~SN explosions, as opposed to if the two groups have completely different explosion and powering mechanisms. The higher ejecta masses will serve to slow the evolution of the SED as it will take longer for the optical depth to drop. Indeed, we also observe that the broad SNe take longer to become nebular compared to ordinary SE~SNe (Sect. \ref{sec:spec}). 

We also compared the nebular line flux ratio of [\ion{Ca}{ii}] to [\ion{O}{i}]. A lower value is associated with high ejecta masses in models \citep{Fransson1987,Jerkstrand2015}. \citet{Terreran2019} found a very low value of this ratio for SN~2016coi, which had radio and X-ray observations showing high mass-loss rates associated with a high-mass progenitor. It also had a broad lightcurve, and lightcurve modeling indicating a high ejecta mass. A low value of [\ion{Ca}{ii}] to [\ion{O}{i}] was also associated with broader lightcurves and more massive ejecta in a study of many SE~SN nebular spectra by \citet{Fang2019, Fang2022}, but see also \citep{Dessart2021,Ergon2022}. We found that all four of our broad SE~SNe with clearly-nebular spectra had some of the lowest values of [\ion{Ca}{ii}]/[\ion{O}{i}] when compared to the literature (Fig.~\ref{fig:CaIIOI}). This suggests that our broad SE~SNe come from high-mass progenitors with large ejecta masses.

Looking at their environments, the broad sample seemed to be preferentially found in high sSFR and low-metallicity host galaxies. High sSFR is associated with massive stars. We also showed that the broad sample was found in environments typical for superluminous and Type Ic-BL SNe, which are both thought to come from high-mass stars. Low metallicity has been discussed as a way to have high ejecta mass SE~SN explosions \citep{Yoon2015}, which fits the picture that the broad SE~SNe have higher ejecta masses. 

Considering the totality of the evidence, commonly used indicators form a consistent picture suggesting that our broad sample comes from a population of high ejecta-mass SE~SN explosions with high-mass progenitors. Our results are thus good evidence for the lightcurve stretch value being a good proxy for ejecta mass. 

\subsubsection{Ultra-stripped SNe}
If stretch is a good proxy for ejecta mass, then the only known ultra-stripped SN from (i)PTF, the Type Ic SN iPTF14gqr \citep{De2018a}, should be among the lowest stretch value SE~SNe, as ultra-stripped SNe have the lowest ejecta masses among SE~SNe. In fact, we identify both iPTF14gqr ($s_{Ibc}^{r} = 0.44\pm0.01$) and a new ultra-stripped candidate, PTF12fgw ($s_{Ibc}^{r} = 0.40\pm0.04$), as the most rapidly evolving SNe. Their stretch values are consistent with each other and more than three sigma lower than the next lowest (PTF10hie, $s_{Ibc}^{r} = 0.63\pm0.02$). The pre-detection upper limits and $r$-band lightcurve of PTF12fgw reveal a one week rise time similar to that of iPTF14gqr. The correction factor for the lightcurve duration bias (Sect.~\ref{sec:lcd}) for the count rate of these two events is $\approx1.45$, which brings the count rate of ultra-stripped SNe given Poisson noise to $3 \pm 2$ in our sample of (i)PTF, which is now fully consistent with the four predicted by \citet{Hijikawa2019}. The high and low extremes of our stretch distribution are thus directly linked to SE~SNe with the highest and lowest ejecta masses.

\subsection{Ejecta mass distribution of (i)PTF SE~SNe}

Based on the evidence linking the distribution of lightcurve stretch values to that of ejecta mass, we use stretch as a proxy to investigate the ejecta mass distribution of SE~SNe from the untargeted (i)PTF surveys and compare it to the literature results with SNe mostly from targeted surveys. \citet{Prentice2018} collated most of the literature results in a consistent manner. They find that the ejecta mass distribution of SE~SNe is unimodal, with the vast majority having ejecta mass $\lesssim 4$ $M_\sun$. With the Type Ic-BLs removed from their sample (SNe Ic-3/4 in their classification scheme), the only two high-ejecta mass events are the Type Ic SN~2011bm (11 $M_\sun$) and Type IIb SN~2013bb (4.8 $M_\sun$), both of which we also recover using lightcurve broadness. The well studied SN~2011bm is PTF11bov in our sample, and SN~2013bb is iPTF13aby with a stretch value of 1.85 in the $r$ band. 

In comparison, GMM fits to the stretch distribution of lightcurve stretch values in our (i)PTF sample prefer a bimodal distribution composed of the ordinary and broad SE SNe. Bolometric lightcurve modeling and evidence from nebular spectra indicate that the broad SE~SNe have ejecta masses significantly above that typical for most SE~SNe. As seen in Figs.~\ref{fig:hist_norm_stretch} and \ref{fig:kmeans_broad}, the main body of Type Ibc SN stretch values cluster around 1.0 and seem to be normally distributed. The high-stretch/mass tail, ${\sim}13\%$ of the total, is more than five times over-represented for a Gaussian distribution centered at 1.0 with $\sigma \approx 0.25$, which is not compatible with a unimodal distribution.

Our estimate of the real fraction of broad SE SNe represents the ratio between the broad and ordinary plus broad categories of the bimodal ejecta mass distribution. We have estimated that a not-insignificant fraction (${\sim6\%}$) of SE SNe have high ejecta masses and could be coming from high mass progenitors. 

\subsection{Progenitor implications of the ejecta mass distribution}
The observed ejecta mass distribution has implications for the high mass progenitors and formation channels of SESNe. Whether very high-mass stars explode as SE~SNe or fail and form black holes is an ongoing area of research \citep{Smartt2015,OConnor2017}. The confirmation of a not-insignificant population of high-mass SE~SN progenitors offers a possible resolution to problems posed by the case for the missing high-mass stars \citep{Smartt2009}. $6\%$ of SE~SNe having high ejecta masses could account for approximately half the missing oxygen problem (\citealp{Suzuki2018}; A. Suzuki, priv. comm.). Additionally, successful CC SN explosions of stars with $M_{ZAMS}>25~M_\sun$ at sufficient rate are likely needed to form the observed high-mass neutron stars \citep{Raithel2018}.

Similar to our observed distribution, in the binary population synthesis simulations of \citet[their figure 5]{Zapartas2017}, the final mass of SE~SN progenitors shows a distinct double peaked structure of low and high mass (and a gap), which they identify as coming from two separate formation channels of primarily binary or wind-stripped, respectively. However, since their work only focused on SE SNe with low mass companions, or no companions, we redo the same calculation independent of the possible binary companion mass using their binary populations synthesis model. Assuming a realistic mix of single and 70\% initially binary stellar systems \citep{Sana2012}, our only difference is that we follow \citet{Fryer2012} in order to take into account the possibility of weak or failed explosions due to direct collapse onto a black hole \citep[e.g.,][]{OConnor2011}. We do the calculation for a stellar and binary population of $Z=0.004 = 1/4~Z_\sun$ metallicity, close to the mean metallicity found in our work (corresponding to $\log(O/H) = 8.2$, see Fig.~\ref{fig:hostmetal}).

We find no progenitors with $M_{\rm ej}> 6 M_\sun$ that are stripped through binary mass transfer at the given metallicity. Instead, they must be primarily wind-stripped. This ejecta mass threshold is consistent with the ejecta masses we find in our (i)PTF sample (see section 5.4 and Fig~\ref{fig:EjNi}), indicating that the observed bimodality in the broadness of SE~SN lightcurves may originate from differences in progenitor evolutionary channels. In this exercise, we calculated that SESNe with $M_{\rm ej}>6 M_\sun$ are $\sim 20\%$ of all SESNe, higher than our corrected Broad SE~SN fraction. However, since our observed sample comes from a range of metallicities, while this exercise was only done for a single typical metallicity and makes simplifying assumptions of which progenitors produce a SE~SN, we find the fractions close enough to merit comparison. 

The higher ratio of wind-stripped progenitors in the \citet{Zapartas2017} models may point towards lower wind mass loss rates during the main-sequence and red supergiant evolutionary phases than typically assumed \citep[e.g.,][]{Smith2014, Beasor2020}, especially at low metallicity. Alternatively, a higher dominance of failed SNe for these high mass stripped stars compared to the assumed \citet{Fryer2012} prescription \citep[e.g.,][]{Sukhbold2016,Patton2020,Zapartas2021} may be responsible. In the comprehensive work by \citet{Sukhbold2016}, neutrino driven CC SN simulations can explode lower mass stars, and some higher mass stars in the so called islands of stability, but struggle especially with intermediate mass stars around ZAMS masses ${\sim}20-25~M_\sun$. The high stretch SE~SNe we identified in this paper are natural candidates for high mass SN explosions located in these islands, while the gap could represent the dearth of explosions in the intermediate-mass range. However, the ``explodability`` of high mass stars remains a major uncertainty affecting our comparison. Due to the uncertainties, an in-depth investigation of the theoretically predicted progenitors with high ejecta masses, for different set of assumptions and metallicities, is out of the scope of this work.

\subsection{The large nickel mass problem \label{sec:nickel}}

\citet{Kushnir2015} and \citet{Anderson2019} compiled literature nickel masses and found that the masses for SE~SNe were higher than for Type II SNe, despite the expectation that SE~SNe stripped in intermediate mass binaries should share the same progenitor mass-range as proposed for Type II SNe, and hence have similar core and nickel masses. This result was verified by \citet{Meza2020} focusing on SE~SNe. \citet{Ouchi2021} proposed that observational biases enumerated in this paper, namely Malmquist and duration biases, could account for the difference between Type II and SE SNe nickel masses, and issued an observational challenge for more bias free samples. Subsequently, using the ZTF Bright Transient Survey, \citet{Sollerman2022} showed that many SE~SNe indeed are more luminous than predicted by explosion models.

Our broad sample, being also brighter on average, naturally represents the extrema of the nickel mass range for SNe Ibc with an average value of ($0.42\pm0.08$ $M_{\sun}$). As pointed out by \citet{Ouchi2021}, SE~SNe with such large nickel masses could be coming from a separate progenitor population of high mass stars, which also fits the large ejecta masses. As highlighted in the above works, the \emph{relative} nickel masses we obtain are likely correct despite shortcomings in the Arnett model, meaning that our SNe have more nickel than the average SE SNe. Both our nickel masses, and those of bright SE~SNe in the literature, are in excess of values from detailed simulations of neutrino powered explosions (${\sim}0.2~M_\sun$ \citealp{Anderson2019,Ertl2019}). 

\subsection{Alternative powering mechanisms \protect\protect\label{sec:multipm}} 

While we favor the traditional scenario of 
nickel-powered lightcurves, we here discuss alternative models such as circumstellar material interaction (CSMi) and magnetars as possible explanations for the broad SE~SNe. Given that, on the whole, our sample seems to be relatively similar to other SE~SNe, attributing the consistently large ejecta masses for our broad objects compared to ordinary SE~SNe as due to the simplicity of the Arnett model seems contrived.
Certainly, assumptions underlying the model, such as constant effective opacity, centrally located nickel, and spherical symmetry are likely violated by SE~SNe \citep[e.g.,][]{Hoeflich1998,Ergon2015a,Dessart2016,Taddia2018d,Anderson2019}. Despite this, statistical approaches relying on the distribution-of or trends-in explosion parameters can overcome some of the uncertainty (for example, asymmetry and observer viewing-angle effects should be averaged out in larger samples, assuming homogeneity).

For an alternate model to be responsible for the broad lightcurves observed in our SNe and significantly affect our results, it needs to be as strong as radioactive decay powering of ordinary Type Ibc SNe during the bell-shaped part of the lightcurves and contribute to the luminosity output during a similar timescale. Indeed, alternate powering mechanisms may be a feasible explanation for the undulations and late-time luminosity seen in some of our lightcurves, such as in PTF11mnb, iPTF15dtg, and iPTF16flq, or may be present in all SE~SNe. If, as proposed by some, SE~SNe lightcurves all are dominated by alternative powering mechanisms (see \citealp{Dessart2021} for CSMi, \citealp{Ertl2019} for magnetars, and \citealp{Soker2019} for jets, among others), the broad SE~SNe may be associated with simple differences in progenitor conditions (such as higher initial mass), since their long duration lightcurves and otherwise ordinary slower evolution still need to be accounted for. 

Starting with CSMi, the broad SE~SNe do not display the typical observables of SNe powered by CSMi such as narrow emission lines with strong electron-scattering wings, unusually blue colors, flat-topped line profiles at late-times, or low velocity absorption. CSMi almost certainly occurs in many CC SNe and can become significant at late times even when early observable are hidden. However, the progenitors of Type Ibc SNe likely have wind speeds in excess of $10^3$~km~s$^{-1}$. Meaning that, compared to Type II SNe, the CSM needs to be at significantly higher density, and hence the progenitors require even larger mass-loss rates, to drive a radiatively-powered shock strong enough to power the main lightcurve. In fact, many SNe displaying signs of interaction with hydrogen-poor CSM have been discovered as Type Ibn and Icn SNe, and they do not observationally resemble our Broad sample \citep{Hosseinzadeh2017,Perley2022}.

In the few known examples of Type Ibc SNe with strong CSMi hidden at early times, there was either significant rebrightening, or appearance of CSMi signatures, or both \citep{Milisavljevic2015,Sollerman2020}. Furthermore, the main peaks of these SNe are still thought to be primarily nickel powered. Given that we neither see any common indications of CSMi nor have evidence for hidden CSMi, we do not favor this scenario. For PTF11rka, CSMi powering was proposed by \citet{Pian2020} based on low absorption velocities compared to SLSNe. However, our work shows that the broad SNe are better compared to SE~SNe, among which their velocities are completely normal. 

Magneters have been invoked to explain extraordinary Type Ibc SNe, such as the double peaked SN~2005bf \citep{Folatelli2006}, or PTF11mnb and iPTF15dtg \citep{Taddia2016,Taddia2018e,Taddia2019a}. We do not favor magnetars over standard radioactive powering since, unlike CSMi, magnetar models lack strong observable markers, and unlike the analogous case with SLSNe, the broad SNe do not require them. 

If nickel is mixed further out into the ejecta, via a mechanism such as highly asymmetric ejecta or jets, then brighter lightcurves can be expected. However, asphericity is not thought to be associated with lightcurve broadness \citep{Hoeflich1998}. In comparison, jet-driven explosion scenarios make possible significantly asymmetric explosions, high nickel mass yields with large mixing, and may produce luminous SNe \citep[see e.g.,][]{Khokhlov1999,Barnes2018,Soker2019,Soker2022}. 

\section{Conclusions and outlook}

We discovered that $\sim13\%$ of the observed SE~SNe in a magnitude limited sample, and approximately $\sim6-10\%$ in a bias corrected sample, have optical lightcurves that are significantly broader than average. These broad SNe have spectra, velocities, color evolution, and lightcurve shape which are similar to ordinary SE~SNe, with the evolution timescale mainly modulated by the lightcurve broadness, which we measured with a template stretch parameter. We also showed via bolometric lightcurve modeling that if they are radioactively powered, the SNe in the broad sample have ejecta and nickel masses that are significantly larger than typical for ordinary SE~SNe. Therefore, unlike most SE~SNe which are thought to originate from intermediate mass stars in binaries, the broad sample is compatible with having high-mass progenitors. Also, their nebular spectra and host environments show a preference for high-mass progenitors.  

Our work represents a first attempt at obtaining a bias-corrected estimate of the fraction of high-mass stars from lightcurve studies of SE~SNe; correcting for biases such as lightcurve duration, sample selection, and Malmquist bias while using an untargeted sample. Growing sample sizes from surveys such as ZTF and LSST will increasingly make this type of statistical study more powerful for SN science. To act as a guide, we have demonstrated the first order difficulties in obtaining unbiased samples in a quantitative manner from a large survey, using the combined (i)PTF dataset (Appendix A). We also demonstrated a method for correcting for the Malmquist bias in a third (correlated) parameter in a magnitude limited survey, using truncated regression models (Appendix C).

As sample sizes grow, using observational parameters to study statistical properties of SN populations will become an increasingly viable way to answer long standing questions in the field \citep{Perley2020}. We showed that a simple observational parameter, lightcurve stretch, can be used as a good proxy for underlying physical quantities, such as ejecta mass, at least in a sample of SE~SNe with "normal" spectra. However, we could not completely rule out the contribution of alternate powering mechanisms such as CSM interaction. Our simple approach also uncovered evidence for low-stretch, likely ultra-stripped, SNe Ic, as well as a bimodality in stretch values, which may be evidence of a separate progenitor population for the broad SE~SNe.

While a few SE~SNe with broad lightcurves could simply have been uncommon outliers, by quantifying and correcting for observational biases, as well as constructing the sample in a reproducible way, we show that the observed and real fraction is non-zero. Therefore, massive progenitors, or an alternative explanation, can not be ignored when considering the progenitor scenarios of SE~SNe. 

Without being exhaustive, we highlighted implications of the discovery of a small (but significant) population of massive star progenitors. If confirmed with greater sample sizes, the broad SE~SNe may offer answers to the missing high-mass stars problems, such as oxygen abundance and the existence of high-mass neutron stars. More precise numbers divided by SE~SN subtype could offer insights into progenitor formation channels of SE~SNe, such as binary and single star, the mass-loss histories of massive stars, and even the rates of direct collapse to stellar mass black holes via comparison to population synthesis simulations. 

\begin{acknowledgements}
We gratefully acknowledge support from the Knut and Alice Wallenberg Foundation. E.K., F.T. and M.S. are supported in part by grants from the VILLUM FONDEN (grant number 28021) and  the Independent Research Fund Denmark (IRFD; 8021-00170B). The Oskar Klein Centre is funded by the Swedish Research Council.
The intermediate Palomar Transient Factory project is a scientific collaboration among the California Institute of Technology, Los Alamos National Laboratory, the University of Wisconsin (Milwaukee), the Oskar Klein Centre, the Weizmann Institute of Science, the TANGO Program of the University System of Taiwan, and the Kavli Institute for the Physics and Mathematics of the Universe. 
This work was supported by the GROWTH project funded by the National Science Foundation under Grant No 1545949.

Some of the data presented herein were obtained at the W. M. Keck Observatory, which is operated as a scientific partnership among the California Institute of Technology, the University of California, and NASA. The Observatory was made possible by the generous financial support of the W. M. Keck Foundation. The authors wish to recognise and acknowledge the very significant cultural role and reverence that the summit of Maunakea has always had within the indigenous Hawaiian community. We are most fortunate to have the opportunity to conduct observations from this mountain.
Partly based on observations at Kitt Peak National Observatory (KPNO), National Optical Astronomy Observatory, which is operated by the Association of Universities for Research in Astronomy (AURA) under cooperative agreement with the NSF. The authors are honoured to be permitted to conduct astronomical research on Iolkam Du'ag (Kitt Peak), a mountain with particular significance to the Tohono O'odham.
The data presented here were obtained in part with ALFOSC, which is provided by the Instituto de Astrofisica de Andalucia (IAA) under a joint agreement with the University of Copenhagen and NOT.

Partly based on observations made with the University of Hawaii’s 2.2-m telescope, at Maunakea Observatory, Hawaii, USA.

Partly based on observations made with the William Herschel Telescope operated on the island of La Palma by the Isaac Newton Group of Telescopes in the Spanish Observatorio del Roque de los Muchachos of the Instituto de Astrof\'{i}sica de Canarias.

Partly based on observations made with the 5.1-m Hale Telescope (P200), at Palomar Observatory, California, USA.

Partly based on observations made with the Kast spectrograph on the Shane 3-m telescope at Lick Observatory, Mount Hamilton, California, USA. Research at Lick Observatory is partially supported by a generous gift from Google.

This work is partly based on observations made with DOLoRes@TNG.

These results made use of the Discovery Channel Telescope (DCT) at Lowell Observatory. Lowell is a private, non-profit institution dedicated to astrophysical research and public appreciation of astronomy and operates the DCT in partnership with Boston University, the University of Maryland, the University of Toledo, Northern Arizona University and Yale University. The Large Monolithic Imager was built by Lowell Observatory using funds provided by the National Science Foundation (AST-1005313). The upgrade of the DeVeny optical spectrograph has been funded by a generous grant from John and Ginger Giovale and by a grant from the Mt. Cuba Astronomical Foundation.

Based on observations obtained at the Gemini Observatory [include additional acknowledgement here, see section 1.2], which is operated by the Association of Universities for Research in Astronomy, Inc., under a cooperative agreement with the NSF on behalf of the Gemini partnership: the National Science Foundation (United States), National Research Council (Canada), CONICYT (Chile), Ministerio de Ciencia, Tecnología e Innovación Productiva (Argentina), Ministério da Ciência, Tecnologia e Inovação (Brazil), and Korea Astronomy and Space Science Institute (Republic of Korea).

Based on observations collected at the European Southern Observatory under ESO programme(s) 090.D-044.

EZ acknowledges funding support from the European Research Council (ERC) under the European Union’s Horizon 2020 research and innovation programme (Grant agreement No. 772086).

TP acknowledges the financial support from the Slovenian Research Agency (grants I0-0033, P1-0031, J1-8136 and Z1-1853). 

MMK acknowledges generous support from the David and Lucille Packard Foundation.

AGY’s research is supported by the EU via ERC grant No. 725161, the ISF GW excellence center, an IMOS space infrastructure grant and BSF/Transformative and GIF grants, as well as the André Deloro Institute for Advanced Research in Space and Optics, The Helen Kimmel Center for Planetary Science, the Schwartz/Reisman Collaborative Science Program and the Norman E Alexander Family M Foundation ULTRASAT Data Center Fund, Minerva and Yeda-Sela; AGY is the incumbent of the The Arlyn Imberman Professorial Chair.

The author would like to thank A. Suzuki for his valuable contributions. The author would like to thank J.O.~Persson at the Statistical Research Group in the Department of Mathematics, Stockholm University, for his contributions to the Malmquist Bias correction. The author would like to thank J. D. Lyman for his helpful comments and suggestions. The author would like to thank P. Hoeflich for his valuable insights. We would like to acknowledge the numerous students, postdocs, observers, engineers, and staff whose hard work was crucial in providing the extensive (i)PTF data used in this paper.

This research made use of Astropy, a community-developed core Python package for Astronomy (Astropy Collaboration, 2018). IRAF, is distributed by the National Optical Astronomy Observatories, which are operated by the Association of Universities for Research in Astronomy, Inc., under cooperative agreement with the National Science Foundation. The python version which was used, PyRAF, is a product of the Space Telescope Science Institute, which is operated by AURA for NASA.

\end{acknowledgements}

\bibliographystyle{aa}
\bibliography{ads_references.bib}

\begin{appendix}

\section{Constructing the broad sample \protect\protect\label{sec:sampleselection}}
Here we describe in detail the steps used to arrive at the final sample of Type Ibc SE~SNe with broad lightcurves. We used a reproducible and quantitative method based on template fitting to obtain the distribution of stretch parameter representing lightcurve broadness. We also applied various tests and corrections to the obtained stretch distribution as described below.

\subsection{Sample selection criteria}
The criteria for inclusion in the sample were based on whether the SNe are ``templatable'' or not, which we describe in detail below. The selection criteria were chosen to ensure that the SN lightcurve can constrain the template fit (Sect.~\ref{sec:LCfit_init}), especially the stretch value that we used to measure the lightcurve broadness.

As a first step, we removed all SNe that do not have data points around peak, i.e., the range of the photometric epochs should cover the characteristic bell-shaped lightcurve of a SN. This is done since we require a peak measurement in order to fit our template. For example, our template fitting process would fail on a monotonically declining lightcurve such as the late-time tail of a SE~SN lightcurve. Some SNe were classified with only a few epochs of photometry, and these are also impossible to fit with a template. A total of 45 Type Ibc do not have data that include the peak, leaving 73 Type Ibc (Table \ref{tab:cuts}). We were conservative in this step, and only removed objects clearly lacking a peak. Since some of the remaining objects also cannot constrain a template fit, we then applied a numerical template selection criteria.

In our ``templatability'' criteria, the typical bell-shaped lightcurve of a SE~SN is split into four intervals, the rising part, the peak, the decline from the peak, and the late-time linearly declining part. Next, the number of points ($N$) in each interval is checked, subject to the following conditions: points on the rising and declining parts must have a separation of at least ${\Delta}t_{rise}$ and ${\Delta}t_{dec}$ days, respectively, while points near the peak must be within the interval ($i$) of $i_{peak}$ days of the peak epoch. Using this method, we establish the ``templatability'' criteria based on the count values of $N_{rise}$, $N_{dec}$, $N_{declate}$, and $N_{peak}$, which themselves are only dependent on ${\Delta}t_{rise}$, ${\Delta}t_{dec}$, $i_{peak}$, the peak epoch, and the definition of the rise, decline, and late-declining intervals. We estimated the peak and the definition of the lightcurve intervals using the template fit\footnote{Alternatively, a low-order polynomial or spline fit could be used.}. The peak of the lightcurve was the peak of the template $\pm i_{peak}$~days, where we set $i_{peak}=7\times s_{sn}$ days and $s_{sn}$ is the stretch value of the template fit.

The numerical criteria for inclusion in the final sample were then as follows: 

1) $N_{rise}$ > 1 and $N_{rise}+N_{dec}+N_{declate} \geq 6$~points, with ${\Delta}t_{rise},{\Delta}t_{dec}=2,4$~days, respectively,

or

2) $N_{rise},N_{peak},N_{dec}>0$ and $N_{rise}+N_{dec}\geq 4$, with ${\Delta}t_{rise},{\Delta}t_{dec}=1$~day.

\noindent The first criterion selects all lightcurves that have at least six points on the rise and decline that are separated by at least two days on the rise and at least four days on the decline. The trend established by a well observed rising and declining lightcurve strongly constrains the template fit, as long as these points are sufficiently separated, even though these lightcurves may not have points exactly at peak. The second criterion selects those lightcurves which have at least 1 point at peak, rise and decline, and a total of at least 4 points on the rise and decline combined separated by at least 1 day. We found that a point at peak, when combined with several points on the rise and decline, also can constrain the template fit even if the coverage is more sparse. We include a SN lightcurve if it passes either criteria. 

These simple numerical criteria enable testing the impact of the selection criteria on the final results, as well as making them reproducible. Another aspect of these numerical template selection criteria is that all objects that were removed in our first cut would have failed these criteria (although we could not check this in an automatic way since they lack a peak). 

After the sample selection criteria, we are left with a total of 68 Type Ibc SNe which are templatable (Table~\ref{tab:cuts}). Of these, 62 have $r$-band lightcurves, 24 have $g$-band lightcurves, whereof 6 have only $g$-band lightcurves. 

\subsection{Absolute magnitude lightcurves \protect\label{sec:absmag0}}

The template fitting was performed on lightcurves converted to absolute magnitudes. 
To obtain the distance modulus, redshifts obtained from their spectra were used \citep{Fremling2018}. These redshifts are obtained from, in priority order, i) spectral redshift of the host from surveys such as the SDSS \citep{Strauss2002}, ii) galaxy lines in the (i)PTF SN spectra, iii) template fits using the SNID \citep[SuperNova IDentification program;][]{Blondin2007}. 
A flat cosmology as measured by WMAP5 \citep{Komatsu2009} was used with $\Omega_m=0.277$ and $H_0=70.2$~km~s$^{-1}$~Mpc$^{-1}$. No adjustment was made for peculiar velocities. The lightcurves were also corrected for time dilation. The absolute magnitudes were corrected for extinction of the MW by an $E(B-V)_{MW}$ value from \citet{Schlafly2011}, using an $R_V$ of $3.1$ and the central wavelength of the respective band with a \citet{Fitzpatrick1999} extinction law. 

\subsection{Template lightcurve fitting \protect\label{sec:LCfit_init}}

The absolute magnitude lightcurves are converted into flux, and then we fit the $g$ and $r$ band restframe lightcurves with template SN lightcurves in the same band. This is done using a three parameter fit of shift, stretch, and scale. The shift is an additive time shift, the stretch is a multiplicative factor applied to the phases to match the lightcurve shape (and is used to measure broadness), while the scale is a multiplicative factor in flux-space (additive in magnitude-space) to match the brightness of the template with the SN. The stretch value $S_{g,r}$, meaning $g$ or $r$, is the key parameter of interest in our fitting.

\begin{figure}
    \centering
    \includegraphics[width=\linewidth]{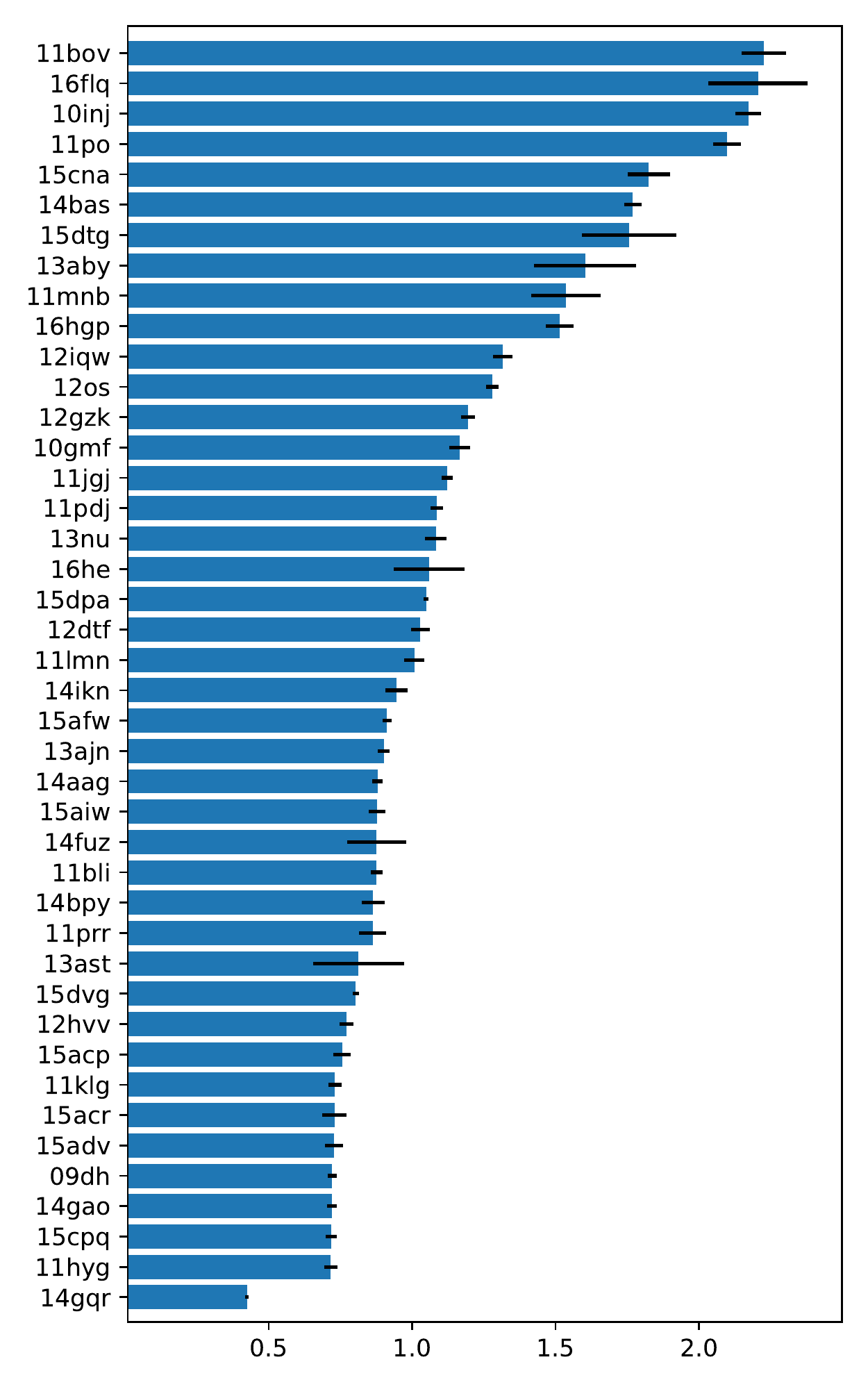}
    \caption{Stretch values of the $g$ band sample with the associated standard errors. Type IIb stretch values have been added to the Type Ibc sample. The stretch values are not corrected to the $r$-band stretch space as in Sect. \ref{sec:sampleselect}.}
    \label{fig:gstretch_bar}
\end{figure}

\begin{figure*}
    \centering
    \includegraphics[width=0.5\linewidth]{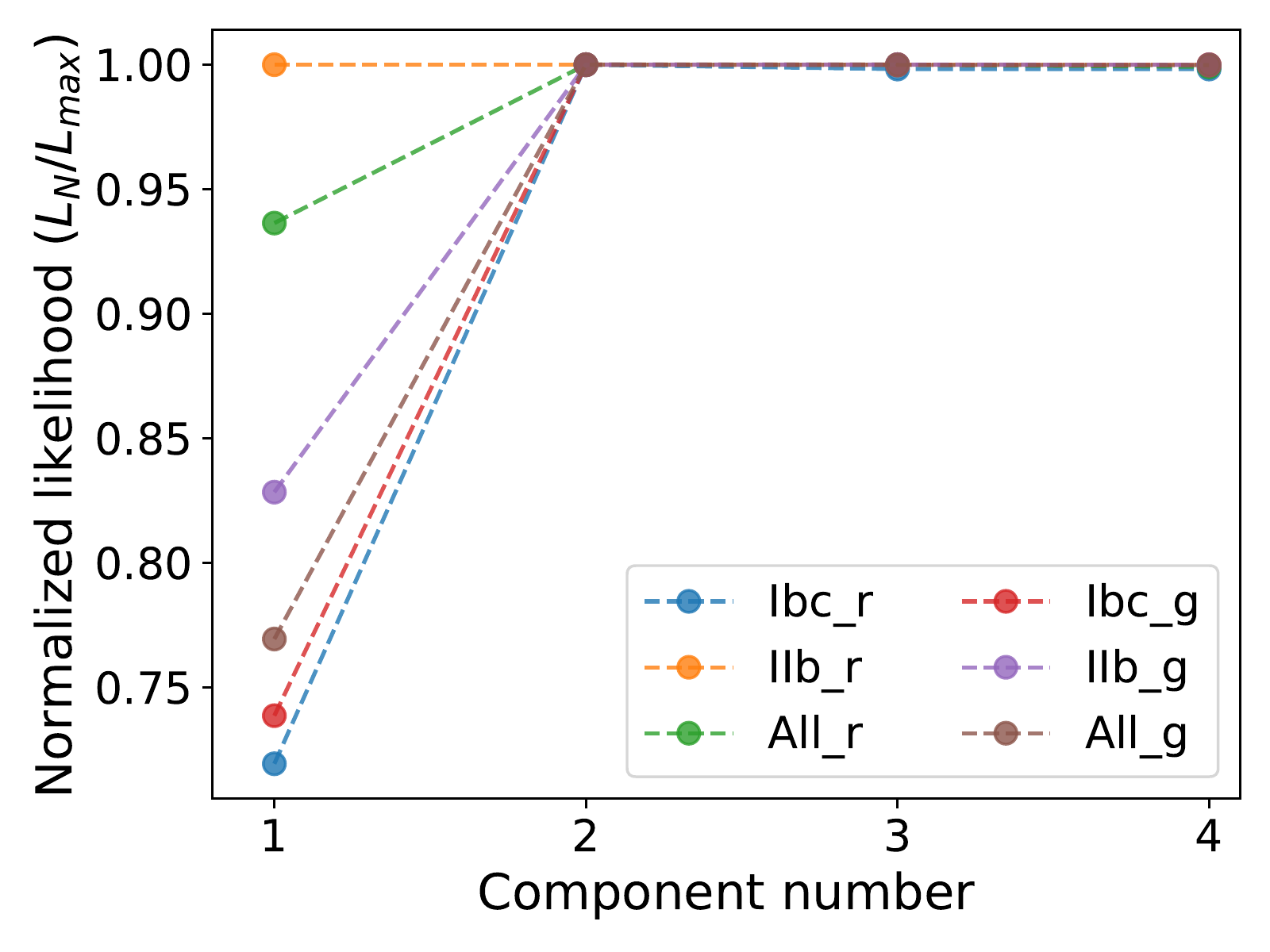}\\
    \includegraphics[width=0.3\linewidth]{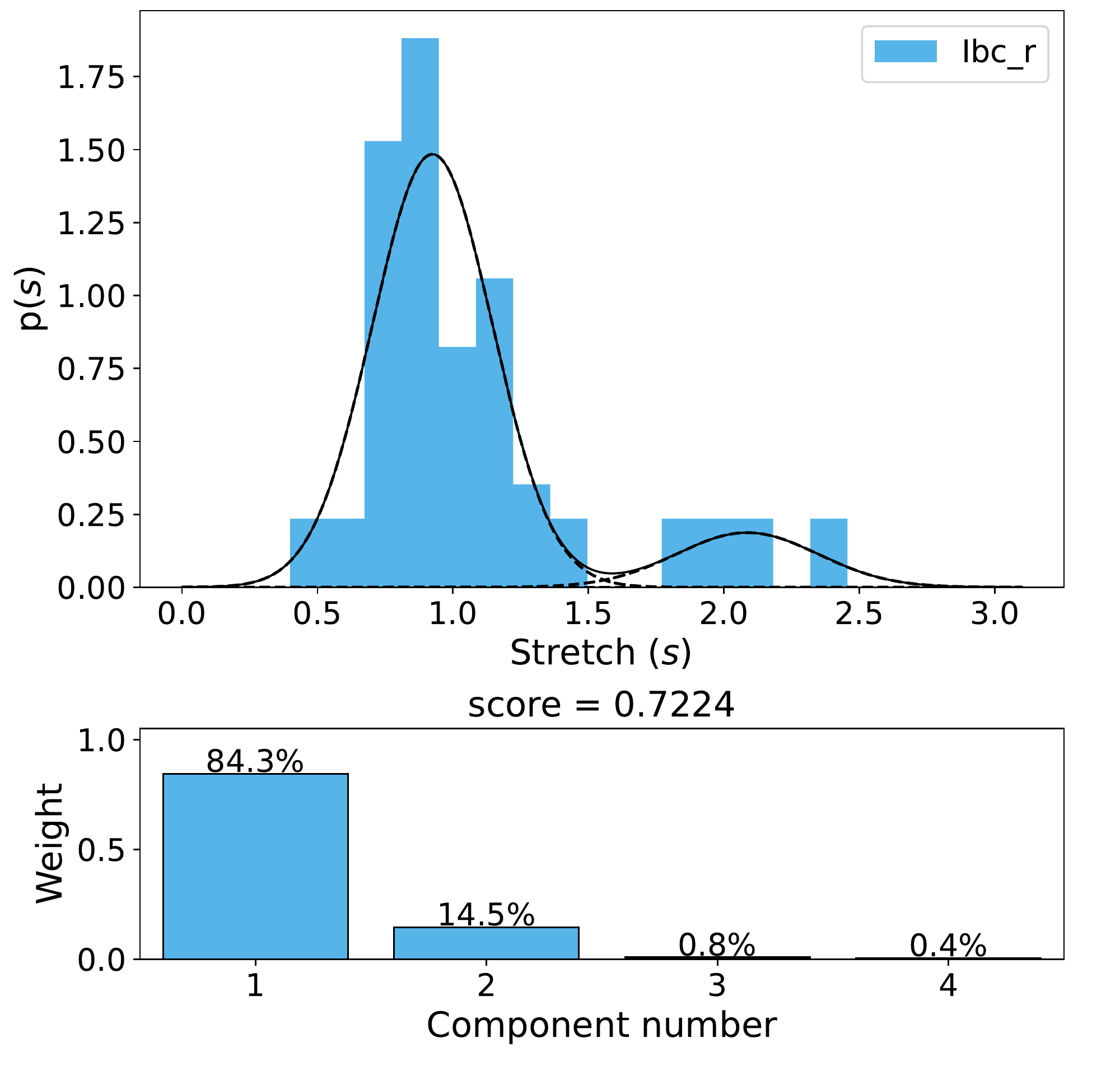}
    \includegraphics[width=0.3\linewidth]{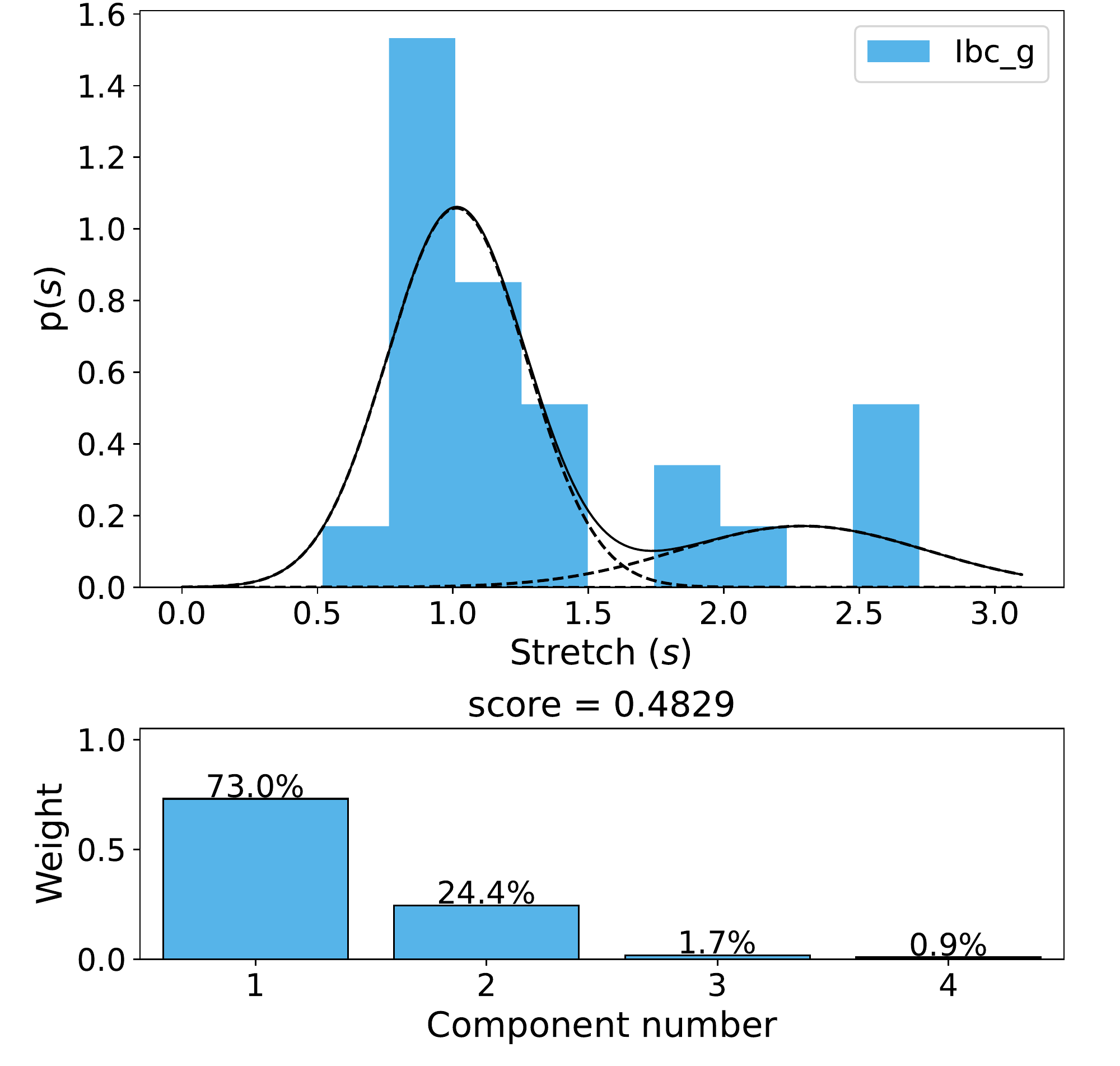}
    \includegraphics[width=0.3\linewidth]{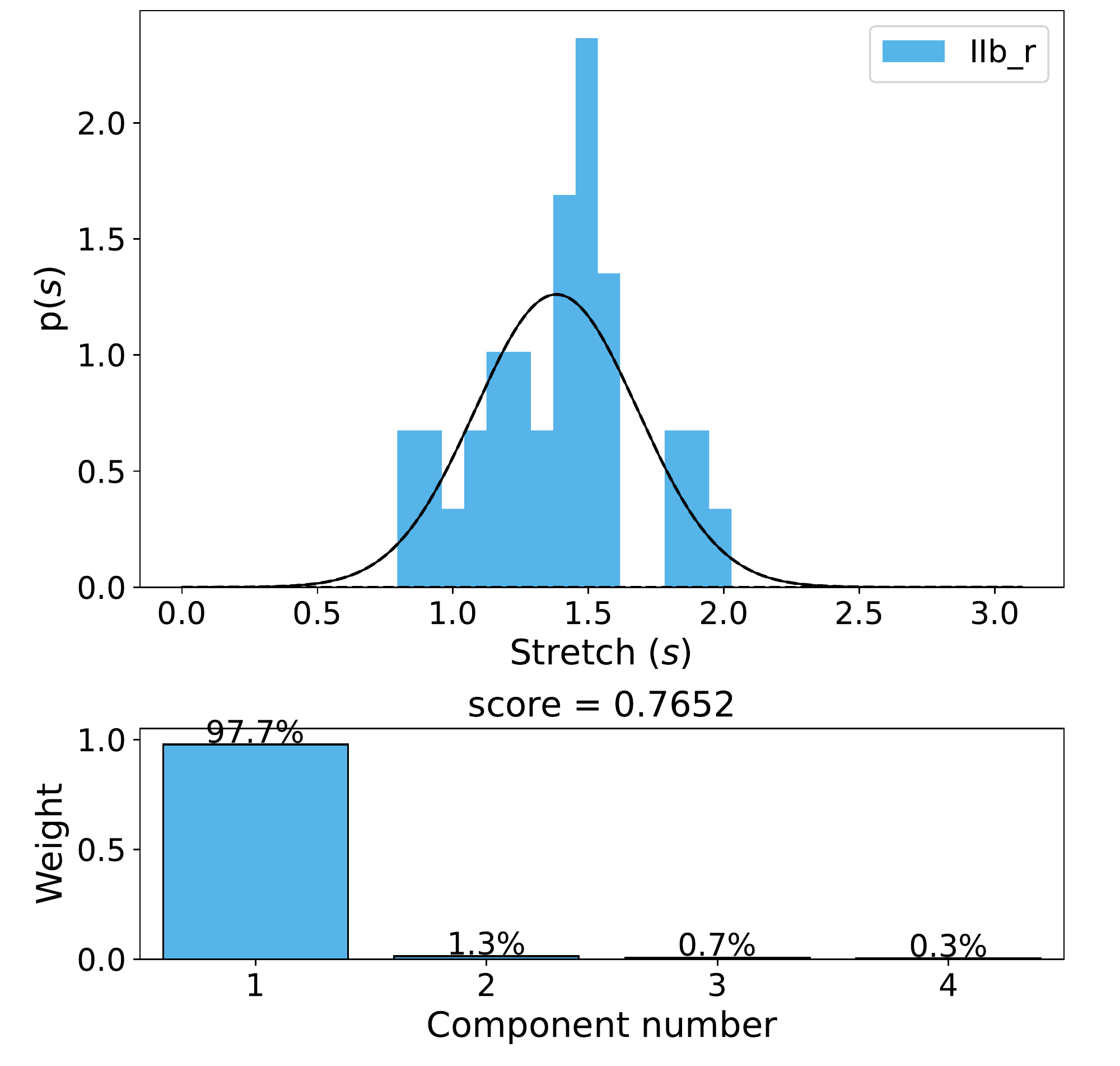}
    \includegraphics[width=0.3\linewidth]{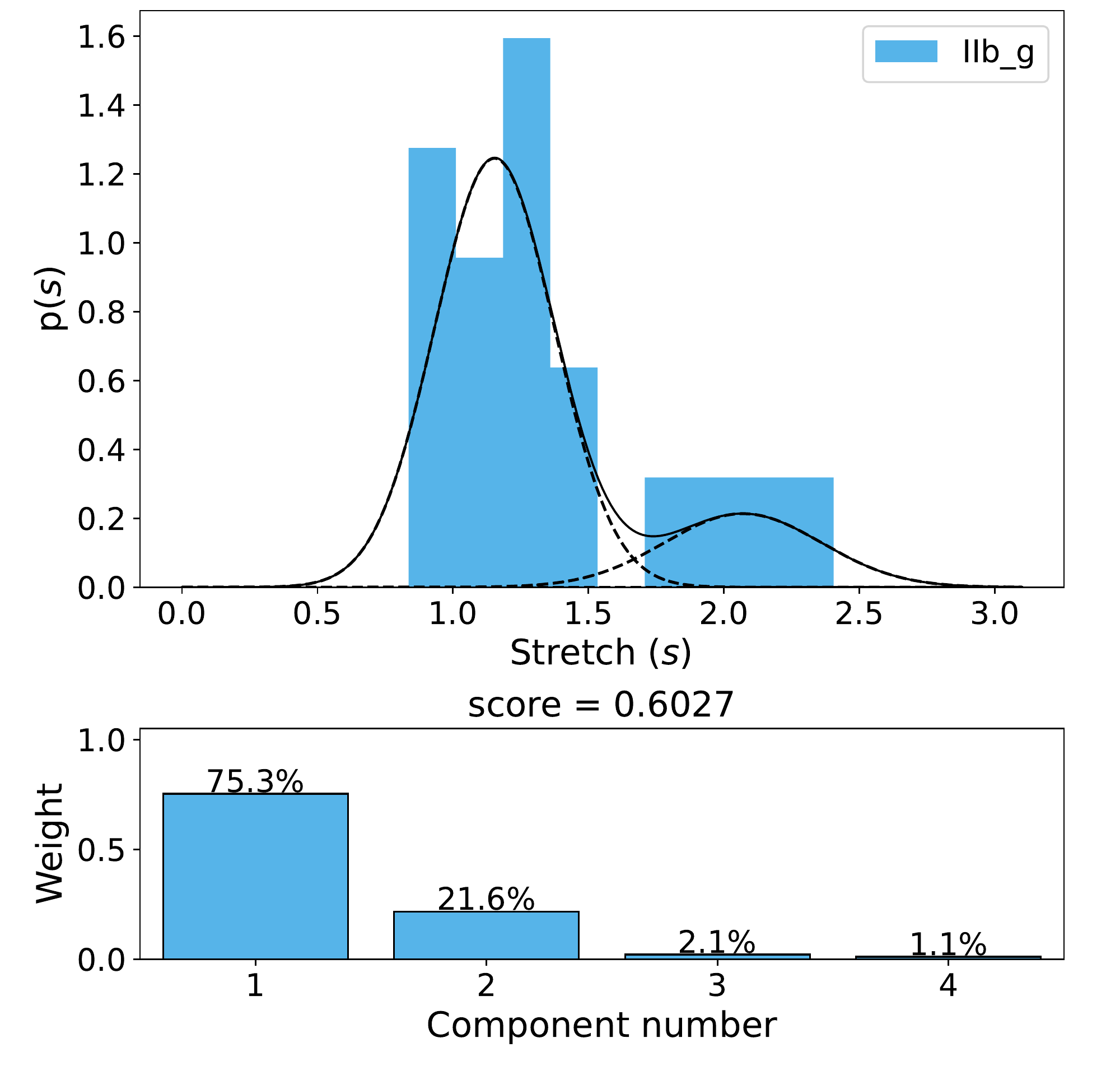}
    \includegraphics[width=0.3\linewidth]{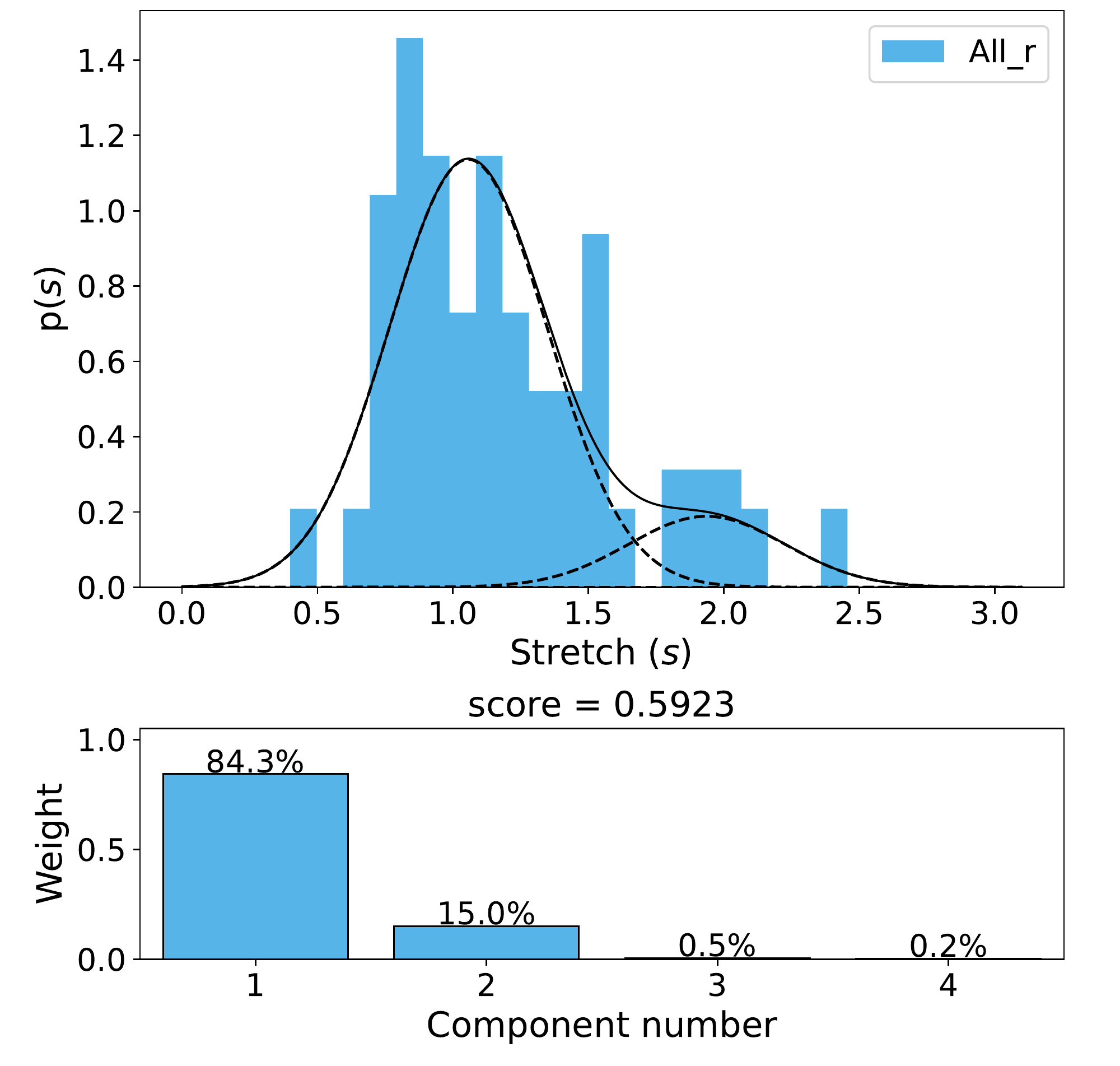}
    \includegraphics[width=0.3\linewidth]{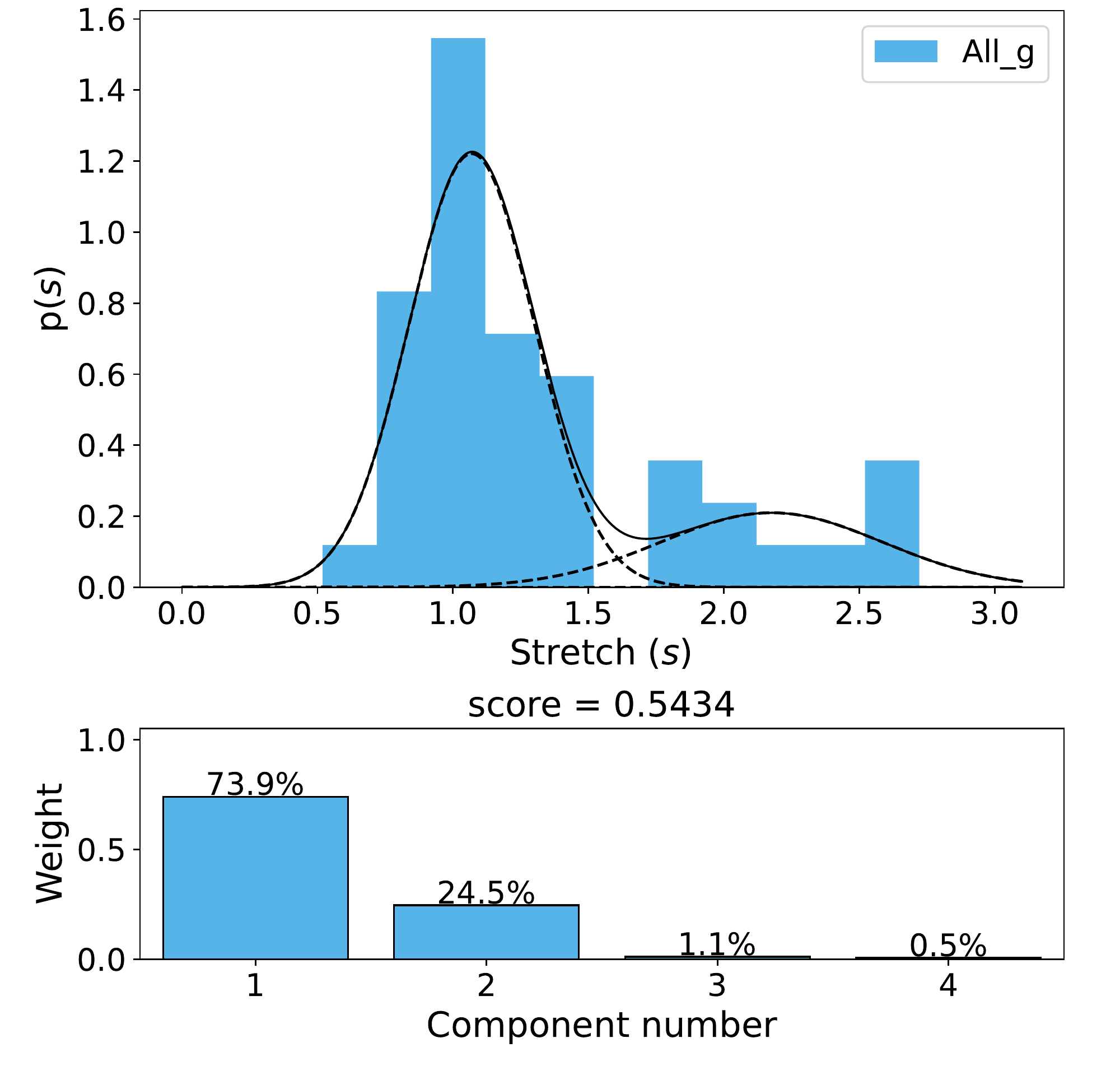}
    \caption{\textit{Top:} Improvement in likelihood based on number (N) of Gaussian components used in the Bayesian GMM fits. All but Type IIb $r$ band reach maximum likelihood after 2 components, indicating the preference for two populations in the stretch values. \textit{Bottom:} Bayesian GMM fits to the stretch distribution with up to 4 components. As can be seen, 2 components are favored for all but Type IIb $r$ band, where 1 component is favored. Higher numbered components have very low weights indicating that they are not favored to explain the observed stretch distribution.}
    \label{fig:GMMs}
\end{figure*}

For Type Ibc SNe we use the template presented by \citet{Taddia2015}. Since the template fitting is done near peak only, the variable early cooling phase seen in a few SE~SNe is not included in the fitting, but also does not have a large effect on the fits. The fitting was done in Python 3, using a least squares, non-linear, gradient-based regression algorithm (Sequential Least-Squares Quadratic Programming, SLSQP), where the template was stretched, shifted, and scaled until the square of the residuals from the observed lightcurve of the SN and the template was minimized. The fits also respect limits from non-detections by ensuring the difference between the template and the limits are positive (i.e., the limit is above the template), which is done by minimizing the sum of any negative differences. As rare exceptions, a small negative value was allowed if overwhelmingly preferred by the fit. This can happen when the limit has unaccounted observational calibration uncertainties or if the lightcurve behaves in a non-standard manner at very early times. In addition, while we performed the fits in flux normalized to the brightest point, in a few cases, allowing a small additive constant yielded much better fits. This constant was kept fixed to zero for all but 4 SNe. 

For further details of the template fitting process and the algorithms used, see \citet[][their section 6.3]{Karamehmetoglu2018}. Since the template lightcurves have well-behaved characteristics, we can use the fit to estimate the explosion epoch of the SNe as well. This method was used to aid the phase determination in \citet{Fremling2018}. Most importantly however, the fits give an estimate of the broadness of a SN lightcurve as a single parameter, the stretch value. We perform the template fitting process for the combined (i)PTF sample with visual verification of the fits to obtain the distribution of lightcurve broadness in $g$ and $r$ bands, measured as this stretch parameter. 

\subsection{Selection of the Broad sample \protect\label{sec:sampleselect}}

The measured stretch parameter in $g$ and/or $r$ band for the Type Ibc sample is plotted as a histogram showing the underlying distribution in the top panel of Fig.~\ref{fig:hist_norm_stretch}. In order to allow a direct comparison between the bands, we fit the $r$-band Type Ibc template lightcurve around peak with a stretched version of the $g$-band Type Ibc template. We thus obtain a stretch correction factor 
$S^{r}_{Ibc}/S^{g}_{Ibc} \approx 1.22$. Multiplying the stretch values measured in the $g$ band by this factor corrects them to the $r$-band template stretch-space, which is what we plot for the $g$ band in the right panels of Fig.~\ref{fig:hist_norm_stretch}. 

Looking at the figure, the distribution of stretch values for Type Ibc SNe seem to indicate the presence of two populations. The majority of SE~SNe cluster around a stretch value of 1.0, with a secondary distribution located at high broadness values ($S^r_{Ibc}\gtrsim1.5$ in stretch value).

It can be seen in the bar plot of $g$-band stretch values plotted in Fig.~\ref{fig:gstretch_bar} that the broad objects seem to be have statistically significant higher stretch values\footnote{This figure includes Type IIb SNe (Appendix~\ref{sec:IIb}).}. Interestingly, there is also an object with a significantly lower stretch value, iPTF14gqr, which was found to be an ultra-stripped SE~SN \protect\protect\citep{De2018a} as discussed in Sect.~\ref{sec:discussion}. We tested the apparent bimodality in lightcurve broadness with two approaches: Gaussian mixture modeling and K-means clustering. We will use the results of these statistical tests to derive our final broad sample. 

\begin{figure*}
    \centering
    \includegraphics[width=0.95\linewidth]{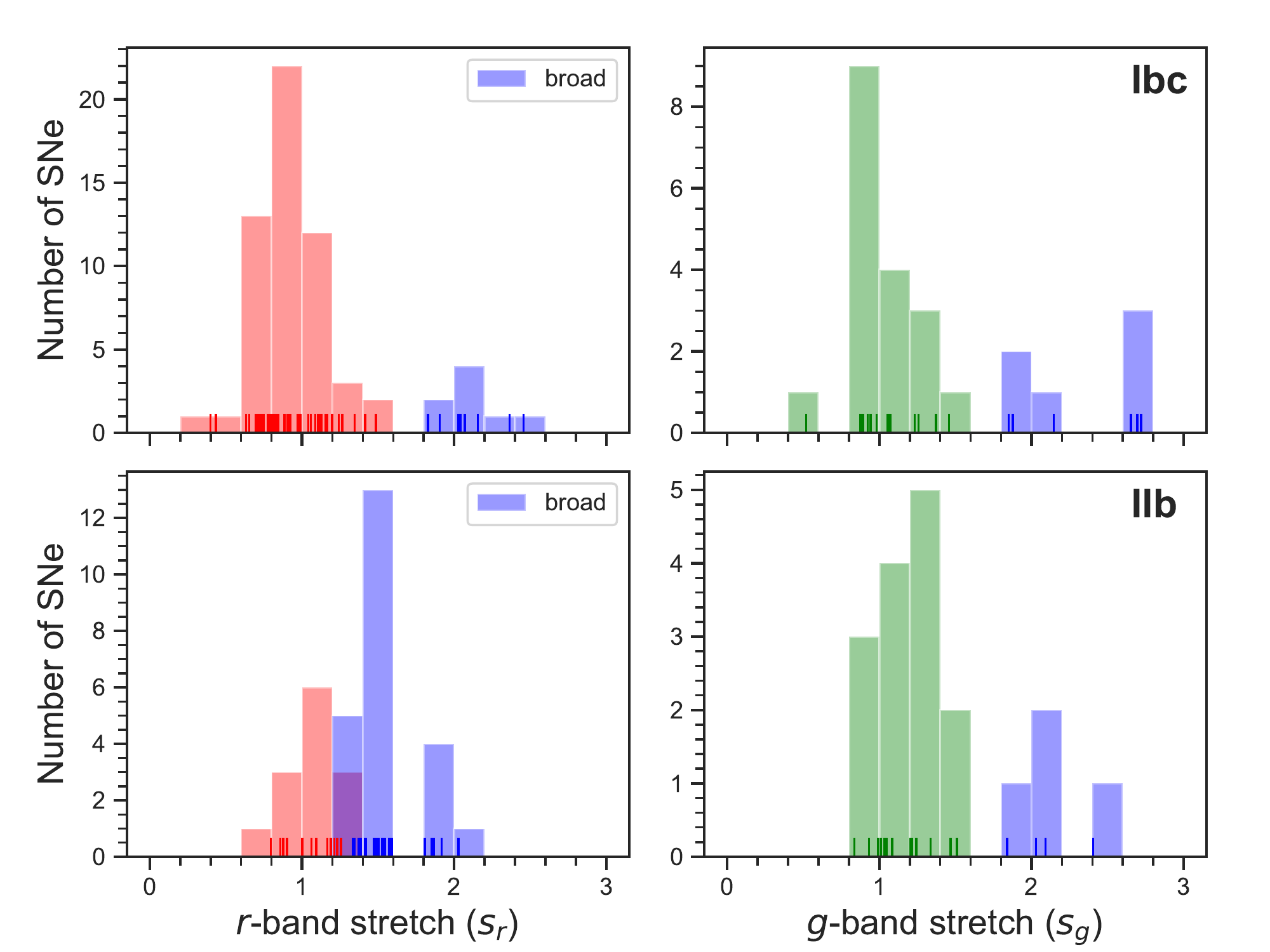}
    \caption{Histogram of the distribution of stretch values plotted after running a k-means clustering algorithm on the one-dimensional stretch parameter dataset. The SNe put in to a second cluster by the algorithm have been plotted in blue, and labeled as broad. The ordinary (non-broad) SNe in each band have been plotted with the respective color of that band ($r$ band left and $g$ band to the right). The histogram was not normalized so the counts represent the number of SNe in each bin. \textit{Top:} Type Ibc SNe, \textit{Bottom:} Type IIb SNe.}
    \label{fig:kmeans_broad}
\end{figure*}

\subsubsection{Gaussian mixture modeling}

To test whether the stretch values prefer one or more distributions, we fit them using multi-component Bayesian Gaussian Mixture Models (GMMs) using the \texttt{SKLearn} library in \texttt{Python 3}. These models allow infinitely large component Gaussian fits, but attach very low weights when further components are not helpful in reproducing the observed distribution. As shown in the upper and lower panels of Fig.~\ref{fig:GMMs}, the Type Ibc stretch values prefer two Gaussian components to explain the distribution, with one component to explain the ordinary broadness objects and another to explain the broad ones. Further components are not found to be necessary, as the likelihood has already reached a maximum with two Gaussian components. 

As an additional test, we made 500 realizations of the stretch distribution by varying all values randomly within the standard errors. We fit each of these realizations using ordinary multi-component GMMs and select the preferred model (i.e. the preferred number of Gaussian components) based on the lowest value of the BIC goodness-of-fit indicator. Based purely on the lowest BIC, we find that 100\% of these fits prefer $\geq 2$ components. In the $r$ band, $99.6\%$ prefer exactly two, while this is $63.6\%$ in the $g$ band, with the remaining primarily preferring three components. If adjusted for the ``rule of thumb'' that models within $\Delta BIC\leq 2.0$ are not worth distinguishing, the percentage of $g$ band two-component favored fits increases to $80.4\%$. Single component fits are still never favored, neither in the $g$ nor the $r$ band. 

That a certain number of Gaussians is preferred means that this combination best represents the data compared to any other combination of Gaussians. The resulting GMM fit is then a good representation of the data under the assumption that the underlying distribution of the data is a combination of Gaussians. Based on these results, we choose to split the stretch distribution into two components and identify the broad Type Ibc SNe using the component centered around higher stretch values. 

\subsubsection{K-means clustering \label{sec:kmeans}}

In order to obtain the broad sample, we also ran a k-means clustering algorithm\footnote{Using the \texttt{SKLearn} package in \texttt{Python 3}.} with two initial seeds that will form two clusters in the data. These seeds are placed pseudo-randomly and grow by acquiring their nearest neighbors while trying to minimize the sum of the distances within a cluster. The process is repeated to find the best overall fit. In this manner, the algorithm will yield the two most tightly bound clusters possible in this distribution. The one dimensional case used here is trivially solved, so while the k-means clustering algorithm is not a great statistical arbiter, it does show the natural clustering present in our dataset. A benefit of this method is that we do not have to assume any underlying distribution, as is the case for GMM based tests. 

As shown in Fig.~\ref{fig:kmeans_broad}, and mirroring the result from the GMM fits, we find that the data readily clusters into two distributions, one centered around a stretch value of 1.0 and the other located at stretch values $>1.5$, which we label ``ordinary'' and ``broad'', respectively. The result indicates that a hard cut at a stretch value between 1.5--1.8 is optimal for splitting the Type Ibc sample into two distributions, and we pick the mid-point value of $1.65$. Several SNe have both $g$ and $r$-band lightcurves. All of our SNe with multiband stretch values show up in the ``ordinary'' or ``broad'' categories regardless of the band. Even when taking into account 1-sigma fitting errors, the SNe cannot move categories when using $1.65$ as the cut-off value.

\section{Type IIb SNe \label{sec:IIb}}
In this paper, we developed a method of finding SE~SNe with broad lightcurves, and applied it to Type Ibc SNe. Studying the resulting broad SE~SNe in detail, we found them to be likely associated with massive stars, at least when applying the Arnett formalism (however, see Sect. \ref{sec:discussion}). Then, we estimated the fraction of broad Type Ibc SNe to be ${\sim13}\%$, which corresponds to ${\sim6}\%$ after correcting for observational biases\footnote{Using the $r$ band to account for the follow-up bias as discussed in Sect.~\ref{sec:biases}.}. In this section, we show that Type IIb SNe in (i)PTF also hint at having a similar fraction of broad (and likely high-mass) SNe. However, a detailed investigation of Type IIb SNe, especially broad Type IIb SNe, is beyond the scope of the paper.

\subsection{Template fitting for Type IIb SNe}
Following exactly the same steps as with Type Ibc SNe, we fit template lightcurves to the $g$ and $r-$band lightcurves of the (i)PTF Type IIb sample. For Type IIb SNe, we use the well-observed and well-behaved SN 2011dh \citep{Ergon2014,Ergon2015a} as a template. We are confident in using a single SN as a template since \citet{Fremling2016} showed a remarkable similarity between the lightcurves of two ordinary Type Ib and IIb SNe that exploded in the same galaxy at the same distance close in time. The template fitting is done near peak only, so the variable early cooling phase seen in some Type IIb SNe does not have an effect on the fits. 

We use the same templatability criteria as for Type Ibc SNe, with the exception that we use the Type IIb template. As reported in Table~\ref{tab:cuts}, there are 58 Type IIb SNe, which reduce to 39 after all cuts on templatability. These 39 Type IIb SNe are fit using a template to obtain the stretch parameter $s^{g/r}_{IIb}$ for each SN. Just like for the Type Ibc SNe, we verify the fits visually. We noticed that it was not possible to find a good fit to PTF10pzp, although it technically barely passes the templatability criteria. This object has a stretch value of $1.6\pm0.1$ in the $r$ band. Although we nominally find it to be in the broad category (see below), we are not confident of the stretch value. Therefore, the broad fractions we report in this section can be 20\% lower if this object is excluded. Out of all 107 Type Ibc and Type IIb SNe, this object is the only one with an obvious visually-poor fit.

In order to be able to compare the resulting distributions, we also fit the Type IIb $g$ and $r$-band template lightcurves to the Type Ibc equivalent and obtain correction factors of $1.202$ for the $r$ band and $1.147$ for the $g$ band. The resulting distribution is plotted in the middle panel of Fig.~\ref{fig:hist_norm_stretch}. These distributions have all been corrected to the space of $r$ band Type Ibc stretch values.

\begin{table}[]
    \centering
    \begin{tabular}{c|ccc}
        \toprule
        band     & observed & duration  & Malmquist  \\
        \midrule
        $r$ & $0.139^{+0.06}_{-0.06}$ & $0.132^{+0.06}_{-0.06}$ & $0.061^{+0.04}_{-0.03}$\\ [8pt]
        $g$ & $0.222^{+0.07}_{-0.08}$ & $0.208^{+0.08}_{-0.09}$ & $0.081^{+0.06}_{-0.05}$\\ [8pt]
        combined & $0.154^{+0.06}_{-0.06}$ & $0.146^{+0.06}_{-0.06}$ & $0.068^{+0.04}_{-0.03}$ \\ 
        \bottomrule
    \end{tabular}
    \caption{The fraction of broad Type IIb SE~SNe after correcting for observational biases, reported with 90\% Poissonian uncertainties.}
    \label{tab:IIbFrac}
\end{table}

We find that Type IIb SNe also show tentative evidence of two clusters: one located around a stretch value of ${\sim}1.2$ in the $r$ band and ${\sim}1.0$ in the $g$ band (before correcting to the $r$-band stretch), with the other at a higher stretch value. Interestingly, applying the corrections that were simply derived on the templates, not on the SNe, leads us to seeing a possible gap in stretch values located exactly around $s=1.65$, the same value as for Type Ibc SNe. However, the gap is much less clear for the Type IIb SNe. This can also be seen in the result of the K-means test plotted in the lower panel of Fig.~\ref{fig:kmeans_broad}. In Fig.~\ref{fig:GMMs}, the GMM test for multiple distributions in Type IIb SNe favors one distribution for the $r$ band and two distributions for the $g$ band. Thus, while the stretch distribution in Type IIb SNe also seems to indicate a population of broad events, it is more difficult to determine an exact cut-off using our methodology from Type Ibc SNe to select a broad sample. The lower number of Type IIb SNe, and the smaller range of stretch values inhabited, exacerbates the difficulty of selecting a broad sample especially in the $r$ band.

iPTF13aby (SN~2013bb), the most massive Type IIb SNe discovered to date \citep{Prentice2018}, has a stretch value of ${\sim}1.85$ in the corrected $g$ and $r$-band stretch values, making it one of the broadest Type IIb SNe in our sample. Interestingly, there are several other Type IIb SNe with similarly broad lightcurves, which suggests that many more broad and likely massive Type IIb SNe might exist in the (i)PTF database. Therefore, the broad sample of (i)PTF Type IIb SNe should include iPTF13aby (and others like it).

First, we consider simply using the same cut-off as with the Type Ibc SNe, since we see an indication that there is an apparent gap at this value, and since SE~SN lightcurves behave rather similarly \citep{Fremling2016}. This would also correctly label iPTF13aby as a broad SN. However, the GMM and K-means based methodologies we employed for Type Ibc SNe (Sect.~\ref{sec:kmeans}) are not fully consistent in this simple approach. GMM fits to the Type IIb $r$-band stretch distribution prefer a single Gaussian to explain the observed distribution. When we fit multi-component GMMs to 500 realizations of the Type IIb $r$-band stretch distribution obtained by randomly varying the stretch values within the errors, a two-component fit was favored over a single component fit in only ${\sim}3\%$ of those realizations, as determined by the lowest Bayesian information criteria (BIC) of the fit. In those $3\%$, a stretch cut of $1.65$ would be sufficient to separate the broad SNe from the ordinary, matching the value found for Type Ibc SNe in either band as well as for the $g$-band Type IIb SNe. In the remaining $97\%$ of realizations of the GMM fits, a single Gaussian component fit is preferred. 

This fact can also explain the unclear result from our K-means test. We are searching for two clusters when only one seems to be present. Therefore, we consider the stretch cut implied by our K-means test on the Type IIb $r$ band to be a lower limit (see Fig.~\ref{fig:kmeans_broad}). The most conservative upper limit is established by considering all $r$-band Type IIb SNe as ordinary (including those that would in the broad category in $g$ band), leaving only iPTF15cna as broad, which has a $g$-band stretch of $2.1$. While iPTF15cna is only the second broadest after PTF11po, iPTF15cna lacks an $r$-band peak that might have disqualified it from being labeled broad. A more sensible upper limit might instead be derived using the well-determined stretch value of iPTF13aby from \citet{Prentice2018}, which is a known broad and massive object, and then considering those that have a stretch value greater than or equal to that of iPTF13aby within 1-sigma errors. Interestingly, this yields the same result as using a stretch cut of $1.65$.

Therefore, in our limited investigation, we use a stretch cut of $1.65$ also for Type IIb SNe, even though the presence of a well-separated second population of high stretch events is less certain, especially in the $r$ band. Nevertheless, we do know of several broad events that become correctly labeled using this division, including iPTF13aby.

\subsection{Corrected fraction of broad Type IIb SNe}

Based on a stretch cut of $1.65$, we are left with 6 Type IIb SNe out of 36 being broad, meaning an observed fraction of $15^{+6}_{-6}\%$. The numbers for each band are reported in Table~\ref{tab:cuts}, while the broad fraction is reported in Table~\ref{tab:IIbFrac}. After correcting for observational biases in the same way as for Type Ibc SNe, we calculate the broad fraction in the $r$ band as $6^{+4}_{-3}\%$ and in the $g$ band as $8^{+6}_{-5}\%$, reported with $90\%$ confidence intervals. 

Since we correct the stretch distribution to that of the Type Ibc $r$-band template, we are able to use the same calculation for the lightcurve duration bias. For the Malmquist bias, we follow the steps outlined in Appendix~\ref{sec:malmappendix} using values from the Type IIb sample. However, we have kept the same $z_{max}$ as with Type Ibc SNe for consistency, even though the most distant Type IIb SN is located at redshift $0.099$. Using $z_{max}=0.1$ would instead yield a corrected fraction of $10\%$ in the $r$ band and $13\%$ in the $g$ band. Similar to Type Ibc SNe, there also seems to be a follow-up bias towards broad SE~SNe in the $g$ band.

\subsection{Broad Type IIb SNe}
PTF10qrl, PTF10pzp\footnote{We remind that PTF10pzp is the only lightcurve in our sample where visual inspection shows a poor fit. Thus, its identification as a broad SN could be incorrect.}, PTF11po, iPTF13aby, iPTF14bas, iPTF15cna are found to be broad Type IIb SNe, for which we briefly discuss some early indications. iPTF13aby (SN~2013bb) was studied by \citet{Prentice2018} and found to be the most high-mass Type IIb SN discovered to date. iPTF14bas was identified as an unusually long rising Type IIb SN by \citet{Rubin2016}. 

iPTF15cna has late-time Keck spectra which can be used to estimate the metallicity as in Sect.~\ref{sec:hosts}. It also seems to be located in a dwarf host galaxy with high SFR \citep{Schulze2021}. Using the late time Keck spectrum, we calculate the metallicity at the location of iPTF15cna using the O3N2 method and find log(O/H)+12=${\approx}8.2$~dex in both the PP04 and M13 methods. It seems that at least some broad Type IIb SNe, like iPTF15cna, are located in low-metallicity environments similar to the broad Type Ibc SNe. 

Using the nebular Keck spectrum, we also estimate the $L_{[\ion{Ca}{ii}]}/L_{[\ion{O}{i}]}$ for iPTF15cna and obtain ${\sim}0.5$. This hints that iPTF15cna might be similar to iPTF13aby in having a high-mass origin. Thus, in addition to the fraction of broad Type Ibc SNe (Table~\ref{tab:IbcFrac}), we also report the fraction of all broad SE~SNe including Type~IIb SNe (see Table~\ref{tab:broadFrac}).

\section{The Malmquist Bias \label{sec:malmappendix}}

Correcting for the Malmquist bias is not a straightforward endeavor. Typically, there are three conceptually distinct ways to correct for it. The first and easiest method is to use the distribution only out to a distance where the entire brightness range is not truncated, thereby converting a magnitude-limited survey into an effectively volume limited survey. A second way is to use an empirical $1/\text{V}_\text{Max}$ correction based on the distance \citep{1968ApJ...151..393S}. Finally, a more complicated correction that uses the distribution observed in the lower distance (non-biased) part, or knowledge about the underlying distribution, in order to correct the higher distance (Malmquist-biased) part can be employed \citep[see e.g.,][]{Richardson2014}. The last of these methods makes the most use of the data, but cannot be used in our case because we are interested in the distribution of a correlated third parameter, stretch, whose inherent distribution we have no knowledge of. Instead, in this paper, we employed a novel method to correct for the peculiar case of the Malmquist bias on two parameters, which are correlated with a third parameter.

\begin{figure}
    \centering
    \includegraphics[width=\linewidth]{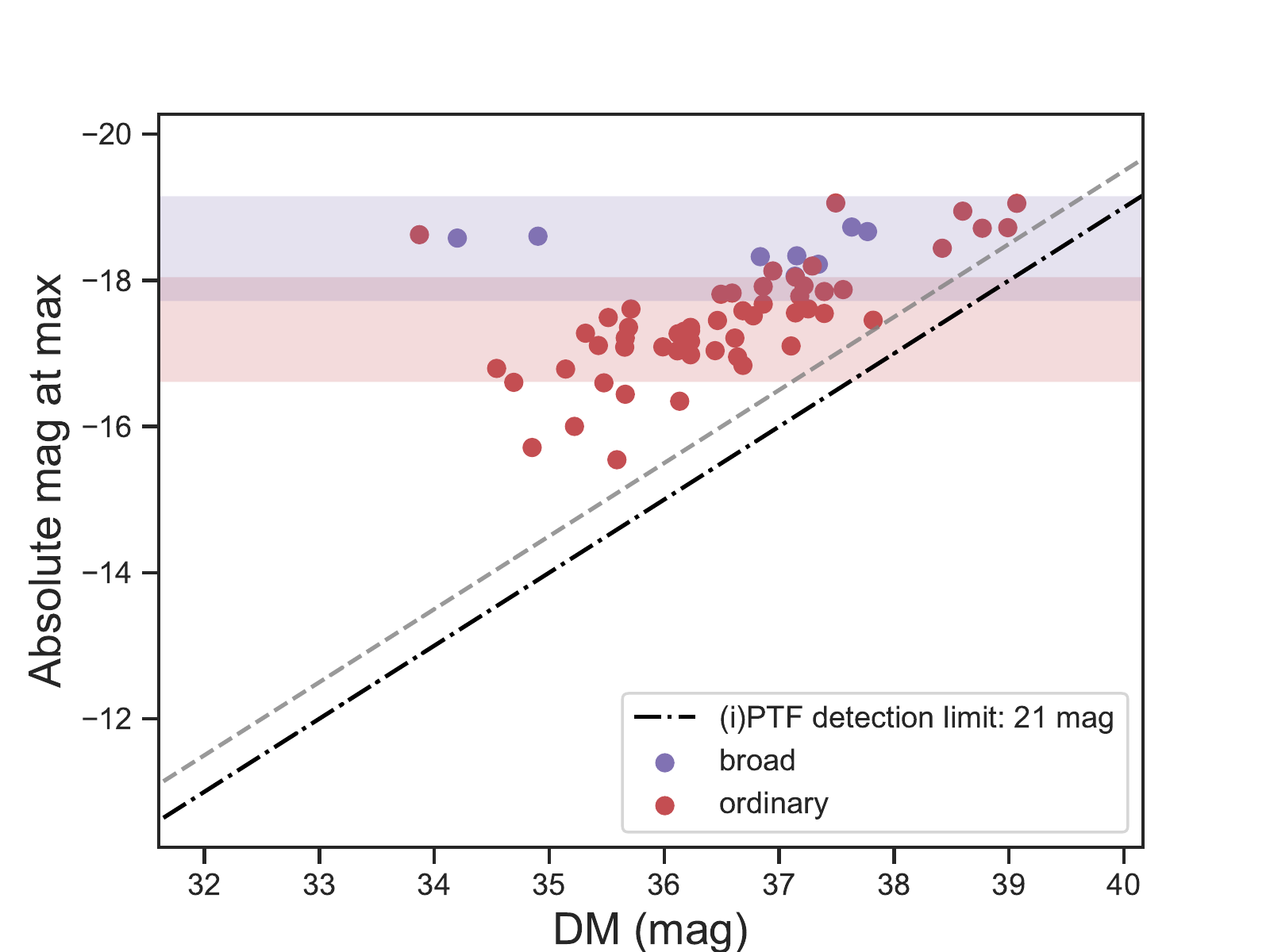}
    \includegraphics[width=\linewidth]{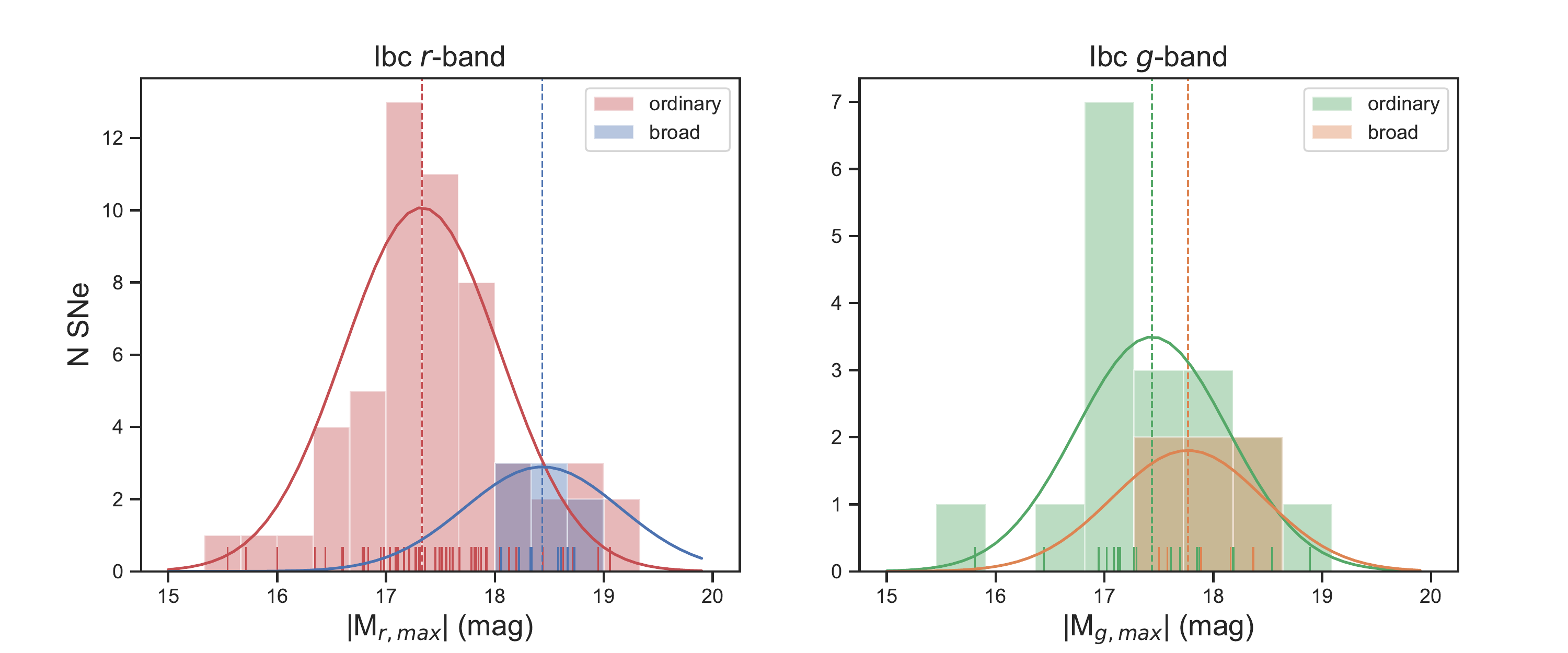}
    \caption[Malmquist bias]{\textit{Top:} A plot of peak absolute magnitude in $r$ band versus distance modulus (DM) for ordinary (red circles) and broad (purple circles) SE~SNe, which display the Malmquist bias associated with the (i)PTF detection limit thresholds. The detection thresholds at $r = 20.5$~mag and $r = 21.0$~mag are plotted as grey dashed and black dot-dashed lines, respectively. They are approximate estimates of the detection threshold for (i)PTF during bright and dark times, respectively. This plot shows that the broad SNe are preferentially brighter than the ordinary SE~SNe, meaning that they are detected at a higher fraction at higher redshifts. \textit{Bottom:} Histogram of peak magnitude with the gaussian fits overlaid for the broad and ordinary groups in $g$ and $r$ bands. This type of distribution is often referred to as a truncated distribution \protect\citep{feigelson_babu_2012}, and correcting for this truncation is a complicated but necessary step for all magnitude limited surveys.}
    \label{fig:malmq}
\end{figure}

Since the luminosity function of our SNe is not known, especially for the broad class, we attempt to estimate and correct for this before calculating the ratio of broad versus ordinary SE~SNe. The distribution resulting from the Malmquist bias is a truncated distribution \citep{feigelson_babu_2012}, more specifically right or left truncated depending on if the peak brightness is reported in magnitudes or flux.

The method works as follows. To perform the fit we assumed that the peak brightness for each class (ordinary and broad) was normally distributed with some unknown mean and variance. Using a constant mean peak brightness is justified if we do not expect cosmic evolution of the brightness with respect to distance, which we do not for the low redshifts of our SNe. Thus, we fit a truncated linear regression model\footnote{Using the {\tt truncreg} function in Stata version 16.0, with a left truncation on the absolute value of the peak absolute magnitudes.} to the peak magnitude for ordinary and broad SE~SNe separately, to estimate the mean of the peak brightness of each sample. We did this separately for each band and obtained the values reported in Table~\ref{tab:brightness} where $\mu_0$ and $\mu_1$ are the mean values of the ``ordinary'' and ``broad''  samples, respectively, while $\sigma$ is the standard deviation (assumed to be the same for simplicity). The resulting gaussian fits are plotted in the lower panel of Fig.~\ref{fig:malmq} for the $g$ and $r$ bands of the Type Ibc stretch distribution.

\begin{table}
    \centering
    \begin{tabular}{c|c|c|c}
        Band & Parameter & Estimate & Standard Error \\
        \hline
        $r$ & $\mu_0$ & 17.33 & 0.107 \\
          & $\mu_1$ & 18.44 & 0.256 \\
          & $\sigma$& 0.717 & 0.069 \\
          \hline 
        $g$ & $\mu_0$ & 17.44 & 0.172 \\
          & $\mu_1$ & 17.77 & 0.358 \\
          & $\sigma$& 0.697 & 0.111 \\
    \end{tabular}
    \caption{Results of the truncated regression model fit for Type Ibc SNe. $\mu_0$ and $\mu_1$ are the means of the ``ordinary'' and ``broad'' samples, respectively, while $\sigma$ is the standard deviation (assumed to be the same for simplicity).}
    \label{tab:brightness}
\end{table}

Next we correct the fraction of broad to ordinary SE~SNe by the statistical \textit{exposure}, which in this case represents the volume size of each SN. The Malmquist bias means that brighter SNe had a larger volume size and thus were measured with a longer \textit{exposure}. The formula for this correction is simply:

\begin{align}
    \text{Corrected ratio} = \frac{n_1/E_1}{n_0/E_0 + n_1/E_1},
    \label{eq:cr}
\end{align}

\noindent where $n_x$ is the number of SNe in group $x$, 0 for ordinary, 1 for broad, and $E_x$ is the \textit{exposure} for each group. According to the Malmquist bias the intrinsically brighter group will have a larger exposure. Next we estimate the $E_x$ for each group from our truncated linear regression model results.

To estimate the $E_x$ we need to know the observable fraction for each group. For this we assume that the SNe brighter than apparent magnitude 21, the nominal (i)PTF dark time brightness limit for the standard one minute exposure, are detected. Then we calculate using our normal distributions of peak brightness the observable fraction for each group with the following formula:

\begin{align}
    \text{p}_x = 1-\phi(\frac{dm-21.0-\mu_x}{\sigma}).
\end{align}

\noindent Here $\text{p}_x$ is the observable fraction of group $x$ and $\phi$ is the cumulative normal distribution function. In essence we subtract from unity the normalized sum of the normal distribution that passes the survey brightness threshold. Meaning that we get an estimate of the fraction of SNe in that group which were actually observed. Finally $E_x$ is calculated following

\begin{align}
    E_x = \int^{V_{C,max}}_{V_{C,min}} \text{p}_x dV_C,
\end{align}

\noindent where $V_C$ is the co-moving volume. The differential co-moving volume is integrated from $z=0$ to $z=z_{max}$, where $z_{max}$ is the maximum distance/redshift allowed in to the survey. Ideally, $z_{max}$ should be set ahead of time. At the least, it is crucial to not bias the results by selecting $z_{max}$ to maximize or minimize the desired outcome. In our case, we set $z_{max}$ to be the redshift of the furthest SN in our sample at $z=0.138$. If using a lower cut, the fractions and regression should be updated accordingly before applying the bias correction. We convert $z$ to distance modulus and calculate the differential co-moving distance using the same cosmology as in the rest of the paper based on parameters as measured by WMAP5.
Finally, we use the exposures $E_x$ and Eq.~\ref{eq:cr} to calculate the corrected fraction of broad/ordinary in each band and obtain the values reported in Table~\ref{tab:IbcFrac}. Poissonian uncertainties are calculated by normalizing the exposures to keep the number of detected broad objects the same ($E_1^{\text{norm}} = E_1/E_1 \equiv 1$, $E_0^{\text{norm}} = E_0/E_1$).

In general, our method can be represented algorithmically as follows:
\begin{enumerate}
    \item Determine $z_{max}$ of the sample.
    \item Fit a Gaussian to the peak brightness using truncated regression, with an appropriate left or right truncation, to each group in the sample. (In our case, the groups were ordinary and broad.)
    \item Using the resulting Gaussian mean(s) and standard deviation(s), calculate $\text{p}_x$ as a function of $z$.
    \item $E_x$ at a given $z$ is the integral from $z=0$ to $z$ of $\text{p}_x(z) \times dV_C$. It is important to not neglect the solid angle.
    \item Integrate $E_x$ up to $z_{max}$ to obtain the statistical exposure of each group.
    \item The Malmquist bias corrected fraction for any combination of parameters calculated between the groups is simply obtained by dividing the contribution of each group by its associated exposure.
\end{enumerate}

Malmquist-bias corrected ratios can be obtained for groups created using any measured or derived parameter (e.g., host metallicity or model parameters such as ejecta mass or nickel mass). As a concrete example, if the groups instead represent high- and low-velocity sub-groups within a sample of SNe, the above method allows calculating the fraction of high and low velocity objects. 

It is further possible to generalize this method to use other well-behaved distributions where $\text{p}_x$ can be calculated, instead of having to assume a normal distribution. For example, it would be possible to obtain the Malmquist-bias corrected fraction of Type Ia SNe subtypes versus normal SNe Ia, whose luminosity function usually follow a \citet{Schechter1976} function \citep{Brown2019}.

\section{Supernova classifications \label{sec:SNCLASSIFY}}
\citet{Schulze2021}, henceforth S21, comprehensively classified all (i)PTF CC~SNe from the (i)PTF database. Since we independently constructed and verified our SESN sample from the (i)PTF database, there are some differences with the S21 sample, as can be seen in Table~\ref{tab:cuts}. Generally, the differences were either due to missed classifications from external sources by S21, or due to S21 explicitly excluding or reclassifying the SN different than us, sometimes based on the work of \citet{Modjaz2020}, (M20). However, these minor differences do not impact our results, since most of the differing classifications discussed belong to SNe which do not pass our selection criteria for template fitting anyway. 
\begin{itemize}
    \item PTF09dha: missing from S21; this is classified as a Type Ib SN in the (i)PTF Marshal and by us.
    \item PTF10wg: Excluded by S21 due to uncertain classification; classified as Type Ib/c in (i)PTF Marshal and by us. Later cut in a further step in Table~\ref{tab:cuts} due to lacking a peak but included in our total number.
    \item PTF11bky: Missing from S21; this is a SN Ib/c classified by Central Bureau for Astronomical Telegrams (CBET) 2677 \citep{CBET2677}. 
    \item PTF11gcj: Excluded by S21 as a Ic-BL based on M20 reclassifying it from a Ic; Type Ic SN re-classified again by \citet{Barbarino2021} based on good spectral template matches to several Type Ic SNe.
    \item PTF12mps: Missing from S21; this is a Type Ib/c externally classified by \citep{2012ATel.4461....1H}.
    \item iPTF13doq: Excluded by S21 due to host contamination of the spectrum; we classify it as a Type Ib with an $r$-band stretch value around 0.8.
    \item iPTF13ou: Missing from S21; this is SN~2012fh classified as a Type Ic SN by \citet{2012CBET.3263....1N}. The (i)PTF lightcurve is only significantly after peak. 
    \item iPTF14jhf: Excluded from S21 due to low data quality; Type Ib/c with an $r$-band stretch value around 0.9, also classified by \citet{Fremling2018,Barbarino2021} .
    \item iPTF15cla: Missing from S21; this is a Type Ib/c classified by \citet{2015ATel.7830....1T}. 
    \item iPTF16ewz: Missing from S21; we tentatively classified it as a Type Ic-BL SN as in the (i)PTF marshal, although it was noted as a possible Type Ia or other uncertain classification as well, and hence was excluded from \citep{Taddia2019b}. Later, it was re-classified as a Type Ic by \citep{TNS2021Sollerman}. Due to the uncertain classification, we exclude it from the 183 Type Ibc/IIb SNe in our sample, but it is included in our topline number of 220 SESNe in Table~\ref{tab:cuts}. 

\textit{Excluded from our sample:}

    \item PTF10ysd: Included in S21 as a Type Ic as reclassified by M20; we classify as a Type Ic-BL as in \citet{Taddia2019b,Barbarino2021}.
    \item PTF10aauo: Included in S21 as a Type IIb; excluded by us as Type II with a blue continuum, noted as Type II Blue in the (i)PTF Marshal.
    \item PTF10xfh: Included in S21 as a Type Ib; we exclude it as a peculiar gap transient as in the (i)PTF Marshal.
    \item PTF13dzy: Included in S21 as Type IIb; we exclude it as a Type II as in the (i)PTF Marshal.
\end{itemize}

By including the 9 SE~SNe excluded from the CC SN sample of S21, we update the number of CC SNe in (i)PTF to 897 (from 888 in S21), which includes all peculiar SNe such as SLSN and Type Ibn SNe. Excluding them, the total CC SNe would instead become 833 (from 824 in S21). 

\section{Supporting figures and tables} \label{sec:supportingfigures}

In this section, we provide supporting figures and tables for each SN in our broad Type~Ibc sample. We include plots with apparent magnitude lightcurves, K-corrections, Arnett model fits to the bolometric lightcurves, spectral sequences, and line-velocity evolution. We also include tables with all spectral observations and host metallicity measurements. 
\begin{figure*}
    \centering
    \includegraphics[width=\linewidth]{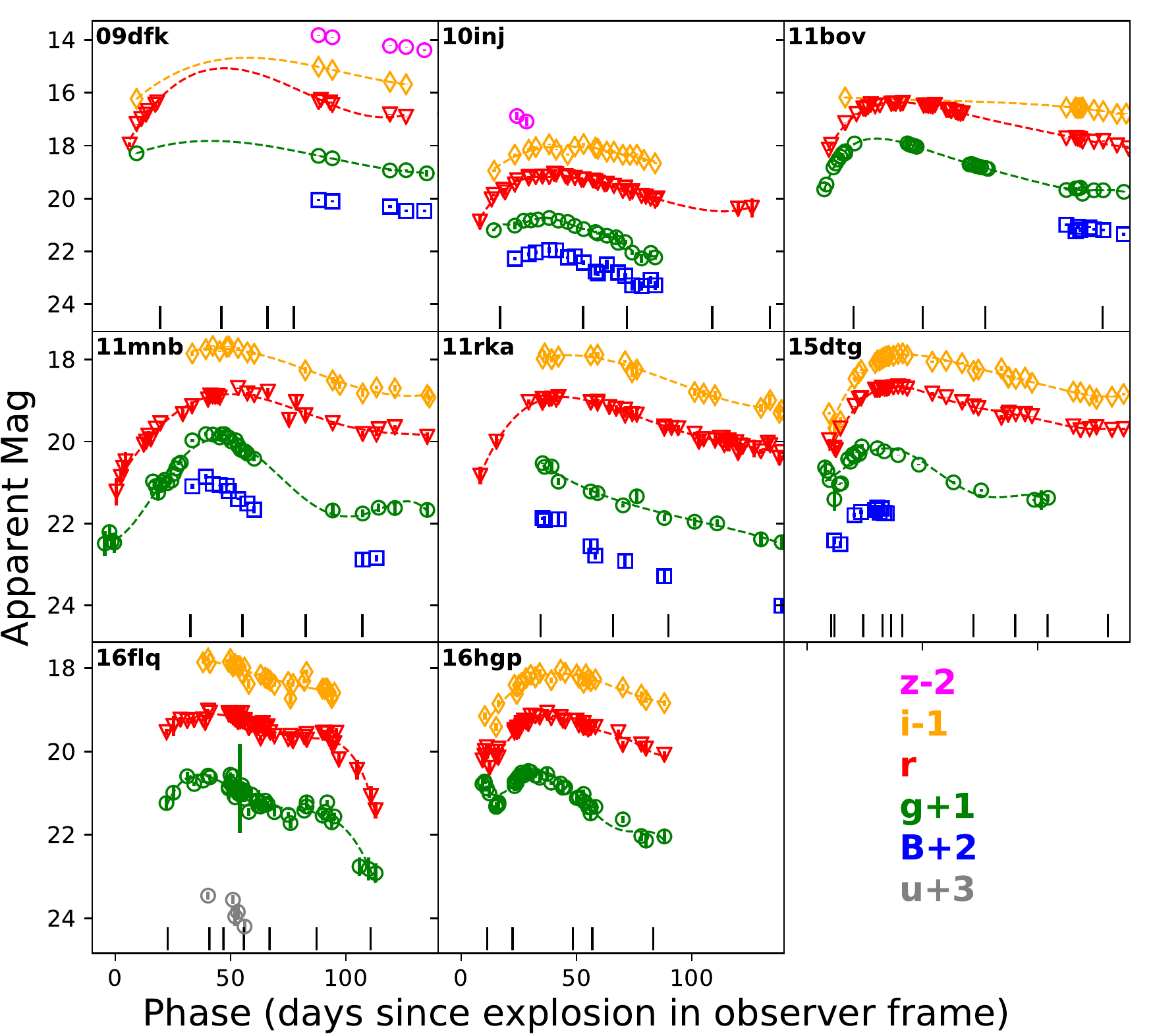}
    \caption{The apparent magnitudes of our broad SE~SNe in the observer frame. No extinction or redshift correction were applied. For clarity, data after 150 days and all upper limits have been omitted. The photometry are available on Wiserep.}
    \label{fig:apparentmags}
\end{figure*}

\begin{figure*}
    \centering
    \includegraphics[width=\linewidth]{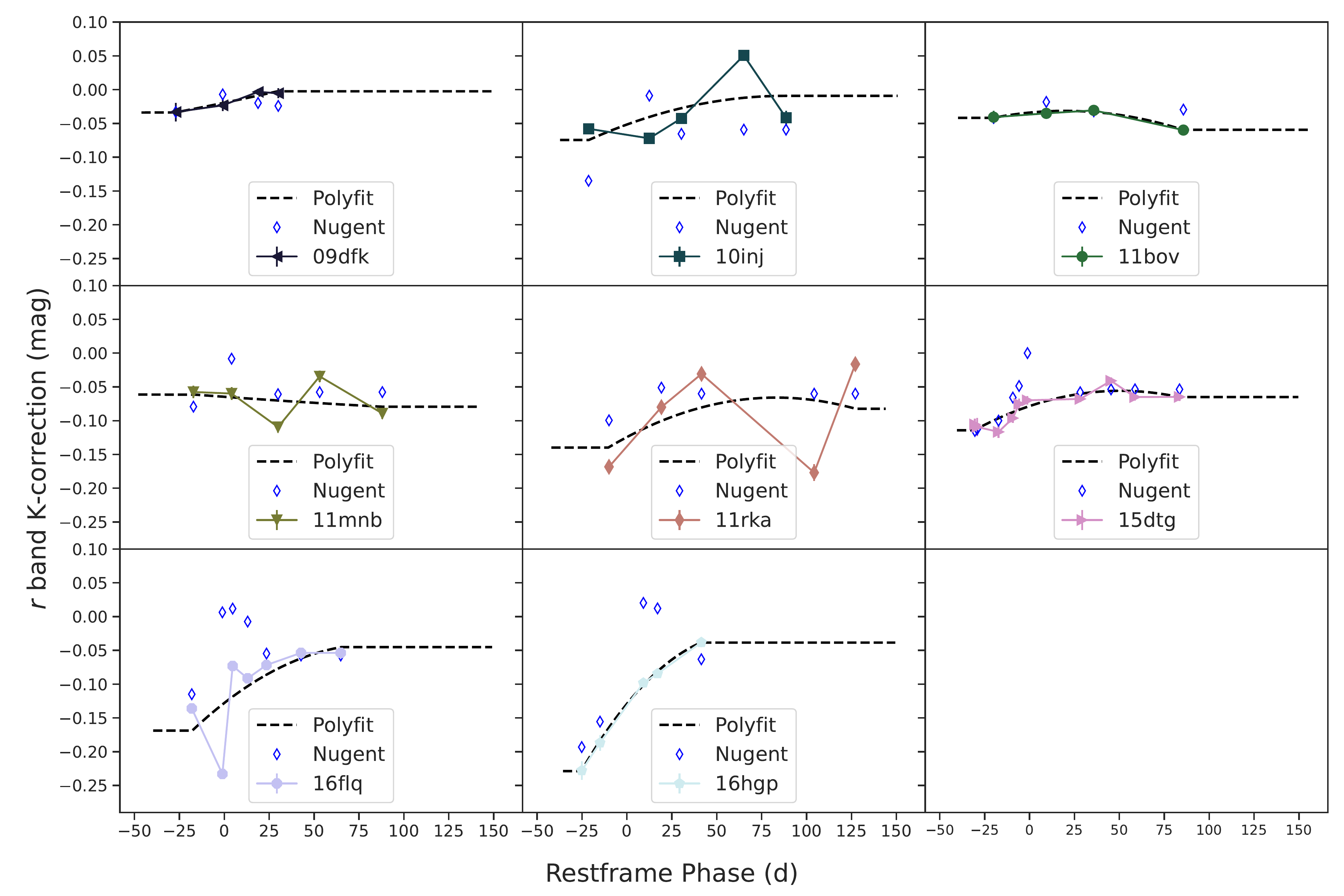}
    \caption{K-corrections in the $r$ band compared to template K-corrections. Open diamonds mark K-corrections calculated from template Type~Ibc spectra, see Sect.~\ref{sec:otherbias}. At the epochs of the existing spectra, colored symbols are K-corrections calculated from the spectra of our broad sample, and the dashed lines represent the actual K-corrections used, which use a second degree polynomial fit for interpolation or a flat extrapolation.}
    \label{fig:kcorr}
\end{figure*}

\begin{figure*}
    \centering
    \includegraphics[width=\linewidth]{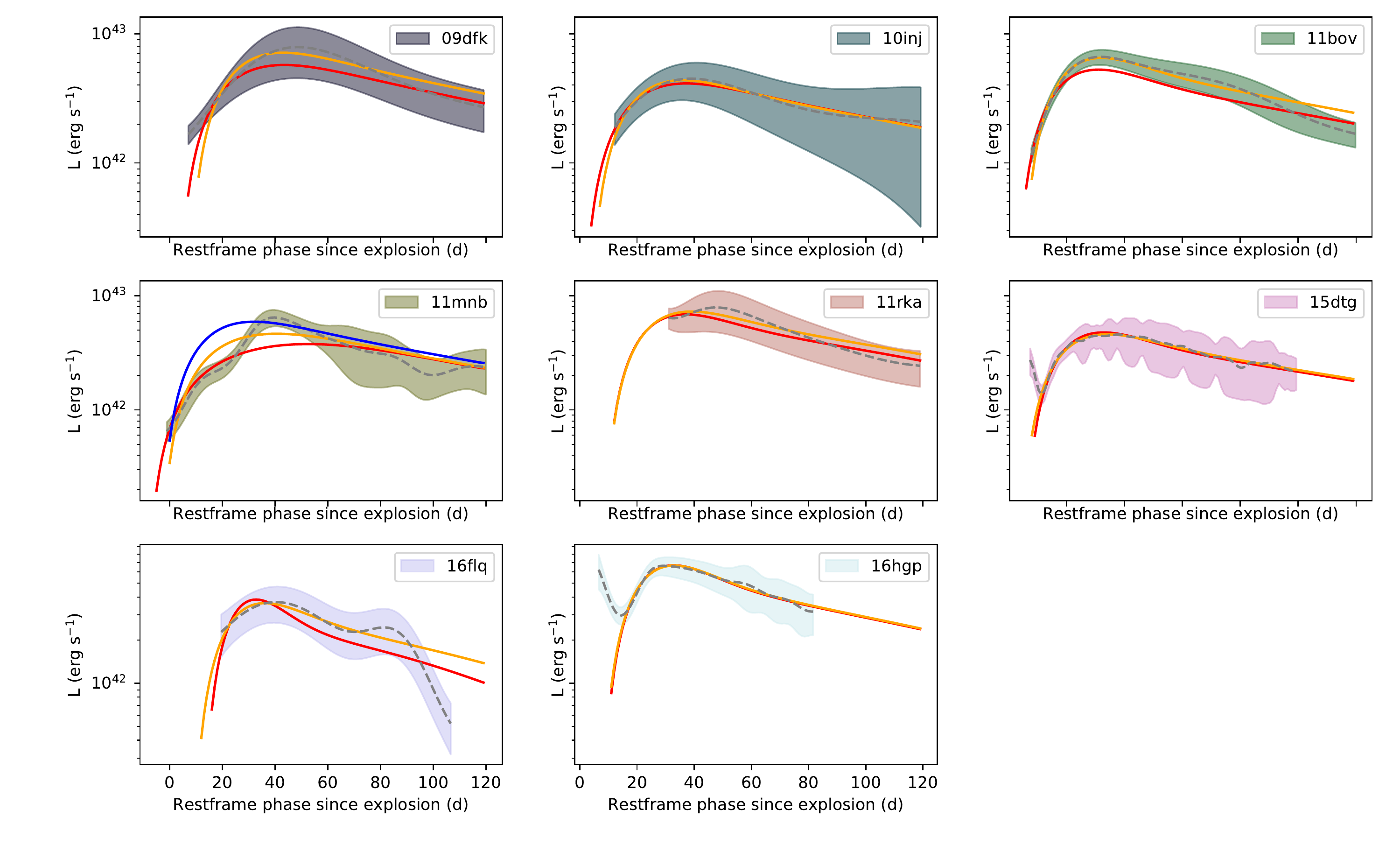}
    \caption{Bolometric lightcurves of the broad SE~SN sample overplotted with weighted (orange) and unweighted (red) Arnett model fits. The fit to the main peak of PTF11mnb is shown with a blue line (see Sect.~\ref{sec:model}).}
    \label{fig:bolomags}
\end{figure*}

\begin{figure*}
    \centering
    \includegraphics[width=\linewidth]{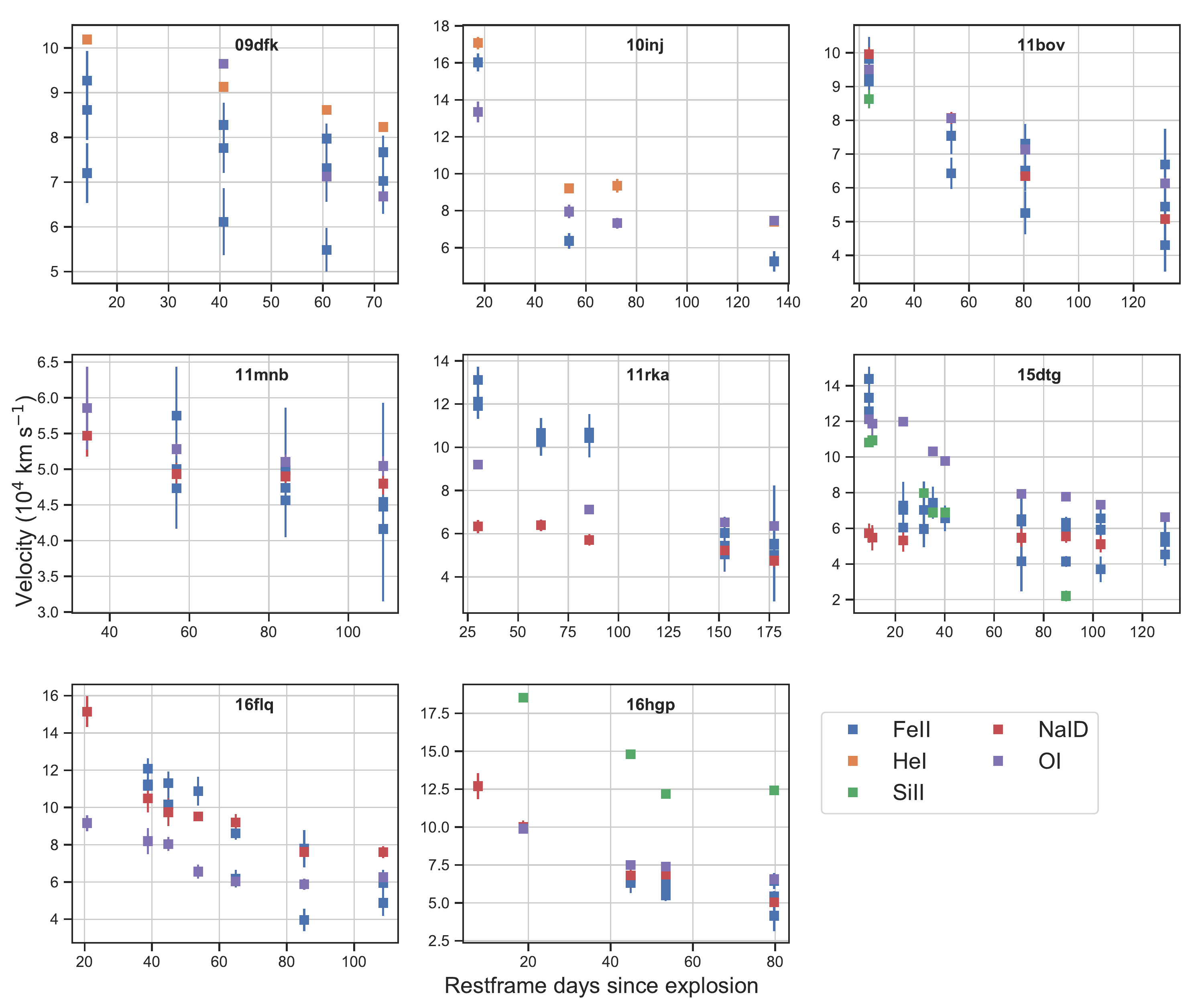}
    \caption{Absorption velocity of \ion{Fe}{ii} $\lambda\lambda 4924, 5018, 5169$, \ion{He}{i} $\lambda 5876$, \ion{Na}{i} $\lambda\lambda 5890, 5896$, \ion{O}{i} $\lambda\lambda 7772, 7774, 7775$, and \ion{Si}{ii} $\lambda 6355$ in each SN.}
    \label{fig:velsbysn}
\end{figure*}

\begin{figure*}
\centerline{
\includegraphics[width=0.5\linewidth,trim=-0cm 0 +2cm 0]{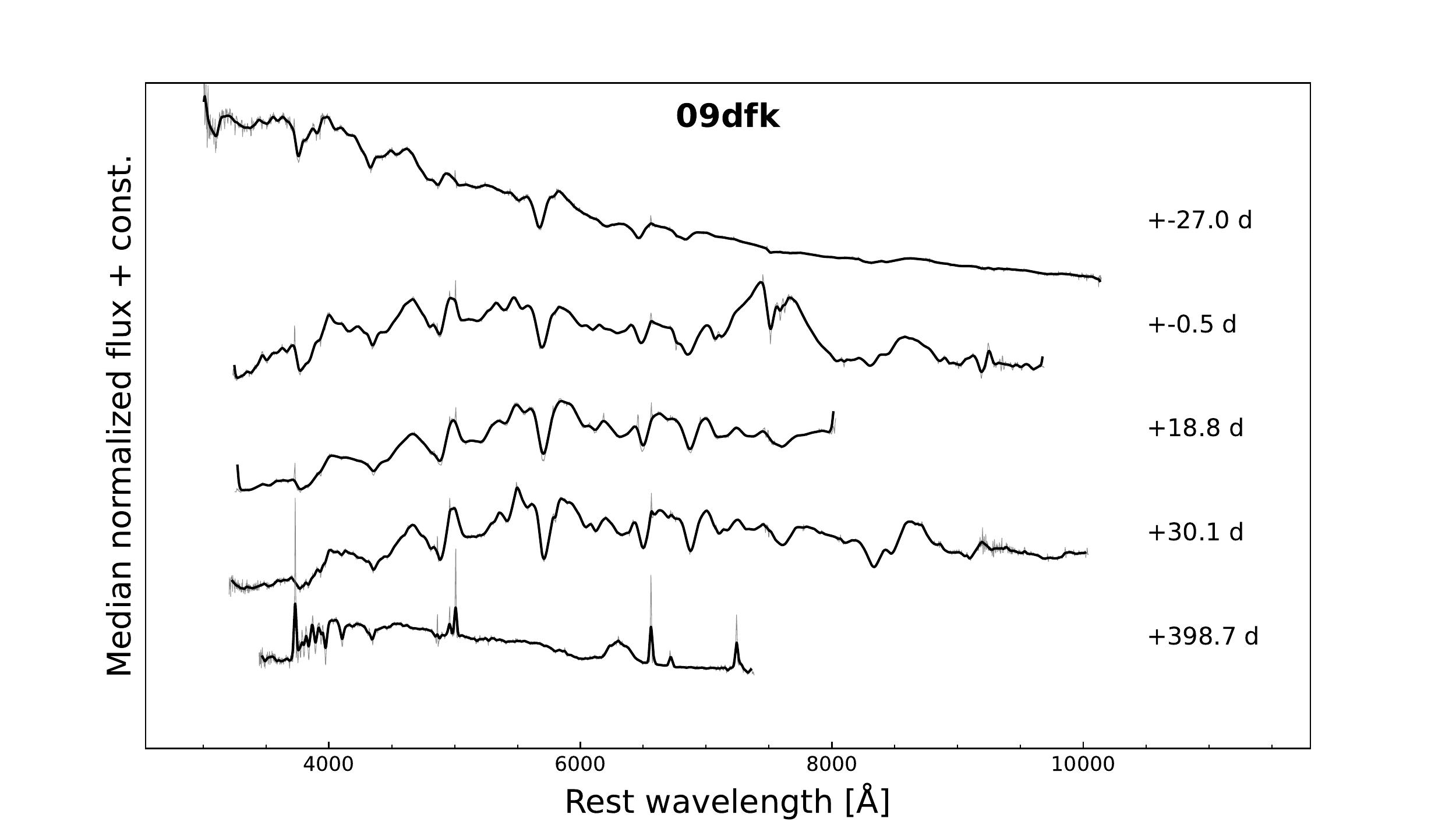}
\includegraphics[width=0.5\linewidth,trim=+2cm 0 0cm 0]{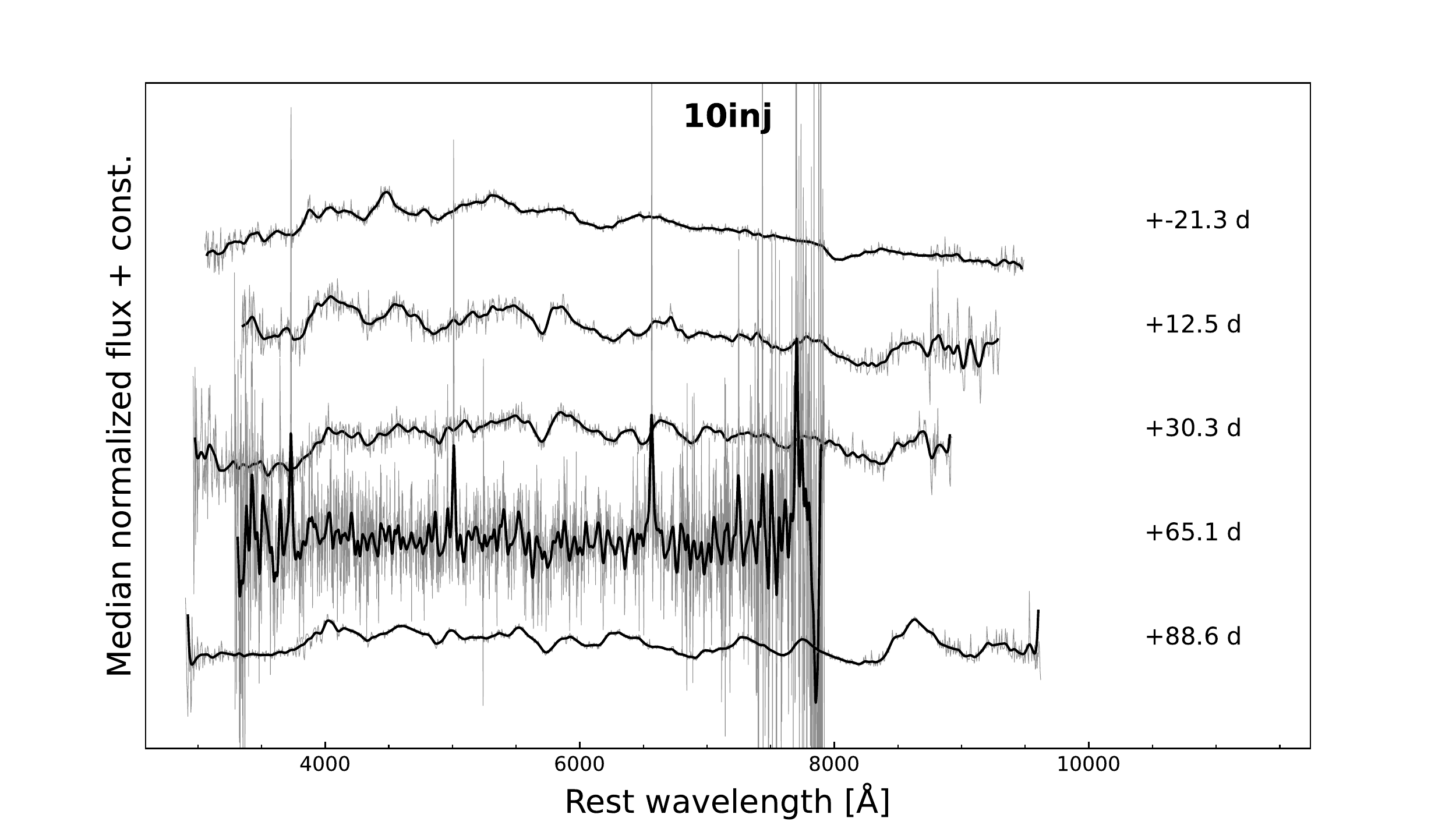}
}

\caption{Gaussian smoothed spectral sequence of PTF09dfk and PTF10inj, labeled with restframe phase.}
\label{fig:spec1}
\end{figure*}

\begin{figure*}
\centering
\centerline{ 
\includegraphics[width=0.5\linewidth,trim=-0cm 0 +2cm 0]{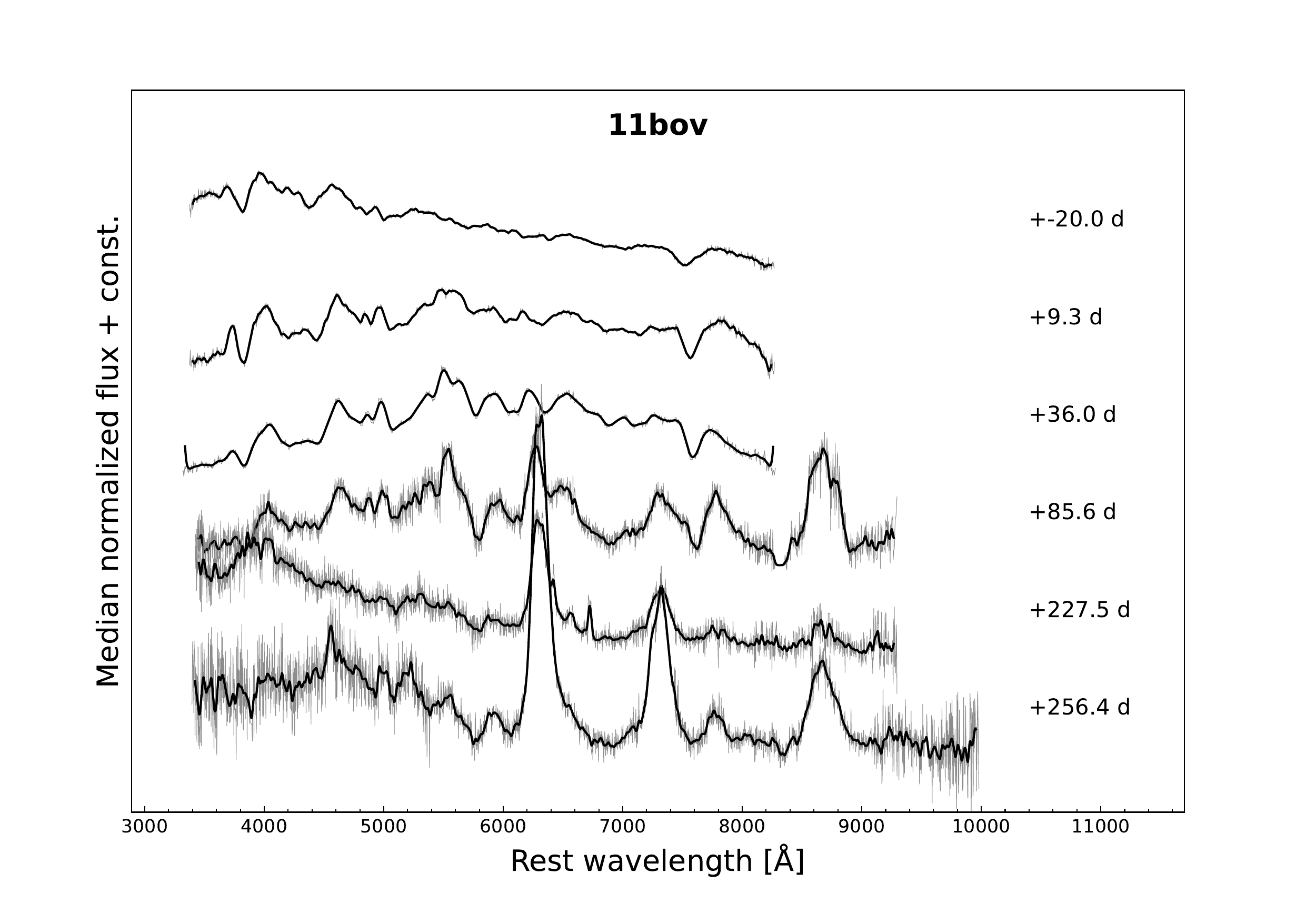}
 \includegraphics[width=0.5\linewidth,trim=+2cm 0 -0cm 0]{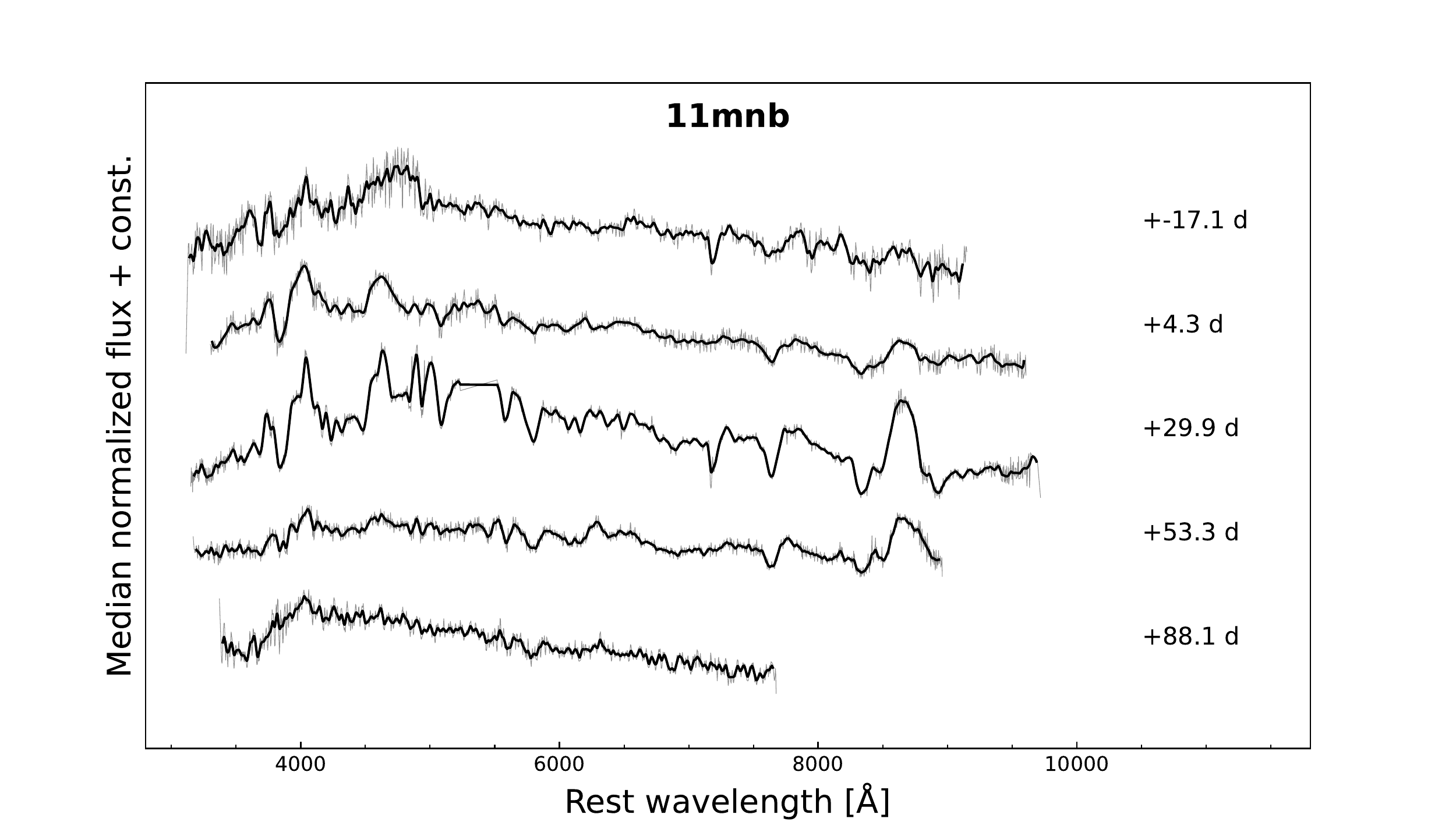}
 }

\caption{Gaussian smoothed spectral sequence of PTF11bov and PTF11mnb, labeled with restframe phase.}
\label{fig:spec2}
\end{figure*}

\begin{figure*}
\centering
\centerline{ 
\includegraphics[width=0.5\linewidth,trim=-0cm 0 +2cm 0]{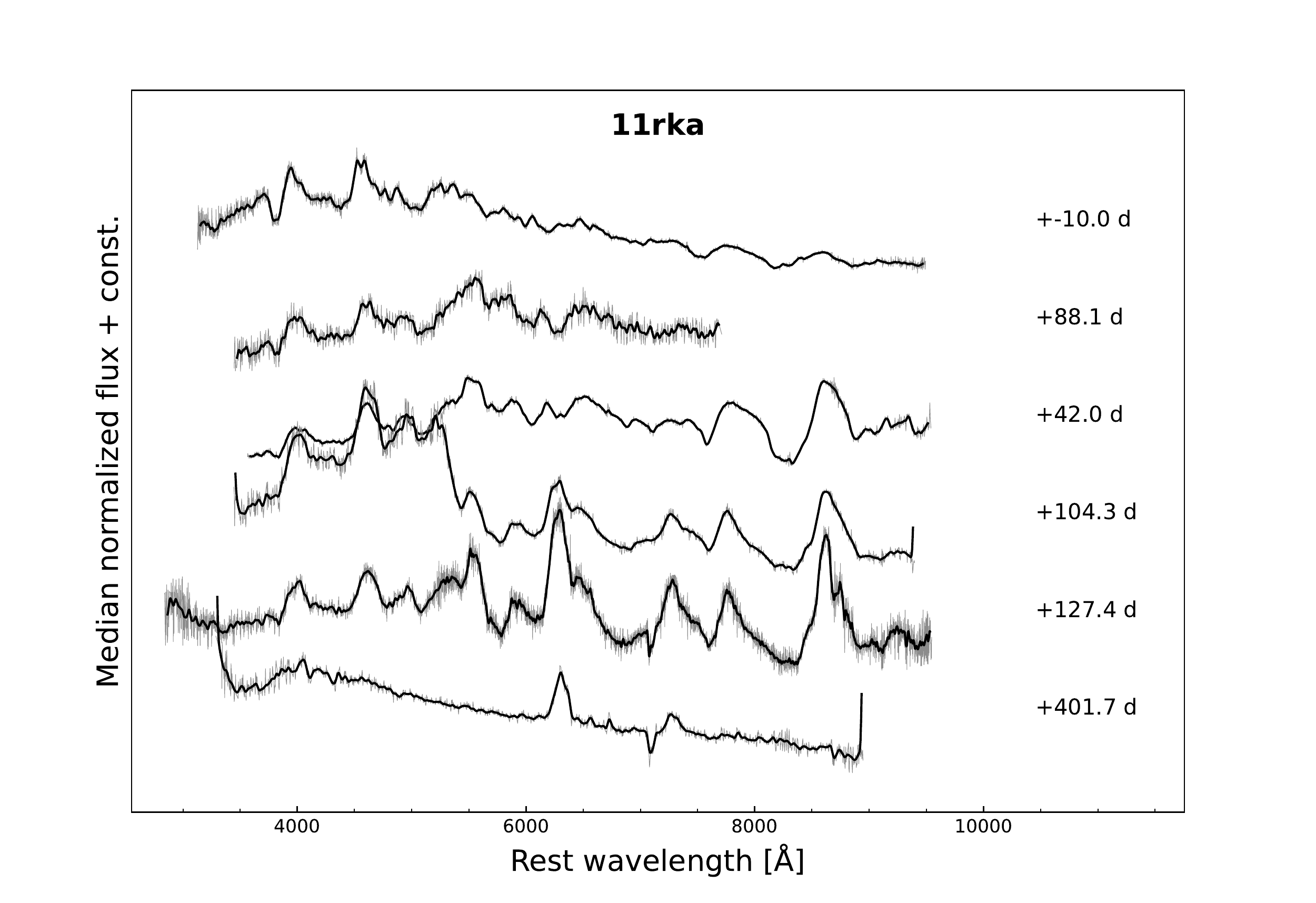}
 \includegraphics[width=0.5\linewidth,trim=+2cm 0 0cm 0]{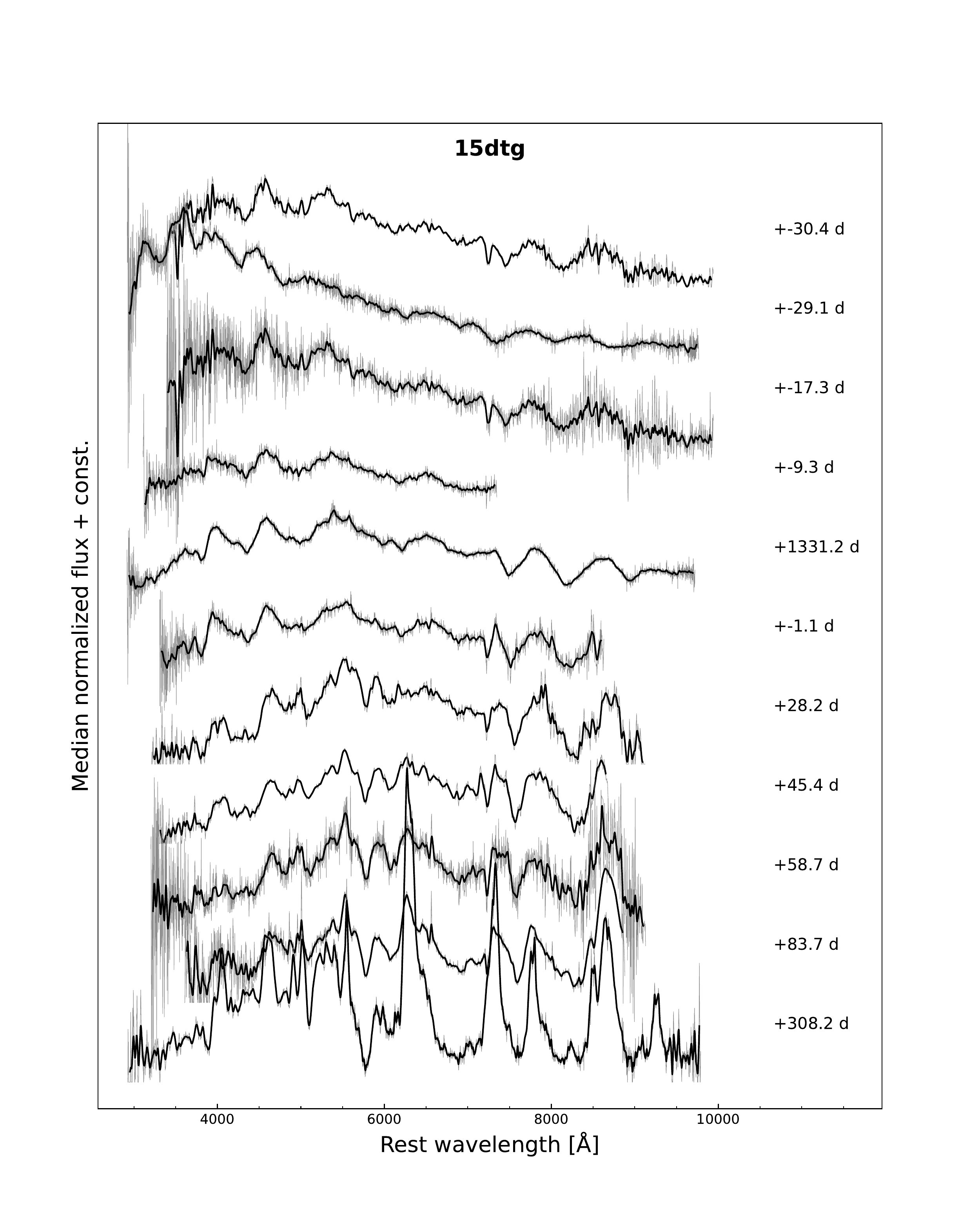}
 }

\caption{Gaussian smoothed spectral sequence of PTF11rka and iPTF15dtg, labeled with restframe phase.}
\label{fig:spec3}
\end{figure*}    
    
  \begin{figure*}
\centering
\centerline{ 
\includegraphics[width=0.5\linewidth,trim=-0cm 0 +2cm 0]{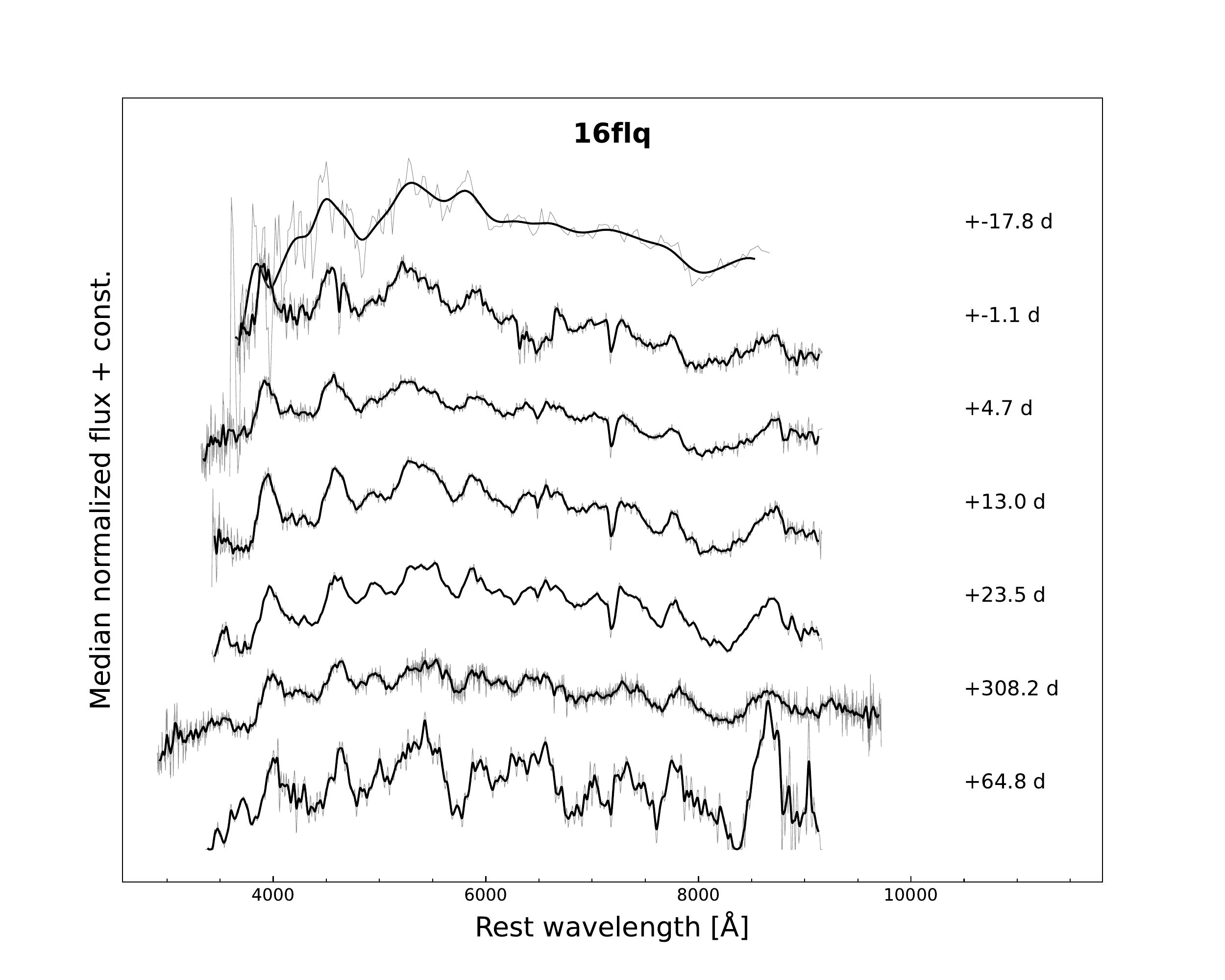}
 \includegraphics[width=0.5\linewidth,trim=+2cm 0 0cm 0]{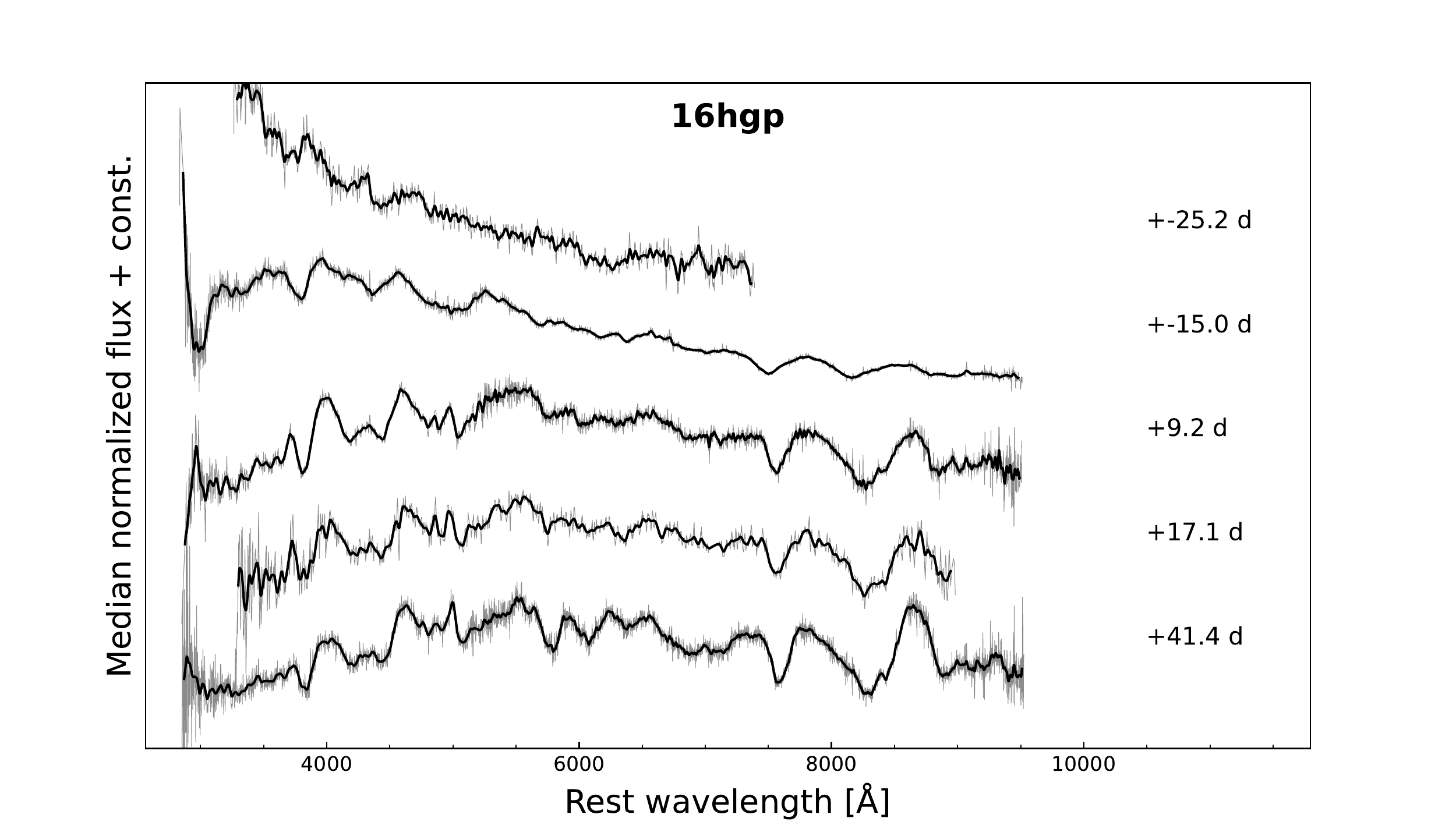}
 }

\caption{Gaussian smoothed spectral sequence of iPTF6flq and iPTF16hgp, labeled with restframe phase.}
\label{fig:spec4}
\end{figure*}

\begin{table*}
\begin{tabular}{llrllr}
\toprule
SN         &         JD &                     Telescope &   Instrument &   Restframe Phase \\
           &            &                               &              &    (days)  \\
\midrule
PTF09dfk  & 2455068.95 &         Keck I 10m &        LRIS &   $-$27.0 \\
          & 2455095.88 &         P200       &        DBSP &    $-$0.5 \\
          & 2455115.50 &           WHT 4.2m &        ISIS &    18.8 \\
          & 2455126.92 &         Keck I 10m &        LRIS &    30.1 \\
          & 2455501.50 &         Keck I 10m &        LRIS &   398.7 \\
\hline
PTF10inj  & 2455355.50 &         Keck I 10m &        LRIS &   $-$21.3 \\
          & 2455391.50 &         P200       &        DBSP &    12.5 \\
          & 2455410.50 &           WHT 4.2m &        ISIS &    30.3 \\
          & 2455447.50 &            KPNO 4m &     RC Spec &    65.1 \\
          & 2455472.50 &         Keck I 10m &        LRIS &    88.6 \\
          & 2457982.73 &         Keck I 10m &        LRIS &  2447.2 \\
\hline
PTF11bov  & 2455662.50 &            KPNO 4m &     RC Spec &   $-$20.0 \\
          & 2455692.50 &            KPNO 4m &     RC Spec &     9.3 \\
          & 2455719.79 &            KPNO 4m &     RC Spec &    36.0 \\
          & 2455770.50 &           WHT 4.2m &        ISIS &    85.6 \\
          & 2455913.74 &                TNG &     DOLoReS &   225.8 \\
          & 2455915.50 &           WHT 4.2m &        ISIS &   227.5 \\
          & 2455945.01 &         P200       &        DBSP &   256.4 \\
\hline
PTF11mnb  & 2455842.01 &               UH88 &       SNIFS &   $-$17.1 \\
          & 2455864.78 &         P200       &        DBSP &     4.3 \\
          & 2455891.91 &         Keck I 10m &        LRIS &    29.9 \\
          & 2455916.68 &         P200       &        DBSP &    53.3 \\
          & 2455953.59 &            KPNO 4m &     RC Spec &    88.1 \\
\hline
PTF11rka  & 2455922.15 &         Keck I 10m &        LRIS &   $-$10.0 \\
          & 2455953.92 &            KPNO 4m &     RC Spec &    19.6 \\
          & 2455977.99 &         Keck I 10m &        LRIS &    42.0 \\
          & 2456044.88 &         Keck I 10m &        LRIS &   104.3 \\
          & 2456069.77 &         Keck I 10m &        LRIS &   127.4 \\
          & 2456364.50 &                VLT &       FORS2 &   401.7 \\
          & 2457363.10 &         Keck I 10m &        LRIS &  1331.2 \\
\hline
iPTF15dtg & 2457336.60 &                TNG &     DOLoReS &   $-$30.4 \\
          & 2457337.98 &         Keck I 10m &        LRIS &   $-$29.1 \\
          & 2457350.46 &                TNG &     DOLoReS &   $-$17.3 \\
          & 2457358.85 &                DCT &  Deveny+LMI &    $-$9.3 \\
          & 2457362.88 &         Keck I 10m &        LRIS &    $-$5.5 \\
          & 2457367.43 &                NOT &      ALFOSC &    $-$1.1 \\
          & 2457398.32 &                TNG &     DOLoReS &    28.2 \\
          & 2457416.37 &                NOT &      ALFOSC &    45.4 \\
          & 2457430.42 &                TNG &     DOLoReS &    58.7 \\
          & 2457456.73 &       Gemini North &        GMOS &    83.7 \\
          & 2457693.02 &         Keck I 10m &        LRIS &   308.2 \\
          & 2457730.53 &                NOT &      ALFOSC &   343.9 \\
          & 2457982.97 &         Keck I 10m &        LRIS &   583.7 \\
\hline
iPTF16flq & 2457628.84 &       Palomar 1.5m &        SEDM &   $-$17.8 \\
          & 2457646.56 &                NOT &      ALFOSC &    $-$1.1 \\
          & 2457652.62 &                NOT &      ALFOSC &     4.7 \\
          & 2457661.48 &                NOT &      ALFOSC &    13.0 \\
          & 2457672.60 &                NOT &      ALFOSC &    23.5 \\
          & 2457692.95 &         Keck I 10m &        LRIS &    42.7 \\
          & 2457716.40 &                NOT &      ALFOSC &    64.8 \\
\hline
iPTF16hgp & 2457683.80 &                DCT &  Deveny+LMI &   $-$25.2 \\
          & 2457694.81 &         Keck I 10m &        LRIS &   $-$15.0 \\
          & 2457720.90 &         Keck I 10m &        LRIS &     9.2 \\
          & 2457729.43 &                NOT &      ALFOSC &    17.1 \\
          & 2457755.78 &         Keck I 10m &        LRIS &    41.4 \\
\bottomrule
\end{tabular}
\label{tab:speclog}
\caption{Log of spectral observations from (i)PTF of the eight broad SE~SNe. Acronyms in addition to those previously introduced: William Herschel Telescope (WHT), Low Resolution Imaging Spectrometer (LRIS), Gemini Multi-Object Spectrograph (GMOS), Large Monolithic Imager (LMI), The Double Spectrograph (DBSP), SuperNova Integrated Field Spectrograph (SNIFS), Intermediate-dispersion Spectrograph and Imaging System (ISIS), SEDMachine (SEDM)}
\end{table*}

\begin{table*}
\centering
\begin{tabular}{lll}
\toprule
SN &                   $M13$ &                  $PP04$ \\
    &  $12 + log(O/H)$      &      $12 + log(O/H)$   \\
\midrule
09dfk &    $8.17^{8.2}_{8.14} $ &   $8.19^{8.22}_{8.14} $ \\[1ex]
10inj &   $8.18^{8.21}_{8.14} $ &    $8.2^{8.23}_{8.14} $ \\[1ex]
11bov &   $8.21^{8.24}_{8.18} $ &   $8.25^{8.28}_{8.18} $ \\[1ex]
11mnb &   $8.18^{8.21}_{8.14} $ &    $8.2^{8.23}_{8.14} $ \\[1ex]
11rka &    $8.0^{8.04}_{7.95} $ &   $7.93^{7.97}_{7.95} $ \\[1ex]
15dtg &    $8.06^{8.1}_{8.02} $ &   $8.03^{8.07}_{8.02} $ \\[1ex]
16flq &   $8.37^{8.39}_{8.34} $ &    $8.48^{8.5}_{8.34} $ \\[1ex]
16hgp &    $8.13^{8.16}_{8.1} $ &    $8.13^{8.16}_{8.1} $ \\[1ex]
\bottomrule
\end{tabular}
\centering
\caption{Metallicity of the host galaxies of the broad sample as measured using the PP04 and M13 line flux ratio methods, with the confidence interval given as the superscript upper and subscript lower bound.}
\label{tab:metal}
\end{table*}

\begin{table*}
\centering
\begin{tabular}{llllll}
\toprule
SN       &           $M_{ej}$ &            $M_{Ni}$ &             $E_k$ &            $E_Exp$ &        $V_{ph}$ \\
         &      ($M_\sun$)    &      ($M_\sun$)     &    ($10^{51} erg$)&       ($restframe days$)   &      ($km s^{-1}$) \\
\midrule
09dfk      &  $ 22.6 \pm 1.8 $ &  $ 0.58 \pm 0.01 $ &  $ 7.3 \pm 2.3 $ &  $ 2.2 \pm 0.9 $ &  $ 7300 \pm 1100 $ \\
10inj      &  $ 26.4 \pm 1.0 $ &  $ 0.39 \pm 0.01 $ &  $ 16.4 \pm 1.7$ &  $ 0.0 \pm 0.3$ &  $ 10200 \pm 500 $ \\
11bov      &  $ 13.0 \pm 1.0 $ &  $ 0.42 \pm 0.02 $ &  $ 4.6 \pm 0.6 $ &  $ 1.5 \pm 0.3 $ &  $  7700 \pm 400 $ \\
11mnb      &  $ 32.4 \pm 5.0 $ &  $ 0.48 \pm 0.05 $ &  $ 5.6 \pm 0.9 $ &  $ -9.3 \pm 0.5$ &  $  5400 \pm 200 $ \\
11mnb\_peak1 &  $ 13.3 \pm 2.5 $ &  $ 0.55 \pm 0.03 $ &  $ 2.3 \pm 0.4 $ &  $ -4.3 \pm 5.3$ &  $  5400 \pm 200 $ \\
11rka      &  $ 20.3 \pm 4.0 $ &  $ 0.54 \pm 0.03 $ &  $ 18.2 \pm 3.6$ &  $ 7.7 \pm 10.2$ &  $ 12200 \pm 200 $ \\
15dtg      &  $ 11.6 \pm 0.4 $ &  $ 0.36 \pm 0.01 $ &  $ 4.0 \pm 0.9 $ &  $ 4.6 \pm 0.2 $ &  $  7600 \pm 900 $ \\
16flq      &  $ 10.1 \pm 2.6 $ &  $ 0.23 \pm 0.01 $ &  $ 9.2 \pm 2.6 $ &  $ 11.7 \pm 2.3$ &  $ 12400 \pm 700 $ \\
16hgp      &  $ 9.7 \pm 0.4  $ &  $ 0.48 \pm 0.01 $ &  $ 3.2 \pm 0.5 $ &  $ 6.8 \pm 0.3 $ &  $  7400 \pm 600 $ \\

\bottomrule
\end{tabular}
\centering
\caption{Arnett model parameters with their statistical fitting uncertainties and measured characteristic ejecta velocities. \emph{11mnb\_peak1} refers to the fit to only the first peak of 11mnb. 
The explosion epochs were fit as described in the text and their shift and uncertainty is listed under the $E_Exp$ column. There is a significant correlation between errors on $E_Exp$
and $M_{ej}$. $M_{Ni}$ could be driven up by host extinction correction, which we do not apply. The distance uncertainty (e.g. random from peculiar velocities or systematic from $H_0$) is not included in the error terms.}
\label{tab:fitting}
\end{table*}

\begin{table*}
\centering
\caption{Line velocities of \ion{Fe}{ii} $\lambda\lambda 4924, 5018, 5169$, \ion{He}{i} $\lambda 5876$, \ion{Na}{iD} $\lambda\lambda 5890, 5986$ (as 5890), \ion{Si}{ii} $\lambda 6355$, and \ion{O}{i} $\lambda 7774$. Phases are in restframe.}
\label{tab:linevels}
\scalebox{0.9}{
\begin{tabular}{llrlllllll}
\toprule
SN &     Date & Phase &   $\lambda4924$ &   $\lambda5018$ &    $\lambda5169$ &   $\lambda5876$ &   $\lambda5890$ &   $\lambda6355$ &   $\lambda7774$ \\
   &          &  (days) &  ($km s^{-1}$) &   ($km s^{-1}$) &   ($km s^{-1}$) &    ($km s^{-1}$) &  ($km s^{-1}$) &   ($km s^{-1}$) &   ($km s^{-1}$) \\
\midrule
09dfk & 20090825 &  $-27.0$ &  $8620 \pm 670$ &  $9270 \pm 660$ &   $7200 \pm 670$ &  $10190 \pm 60$ &              -- &              -- &              -- \\
09dfk & 20090921 &   $-0.5$ &  $7770 \pm 560$ &  $8280 \pm 500$ &   $6110 \pm 750$ &   $9130 \pm 50$ &              -- &              -- &   $9650 \pm 80$ \\
09dfk & 20091011 &   $18.8$ &  $7310 \pm 750$ &  $7970 \pm 340$ &   $5490 \pm 490$ &   $8620 \pm 70$ &              -- &              -- &  $7130 \pm 110$ \\
09dfk & 20091022 &   $30.1$ &  $70.0 \pm 740$ &  $7670 \pm 380$ &               -- &   $8230 \pm 80$ &              -- &              -- &   $6680 \pm 80$ \\
09dfk & 20101101 &  $398.7$ &              -- &              -- &               -- &              -- &              -- &              -- &              -- \\
10inj & 20100608 &  $-21.3$ &              -- &              -- &  $160.0 \pm 480$ & $170.0 \pm 340$ &              -- &              -- & $13340 \pm 570$ \\
10inj & 20100714 &   $12.5$ &              -- &              -- &   $6370 \pm 420$ &  $9210 \pm 280$ &              -- &              -- &  $7950 \pm 370$ \\
10inj & 20100802 &   $30.3$ &              -- &              -- &               -- &  $9350 \pm 370$ &              -- &              -- &  $7330 \pm 290$ \\
10inj & 20100908 &   $65.1$ &              -- &              -- &               -- &              -- &              -- &              -- &              -- \\
10inj & 20101003 &   $88.6$ &              -- &              -- &   $5260 \pm 560$ &  $7410 \pm 270$ &              -- &              -- &  $7470 \pm 210$ \\
10inj & 20170817 & $2447.2$ &              -- &              -- &               -- &              -- &              -- &              -- &              -- \\
11bov & 20110411 &  $-20.0$ &  $9150 \pm 690$ &  $9810 \pm 660$ &   $9420 \pm 610$ &              -- &  $9960 \pm 200$ &  $8630 \pm 270$ &  $9510 \pm 110$ \\
11bov & 20110511 &    $9.3$ &  $7540 \pm 470$ &  $7540 \pm 530$ &   $6430 \pm 460$ &              -- &  $80.0 \pm 180$ &              -- &  $80.0 \pm 110$ \\
11bov & 20110607 &   $36.0$ &  $7310 \pm 580$ &  $6510 \pm 610$ &   $5260 \pm 630$ &              -- &   $6350 \pm 70$ &              -- &   $7140 \pm 80$ \\
11bov & 20110728 &   $85.6$ & $6690 \pm 10.0$ &  $5440 \pm 260$ &   $4310 \pm 790$ &              -- &  $50.0 \pm 160$ &              -- &  $6140 \pm 170$ \\
11bov & 20111218 &  $225.8$ &              -- &              -- &               -- &              -- &              -- &              -- &              -- \\
11bov & 20111220 &  $227.5$ &              -- &              -- &               -- &              -- &              -- &              -- &              -- \\
11bov & 20120118 &  $255.9$ &              -- &              -- &               -- &              -- &              -- &              -- &              -- \\
11mnb & 20111007 &  $-17.1$ &              -- &              -- &               -- &              -- &  $5470 \pm 290$ &              -- &  $5850 \pm 580$ \\
11mnb & 20111030 &    $4.1$ &  $5750 \pm 690$ &  $50.0 \pm 440$ &   $4730 \pm 570$ &              -- &  $4930 \pm 160$ &              -- &   $5280 \pm 90$ \\
11mnb & 20111126 &   $29.9$ &  $4980 \pm 880$ &  $4740 \pm 690$ &   $4570 \pm 350$ &              -- &  $4900 \pm 130$ &              -- &  $5100 \pm 100$ \\
11mnb & 20111221 &   $53.3$ &  $4160 \pm 580$ &  $4480 \pm 340$ &  $4540 \pm 1390$ &              -- &  $4800 \pm 150$ &              -- &  $50.0 \pm 140$ \\
11mnb & 20120127 &   $88.1$ &              -- &              -- &               -- &              -- &              -- &              -- &              -- \\
11rka & 20111226 &  $-10.0$ & $11910 \pm 600$ & $12110 \pm 450$ &  $13110 \pm 620$ &              -- &  $6330 \pm 310$ &              -- &  $9200 \pm 110$ \\
11rka & 20120220 &   $42.0$ & $10420 \pm 380$ & $10670 \pm 470$ & $10530 \pm 10.0$ &              -- &  $5710 \pm 260$ &              -- &   $7120 \pm 70$ \\
11rka & 20120127 &   $88.1$ & $10240 \pm 630$ & $10670 \pm 690$ &               -- &              -- &  $6390 \pm 270$ &              -- &              -- \\
11rka & 20120427 &  $104.3$ &  $5450 \pm 550$ &  $60.0 \pm 740$ &   $50.0 \pm 800$ &              -- &  $5220 \pm 120$ &              -- &  $6530 \pm 170$ \\
11rka & 20120522 &  $127.4$ & $5550 \pm 2680$ &  $5480 \pm 820$ &   $50.0 \pm 400$ &              -- &  $4750 \pm 280$ &              -- &  $6360 \pm 120$ \\
11rka & 20130313 &  $401.7$ &              -- &              -- &               -- &              -- &              -- &              -- &              -- \\
11rka & 20151206 & $1331.2$ &              -- &              -- &               -- &              -- &              -- &              -- &              -- \\
15dtg & 20151110 &  $-30.4$ & $14380 \pm 690$ & $12570 \pm 350$ &  $13320 \pm 930$ &              -- &  $5730 \pm 550$ & $10820 \pm 230$ & $12120 \pm 140$ \\
15dtg & 20151111 &  $-29.1$ &              -- &              -- &               -- &              -- &  $5480 \pm 710$ &  $10940 \pm 60$ & $11870 \pm 810$ \\
15dtg & 20151123 &  $-17.3$ &  $70.0 \pm 740$ & $7270 \pm 1340$ &   $60.0 \pm 700$ &              -- &  $5320 \pm 630$ &              -- & $11990 \pm 160$ \\
15dtg & 20151202 &   $-9.3$ & $70.0 \pm 1590$ & $5960 \pm 10.0$ &               -- &              -- &              -- &  $7980 \pm 150$ &              -- \\
15dtg & 20151210 &   $-1.1$ &              -- &              -- &   $6560 \pm 720$ &              -- &              -- &  $6890 \pm 130$ &  $9780 \pm 100$ \\
15dtg & 20160110 &   $28.2$ &  $6510 \pm 590$ & $6390 \pm 1430$ &  $4150 \pm 1690$ &              -- &  $5470 \pm 500$ &              -- &  $7940 \pm 100$ \\
15dtg & 20160128 &   $45.4$ &  $60.0 \pm 640$ &  $6320 \pm 270$ &   $4140 \pm 300$ &              -- &  $5550 \pm 360$ &  $2210 \pm 300$ &  $7780 \pm 270$ \\
15dtg & 20160211 &   $58.7$ &  $5920 \pm 650$ &  $6550 \pm 530$ &   $3710 \pm 720$ &              -- &  $5110 \pm 470$ &              -- &  $7330 \pm 140$ \\
15dtg & 20160309 &   $83.7$ & $5230 \pm 1330$ & $5530 \pm 1290$ &   $4540 \pm 570$ &              -- &              -- &              -- &  $6630 \pm 130$ \\
15dtg & 20161031 &  $308.2$ &              -- &              -- &               -- &              -- &              -- &              -- &              -- \\
15dtg & 20161208 &  $343.9$ &              -- &              -- &               -- &              -- &              -- &              -- &              -- \\
15dtg & 20151206 & $1331.2$ &              -- &              -- &   $7430 \pm 900$ &              -- &              -- &  $6890 \pm 100$ & $10310 \pm 150$ \\
15dtg & 20170817 & $2447.2$ &              -- &              -- &               -- &              -- &              -- &              -- &              -- \\
16flq & 20160828 &  $-17.8$ &              -- &              -- &               -- &              -- & $15140 \pm 830$ &              -- &  $9160 \pm 430$ \\
16flq & 20160915 &   $-1.1$ & $11150 \pm 650$ & $120.0 \pm 560$ &  $11230 \pm 900$ &              -- & $10480 \pm 740$ &              -- &  $8200 \pm 690$ \\
16flq & 20160921 &    $4.7$ &              -- & $11300 \pm 620$ &  $10160 \pm 600$ &              -- &  $9740 \pm 750$ &              -- &  $80.0 \pm 380$ \\
16flq & 20160929 &   $13.0$ &              -- &              -- &  $10870 \pm 770$ &              -- &  $9520 \pm 260$ &              -- &  $6560 \pm 370$ \\
16flq & 20161011 &   $23.5$ &  $8610 \pm 340$ &              -- &   $6180 \pm 470$ &              -- &  $9190 \pm 460$ &              -- &  $60.0 \pm 220$ \\
16flq & 20161123 &   $64.8$ &  $5940 \pm 710$ &              -- &   $4880 \pm 690$ &              -- &  $7610 \pm 330$ &              -- &  $6260 \pm 200$ \\
16flq & 20161031 &  $308.2$ & $7790 \pm 10.0$ &              -- &   $3960 \pm 600$ &              -- &  $7610 \pm 290$ &              -- &  $5880 \pm 310$ \\
16hgp & 20161022 &  $-25.2$ &              -- &              -- &               -- &              -- & $12700 \pm 860$ &              -- &              -- \\
16hgp & 20161102 &  $-15.0$ &              -- &              -- &               -- &              -- &   $100 \pm 420$ & $18530 \pm 100$ &  $9890 \pm 130$ \\
16hgp & 20161128 &    $9.2$ &  $6640 \pm 720$ &  $6810 \pm 290$ &   $6310 \pm 670$ &              -- &  $6810 \pm 840$ & $14800 \pm 190$ &  $7500 \pm 210$ \\
16hgp & 20161206 &   $17.1$ &  $6520 \pm 440$ &  $5950 \pm 720$ &   $5480 \pm 360$ &              -- &  $6870 \pm 230$ & $12180 \pm 180$ &   $7390 \pm 90$ \\
16hgp & 20170102 &   $41.4$ &  $6450 \pm 530$ &  $5430 \pm 360$ &  $4150 \pm 10.0$ &              -- &  $50.0 \pm 250$ &  $12410 \pm 60$ &   $6560 \pm 80$ \\
\end{tabular}}
\end{table*}

\begin{table*}
\centering
\begin{tabular}{llllll}
\toprule
Cut-off & $s_0$     & $\tau$ & $c$ &  Normalization \\
(mag)   & (stretch) &       &       & \\
\midrule
20.5 & 0.457 & 0.387 & 1.827 & 2.073 \\
21.0 & 0.377 & 0.380 & 1.456 & 1.650 \\
21.5 & 0.311 & 0.358 & 1.167 & 1.313 \\
\bottomrule
\end{tabular}
\centering
\caption{Lightcurve duration bias exponential fits with their normalizations $W_s(1.0)$ as described by Equation~\ref{eq:ws}. The weight $W_s$ obtained can be used to correct the number of (i)PTF SNe by their stretch to obtain the lightcurve duration bias corrected occurrence. Cut-off values are nominal limiting magnitudes for (i)PTF roughly corresponding to bright, grey, and dark time.}
\label{tab:lcd}
\end{table*}

\end{appendix}
\end{document}